\documentclass[pdftex,twocolumn,epjc3]{svjour3}

\pdfoutput=1 

\smartqed
\usepackage{graphicx}
\usepackage{lscape}
\usepackage{caption}
\usepackage{subcaption}
\usepackage{float}	
\usepackage{amssymb}
\usepackage{cite}
\usepackage[T1]{fontenc} 
\usepackage{color}
\usepackage{xcolor}
\usepackage{pagecolor,lipsum}
\usepackage{mdframed}
\usepackage{multirow}
\usepackage{amsmath}
\usepackage{textcomp}
\usepackage{hyperref}
\usepackage{arydshln}
\usepackage{enumerate}
\usepackage{mathrsfs}
\usepackage{verbatim}
\usepackage{slashed}
\usepackage{epsfig}
\usepackage{enumerate}
\usepackage{color}
\usepackage{wrapfig}
\usepackage{amsfonts}
\usepackage{multirow}
\usepackage{tikz}
\usepackage{amsmath}
 \usepackage[normalem]{ulem}

\usepackage{multicol}

 \usepackage[normalem]{ulem}

\newcommand{\ba}{\begin{array}}
\newcommand{\ea}{\end{array}}
\newcommand{\bd}{\begin{displaymath}}
\newcommand{\ed}{\end{displaymath}}
\newcommand{\bsube}{\begin{subequation}}
\newcommand{\esube}{\end{subequation}}
\newcommand{\bea}{\begin{eqnarray}}
\newcommand{\eea}{\end{eqnarray}}
\newcommand{\bal}{\begin{align}}
\newcommand{\ealign}{\end{align}}
\newcommand{\eal}{\end{align}}
\newcommand{\ben}{\begin{enumerate}}
\newcommand{\een}{\end{enumerate}}
\newcommand{\nn}{\nonumber}
\newcommand{\dis}{\displaystyle}

\newcommand{\gev}{\; {\rm GeV} }

\newcommand{\Slash}[1]{{\ooalign{\hfil/\hfil\crcr$#1$}}}

\begin{document}

\title{Twist-2 operators induced  Dark Matter Interactions}

\author{Hrishabh Bharadwaj\thanksref{e1,addr1,addr2} \and Sukanta Dutta\thanksref{e2,addr2}}
\thankstext{e1}{Corresponding~Author, \hfill\break \hspace*{3pt} e-mail: hrishabhphysics@gmail.com}
\thankstext{e2}{e-mail: Sukanta.Dutta@gmail.com}

\institute{\emph{Department of Physics \& Astrophysics, University of Delhi, New Delhi, India.}\label{addr1}  \and
\emph{SGTB Khalsa College, University of Delhi, New Delhi, India.}\label{addr2}}
\date{}
\maketitle

\abstract{We study the  effective interactions of the fermionic,  scalar and vector dark matter (DM)  with leptons and neutral electroweak gauge Bosons  induced by the higher dimensional effective  twist-2 tensor operators. We constrain these lepto-philic, $\tau^\pm$-philic and $U(1)_Y$ gauge Boson B-philic  effective interactions of DM  with the visible world from the WMAP and Planck data. The thermally averaged indirect DM pair annihilation cross-section and the spin-independent  DM - free and/ or bound electron  scatterinng cross-section  are observed to be consistent with the respective experimental data. Constraining  coefficients of the effective operators from the low energy LEP data for the DM  $\le$ 80  GeV,  we further study  their sensitivities in the pair production of 
such DM $\ge$ 50 GeV in association with di-jets  and  mono-photon respectively  at the proposed ILC. We perform the $\chi^2$ analysis to obtain the 99.73\% C.L. acceptance contours in the $m_{\rm DM}-\Lambda_{\rm eff}$ plane from the two dimensional differential distributions of the kinematic observables and find that ILC has rich potential to probe the contribution of such effective operators.}


\section{Introduction}
\label{intro}
\par It is imperative to determine
 the  nature of elusive {\it Dark Matter} (DM) \cite{Bertone:2004pz} candidates, which constitute roughly $\sim$ 23\%  of the energy density of the universe \cite{Bertone:2004pz, galaxy:RubinFord, Moustakas:2002iz, Milgrom:1983ca, Clowe:2006eq, vanUitert:2012bj}  and whose  predicted relic 
 density  is $\sim$ 0.119 \cite{Komatsu:2014ioa, Ade:2015xua}. The most popular proposition for DM theories are  weakly  interacting dark matter particles (WIMPs). Features of DM interactions can be determined  from the direct and indirect detection experiments apart from their direct searches  in the present \cite{Hong:2017avi, Kahlhoefer:2017dnp, Mitsou} and proposed colliders \cite{Dreiner:2012xm, Battaglia:2005ie, Rawat:2017fak}. The
 direct detection experiments like DAMA/ LIBRA   \cite{Bernabei:2013xsa, Bernabei:2018yyw}, CoGeNT
   \cite{Aalseth:2012if}, CRESST   \cite{Angloher:2016rji}, CDMS   \cite{Agnese:2013rvf},
 XENON100   \cite{Aprile:2016swn, Aprile:2017aty}, LUX   \cite{Akerib:2016vxi}
 and PandaX-II   \cite{Cui:2017nnn} are designed to measure the
 recoil momentum of scattered atom or nucleon by DM in the chemically inert medium of the detector, while the     indirect
 detection experiments such as FermiLAT   \cite{Ackermann:2015zua, TheFermi-LAT:2015kwa, Fermi-LAT:2016uux}, HESS   \cite{Abramowski:2013ax}, AMS-02   \cite{Aguilar:2014mma, Aguilar:2016kjl} {\it etc.} are looking for the evidences of the DM pair annihilation to Standard Model (SM)  particles such as photons, $e^+e^-$, $\mu^+\mu^-$, $\tau^+\tau^-$, $\bar b b$ pairs and {\it etc.} 

\par In last few years, many experiments like PAMELA \cite{Adriani:2013uda, Adriani:2008zr} have reported the excess in the positron  flux ( i.e., flux ratio of positron to sum of electron and positron ) without any significant excess in $\bar{p}$ channel (i.e., flux ratio of protons to anti-protons).   The peaks in $e^+ \, e^-$ channel are also observed in ATIC \cite{Panov:2006kf} and PPB-BETS \cite{PPB-BETS} balloon experiments  at around 1 TeV and 500 GeV respectively.  Recently, Dark Matter Particle Explorer (DAMPE) experiment \cite{Ambrosi:2017wek} has also observed a sharp peak around $\sim$ 1.4 TeV favoring the lepto-philic DM annihilation cross-section of the order of $10^{-26}cm^3/s$. The excess in $e^+ \, e^-$  can be either due to astrophysical
 events like high energy emission from the pulsars  or  resulting from DM pair annihilation in our galactic neighborhood preferably to  $e^+\, e^-$ channel.  Since the aforementioned experiments have not  observed any significant excess in anti-proton channel, the DM candidates, if any, appears  to be lepton friendly {\it lepto-philic} and   have  suppressed interaction with quarks at the tree level. 
 
 \par Various UV complete new physics extensions of SM have been proposed essentially to solve the gauge hierarchy problem in the {\it top-down} approach which  include theories like  extra-dimensions 
   \cite{Appelquist:2000nn}, super-symmetry   \cite{Wess:1974tw, Nilles:1983ge, PRoy}, little-Higgs   \cite{Arkani, Cheng}, extended 2-HDM models with singlets as portal of DM intearctions \cite{Dutta:2018hcz}   and {\it etc.}   These models naturally provide the DM candidates or WIMPs, whose  mass-scales are close to that of the electro-weak physics. However, the  Direct  detection  experiments  have shrunk the parameter space of  the  simplified  and popular models  where the   WIMPs are made to interact with the visible world {\it via} neutral scalars and/ or gauge Bosons.

 \par The model independent DM-SM particles interactions  have also been studied in the {\it bottom-up} Effective Lagrangian approach where the mediator of DM-SM  interactions are believed to be much heavier than the mass-scale of the lighter degrees of freedom say, in  our case SM and DM particles \cite{Hong:2017avi, Kahlhoefer:2017dnp, Mitsou, Dreiner:2012xm, Battaglia:2005ie, Rawat:2017fak}. The nature of these interactions are encapsulated in a set   of coefficients  corresponding to limited number of  Lorentz and gauge invariant   higher dimensional effective operators constructed with the light degrees of freedom. The constrained parameters (coefficients) space from various experimental data  then essentially maps  and direct towards the viable UV complete theoretical models. The generic effective Lagrangian for scalar, pseudo-scalar, vector, axial-vector interactions of SM particles with the scalar, vector, spin 1/2 and spin 3/2 dark matter candidates have been studied in literature \cite{Zheng:2010js, Freitas:2014jla, Savvidy:2012qa, Chang:2017dvm, Dutta:2017jfj, Khojali:2017tuv, Khojali:2016pvu}.

\par  Sensitivity analysis for DM-quark effective interactions at LHC have been performed \cite{Kahlhoefer:2017dnp, Mitsou,Boveia:2018yeb,CMS:2012bw,Aad:2014wra,Bell:2014tta,Bhattacherjee:2012ch} in a model-independent way for the dominant (a) mono-jet + $\slash\!\!\!\!E_{\rm T}$, (b) mono-$b$ jet + $\slash\!\!\!\!E_{\rm T}$ and (c) mono-$t$ jet + $\slash\!\!\!\!E_{\rm T}$ processes. Similarly, analysis for DM-gauge Boson effective couplings at LHC have been done by the authors in reference \cite{Cotta:2012nj, Chen:2013gya, Crivellin:2015wva}.  The sensitivity analysis of the coefficients  and detailed analysis of detection cuts flow strategy for lepto-philic operators have also been performed through $e^+e^-\to \gamma + \slash\!\!\!\!E_{\rm T}$ \cite{Chae:2012bq, Chen:2015tia, Fox:2011pm} and $e^+e^-\to Z^0 + \slash\!\!\!\!E_{\rm T}$ \cite{Bell:2012rg, Rawat:2017fak} channels.
  
\par Gross and Wilczek \cite{Gross:1974cs}  analyzed the second rank twist operators appearing in the operator-product expansion of two weak currents along with the renormalization-group equations of their coefficients for asymptotically free gauge theories in the context of deep inelastic lepton-hadron scattering.  Later authors of reference \cite{Drees:1993bu} analysed the effective DM-nucleon scattering  induced by twist-2 quark operators in the context of super-symmetric models where the Majorana  DM particle neutralino being LSP   was assumed to be much lighter than that of the squark masses. This was followed by series of papers \cite{Hisano:2010ct, Hisano:2010yh, Hisano:2011cs, Hisano:2011um, Hisano:2012wm, Hisano:2015bma, Hisano:2015rsa} where the authors have calculated the one loop effect of DM-nucleon  scattering induced by  the twist-2 quarks and gluonic operators for the  fermion, vector and scalar DM respectively. The hadronic matrix elements induced by twist-2 operators can however be identified with the  second moment of parton distribution functions and  thus can be constrained from the available {\it pdfs}. This in turn  constrain  the coefficients of such higher dimensional  effective operators  to estimate the DM-nucleon scattering cross-sections for a suggested DM and squark mass range.
 
In this paper, we undertake the  analysis for spin 1/2, 0 and 1 DM interactions induced by {\it lepto-philic} and {\it $U(1)_Y$ gauge Boson B-philic} effective twist-2  higher dimensional gauge invariant operators. This article is organized as follows:  we formulate the effective interaction Lagrangian for  fermionic, scalar and vector   DM with SM leptons and neutral electro-weak gauge Bosons {\it via} twist-2 operators in section \ref{sec:eff_int}.  In section \ref{sec:DMConstraints}, we  constrain the coefficients of the effective Lagrangian
 from predicted relic density and perform a consistency check {\it w.r.t.} indirect and direct detection experiments. The constraints from the LEP  on the coefficients of the effective lepto-philic and $U(1)_Y$ gauge Boson B-philic operators and the sensitivity analysis of these coefficients at the proposed ILC are discussed in section \ref{sec:collider}. Finally we summarize in section \ref{sec:summary}.
\section{Effective interactions of lepto-philic \& $U(1)_Y$ gauge Boson B-philic DM}
\label{sec:eff_int}
We initiate the construction of the effective operators by writing the contact interaction between any SM leptons $\psi$  and fermionic DM $\chi$ (Dirac or Majorana) with masses $m_\psi$ and $m_\chi$ respectively, assuming that the mediator  mass scale, if any, should be of the order of the cut-off of the effective theory ($\sim \Lambda$), which in general is much heavier than the masses of the SM and DM fields. These contact interactions for example can be motivated from the  super-symmetric  models where spin-independent neutralino-lepton interactions are facilitated by the exchange of heavy sleptons and/ or Higgses, assuming Majorana neutralino to be the LSP. For an illustration of such contact interactions, we write twist-2 Type-1 and Type-2 Lagrangians with the coupling strengths $\alpha_1$ and $\alpha_2$ respectively  as:
\bea
\mathcal{L}_1&=&\frac{1}{m_\chi}\,\frac{\alpha_1}{\Lambda^3} \,\left(\bar{\chi}\gamma^\mu\partial^\nu\chi \right)\left(\bar{\psi}\gamma_\mu
\partial_\nu \psi-\partial_\nu\bar{\psi}\gamma_\mu \psi\right) \,\,\,\,{\rm and} \label{Lfermion1}\\
\mathcal{L}_2&=&\frac{1}{m_\chi^2}\,\frac{\alpha_2}{\Lambda^3}\,\left( \bar{\chi}\partial^\mu\partial^\nu\chi\right) \left(\bar{\psi}\gamma_\mu
\partial_\nu \psi-\partial_\nu\bar{\psi}\gamma_\mu \psi\right) \label{Lfermion2}
\eea
Using the equations of motion for massive Dirac fermions   along-with the Tensor identity
\bea
P_{\mu\nu}Q^{\mu\nu}=(P_{\mu\nu}-\frac{1}{4}g_{\mu\nu}P^\alpha_\alpha)Q^{\mu\nu}\,+\,\frac{1}{4}P^\alpha_\alpha Q^\beta_\beta
\eea
 the Lagrangians given in equations \eqref{Lfermion1} and \eqref{Lfermion2} can be re-written as
\bea
\mathcal{L}_1^\prime&=&\frac{1}{m_\chi}\,\frac{\alpha_1}{\Lambda^3}\, \left( -2\,i\,\mathcal{O}^{(2)}_{\mu\nu}\,\bar{\chi}\gamma^\mu\partial^\nu\chi - \frac{1}{2} \, m_\psi \, m_\chi \bar{\psi} \psi \bar{\chi} \chi\right) \,\,\,\,\nn\\
\label{lagfer1}\\
\mathcal{L}_2^\prime&=&\frac{1}{m_\chi^2}\,\frac{\alpha_2}{\Lambda^3}\, \left( -2\,i\,\mathcal{O}^{(2)}_{\mu\nu}\,\bar{\chi}\partial^\mu\partial^\nu\chi - \frac{1}{2} \, m_\psi \, m_\chi \bar{\psi} \psi \bar{\chi} \chi\right) \nn\\
\label{lagfer2}
\eea
respectively where  $\mathcal{O}^{(2)}_{\mu\nu}$ is defined as trace-less twist-2  operator  
\bea
\mathcal{O}^{(2)}_{\mu\nu}=\frac{i}{2} \left( \bar{\psi}\gamma_\mu\partial_\nu\psi+\bar{\psi}\gamma_\nu\partial_\mu\psi-\frac{g_{\mu\nu}}{2}\bar{\psi}\gamma^\alpha\partial_\alpha\psi \right)
\label{Omunu}
\eea
and  $ \bar{\psi} \psi \bar{\chi} \chi$ appearing in the Eq.\eqref{Omunu} is the scalar operator. The constraints on the four fermionic scalar DM operators has been extensively studied  in the literature \cite{Rawat:2017fak, Zheng:2010js}. However, the contributions of the tensor operators in equations \eqref{lagfer1} and \eqref{lagfer2} respectively are higher than those of the scalar operators due  to the momentum dependence and therefore the coupling constants $\alpha_1$ and $\alpha_2$ would be comparatively more severely  constrained for a given choice Cut-Off. Therefore it is worthwhile to probe the effect of these tensor operators in the DM pair annihilation and DM-electron or DM-Nucleon scatterings.

\par The twist-2 interactions of fermionic DM can also be realized  through the twist-2 operators constructed out of SM charged and neutral electro-weak vector gauge Bosons ${\cal O}^{V^{\rm EW}}_{\mu\nu}\equiv 
{V^{\rm EW}}^{\rho}_{\mu} {V^{\rm EW}}_{\nu\rho}-\frac{1}{4}g_{\mu\nu}^{}
{V^{\rm EW}}_{\rho\sigma}{V^{\rm EW}}^{\rho\sigma}$ and SM scalar Higgs Boson. Further the formalism can be extended to include the twist-2 Type-2 interactions of spin 0 and spin 1 DM particles with SM charged leptons and electro-weak gauge Bosons using equation \eqref{lagfer2}. We analyse all such operators emerging from twist-2 interactions with SM charged leptons and neutral electro-weak gauge Bosons  in this study. 

 There are other lepto-philic DM tensor operators which can have significant effect on DM phenomenology like dipole moments {\it etc.} as shown in references \cite{Ibarra:2015fqa,Herrero-Garcia:2018koq,Hisano:2018bpz}. Since we are interested in operators which  contribute  to the spin independent non-relativistic DM - nucleon scattering process,  we drop all the operators which are suppressed by the velocity of DM and/or nucleon. On using the equations of motions we expand all the interaction terms in the Lagrangian  in powers of the super-weak coupling constant $\alpha_{T_i}$.  The  spin-independent leading interactions in super-weak coupling expansion of the fermionic, vector and scalar DM operators are retained for analysis as explicitly shown in references \cite{Hisano:2010ct, Hisano:2010yh, Hisano:2011cs, Hisano:2011um, Hisano:2012wm, Hisano:2015bma, Hisano:2015rsa} for the case of fermionic, vector and scalar DM operators interacting with quarks and gluons. 

\par  We  enlist   a minimal set   of relevant    twist-2   lepto-philic and $U(1)_Y$ gauge Boson B-philic operators inducing contact interactions  with Dirac fermion $\chi$, real scalar $\phi^0$ and  real vector $V^0_\mu$ DM candidates in the following Lagrangian:
\begin{subequations}
\begin{eqnarray}
 {\cal L}_{\rm eff.\, Int.}^{\rm DM} &=& {\cal L}_{\rm eff.\, Int.}^{\rm spin \,1/2 \,DM} + {\cal L}_{\rm eff.\, Int.}^{\rm spin \,0 \,DM} +
{\cal L}_{\rm eff.\, Int.}^{\rm spin \,1 \,DM} \nonumber\\	
	=\sum_{p= l, B} &&\left[\sum_{T_i \in \rm T_1,T_2} \,\frac{\alpha_{T_i}}{\Lambda^3} \, {\cal O}_{T_i}^p+\frac{\alpha_{S}}{\Lambda^2} \, {\cal O}_{S}^p  +\frac{\alpha_{V}}{\Lambda^2} \, {\cal O}_{V}^p
\right]\nonumber\\
	= \sum_{p= l, B} &&\left[ \sum_{T_i \in \rm T_1,T_2} \frac{{\lambda^\chi_{T_i}}^2}{\Lambda_{\rm  eff}^3} {\cal O}_{T_i}^p + \frac{{\lambda^{\phi^0}_{T_2}}^2}{\Lambda_{\rm  eff}^2} \, {\cal O}_{S}^p + \frac{{\lambda^{V^0}_{T_2}}^2}{\Lambda_{\rm  eff}^2}  {\cal O}_{V}^p\right]\,\,\,\nn\\
	{\rm where,}\\
	\mathcal{O}_{T_1}^{p}&=& \dis \frac{1}{m_\chi}\bar{\chi} i\partial^{\mu} \gamma^{\nu}\chi\,\mathcal{O}_{\mu\nu}^{p}\label{twistT1} \\
	\mathcal{O}_{T_2}^{p}&=& \dis \frac{1}{m_\chi^2}\bar{\chi} i\partial^{\mu} i\partial^{\nu}\chi\, \mathcal{O}_{\mu\nu}^{p}\label{twistT2}\\
\mathcal{O}_{S}^{p}&=& \frac{1}{m_{\phi^0}^2}\,{\phi^0}\,i\partial^\mu\,i\partial^\nu \, {\phi^0} \,\mathcal{O}_{\mu\nu}^{p}  \\
\mathcal{O}_{V}^{p}&=& \frac{1}{m_{V^0}^2}\,{V^0}^{\rho}\,i\partial^\mu\,i\partial^\nu \, {V^0}_{\rho} \,\mathcal{O}_{\mu\nu}^{p}
\end{eqnarray}
\label{MainLag}
\end{subequations}
The twist-2  operators ${\cal O}^{l^\pm}_{\mu\nu}$ for charged leptons   $l^\pm\equiv e^\pm,\,\mu^\pm,\tau^\pm $ and ${\cal O}^B_{\mu\nu}$ for $U(1)_Y$ gauge field $B^\mu$   are defined in terms of the covariant derivative $D_\mu\equiv \partial_\mu +i g^\prime B_\mu$ as   
\begin{align}\label{twist2op}
 {\cal O}^{l^\pm}_{\mu\nu}&\equiv i\,\frac{1}{2}\overline{l}\biggl(
D_\mu^{}\gamma_\nu^{} +D_\nu^{}\gamma_\mu^{}-\frac{1}{2}g_{\mu\nu}^{}
\Slash{D}\biggr)l~, \\
{\cal O}^B_{\mu\nu}&\equiv 
B^{\rho}_{\mu} B_{\nu\rho}-\frac{1}{4}g_{\mu\nu}^{}
B_{\rho\sigma}B^{\rho\sigma}~,
\end{align}
The cut-off scale of the effective theory $\Lambda_{\rm eff}$ is defined as  $(4\pi)^{1/3}\Lambda$ and and $\sqrt{4\pi}\Lambda$  for  fermionic and Bosonic DM respectively. $\alpha_{T_i}$ and $\lambda_{T_i}$ are the strengths 
and couplings of the interactions respectively, where 
$\left\vert \lambda_{T_i}\right\vert \le \sqrt{4\pi}$ and $\Lambda \gtrsim$ 1 TeV.

\par The Lorentz structure of the DM operators characterize the nature of DM pair annihilation and hence its contribution  to the relic density \cite{Kumar:2013iva}.  Annihilation due to all operators given in equations \eqref{MainLag} except fermionic Type 2 \eqref{twistT2} are found to be contributing to both $s$-wave and $p$-wave  partial amplitudes, while  Type 2 induced DM operator contributes to only  $p$-wave amplitudes and hence comparatively suppressed.  It is to be noted that the partial wave analysis of the annihilation processes induced by fermionic DM operators  given in \eqref{twistT1} and \eqref{twistT2} remain same for both the Majorana type and/ or Dirac type because of the contact effective interactions.

\par We are now equipped to analyse and constrain these effective interactions from the DM phenomenology.

\begin{figure*}
	\centering
\begin{multicols}{2}	
		
	\includegraphics[width=0.49\textwidth,clip]{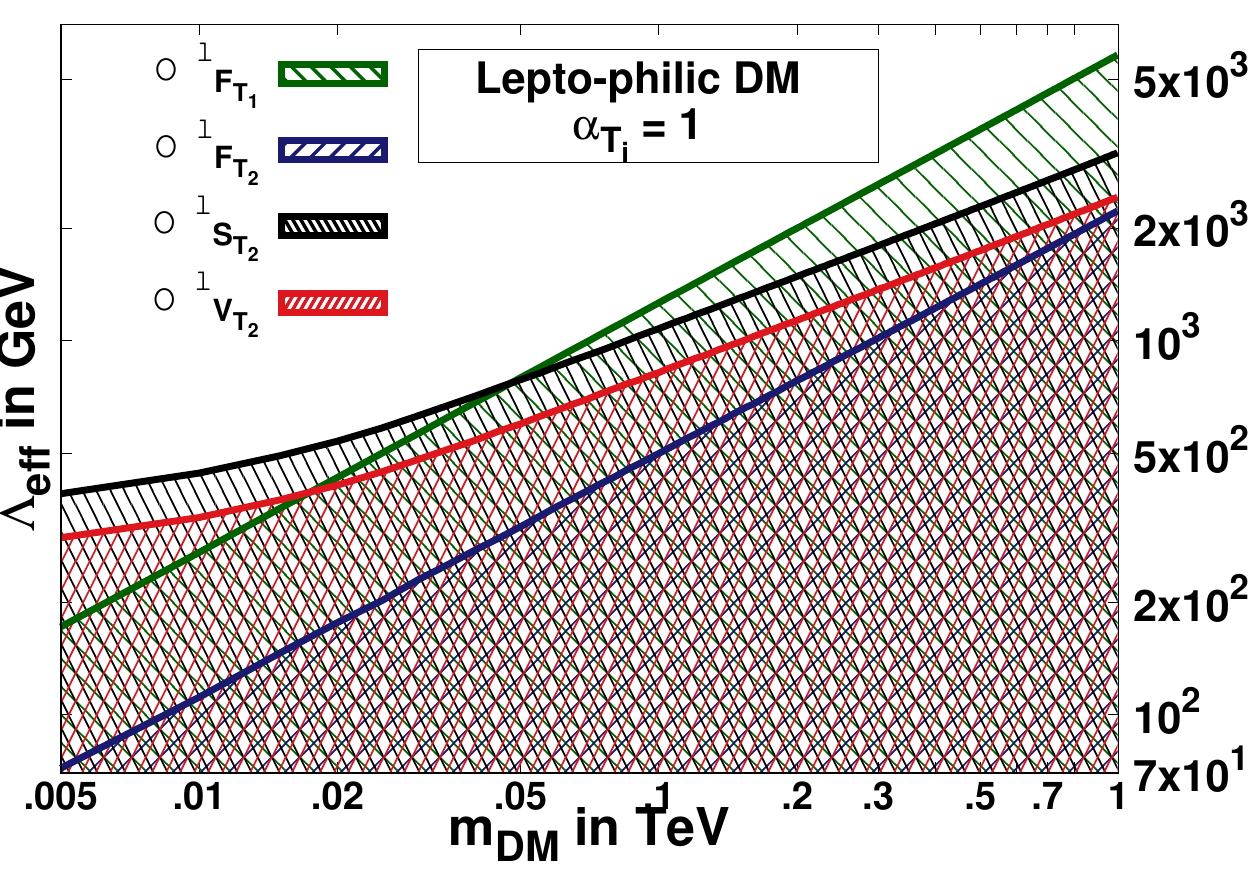} 
	\subcaption{\small \em{lepto-philic DM}}
		\label{leptophilicrelicdensity} 
\columnbreak	
	\includegraphics[width=0.49\textwidth,clip]{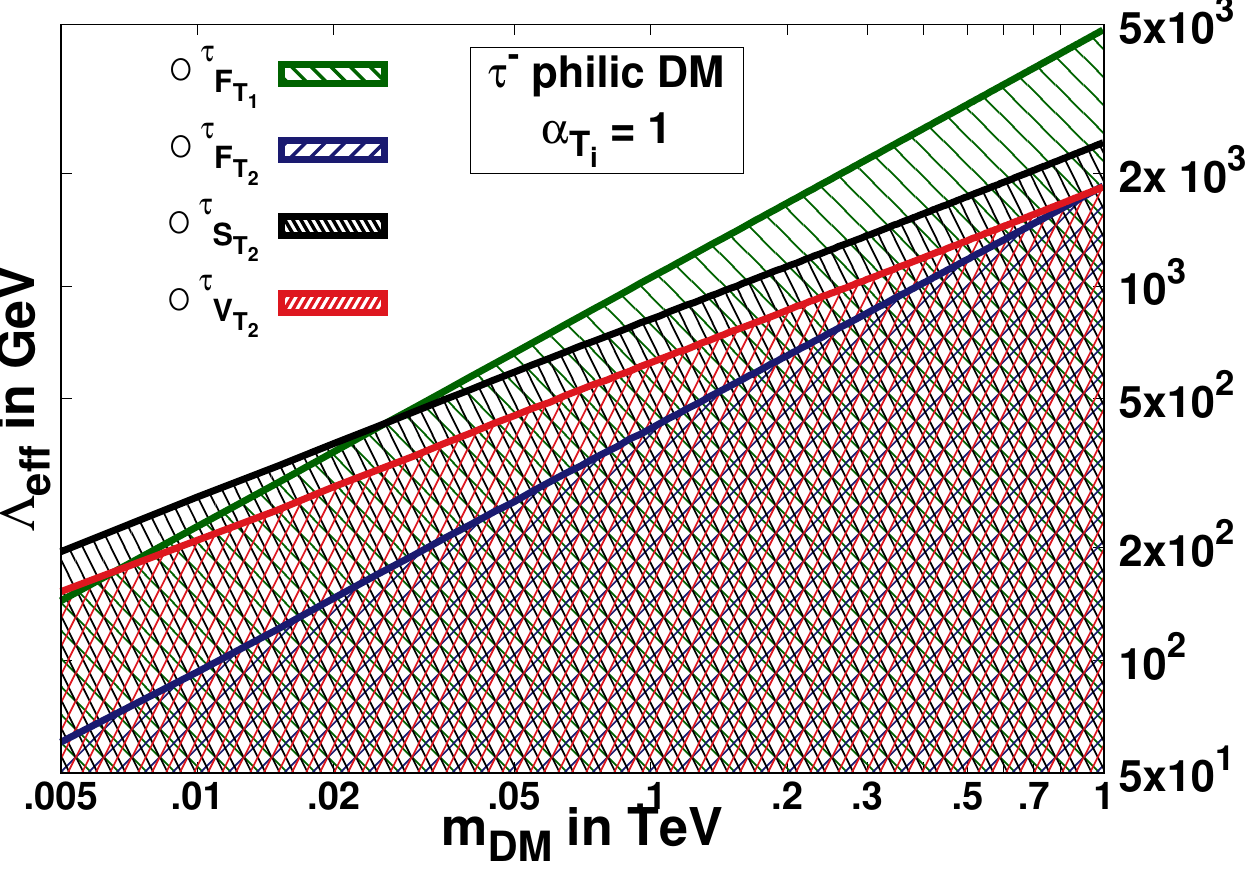}
	\subcaption{\small \em{$\tau^\pm $-philic DM}}
        \label{tauphilicrelicdensity}
\end{multicols}
\begin{center}	
\includegraphics[width=0.5\textwidth,clip]{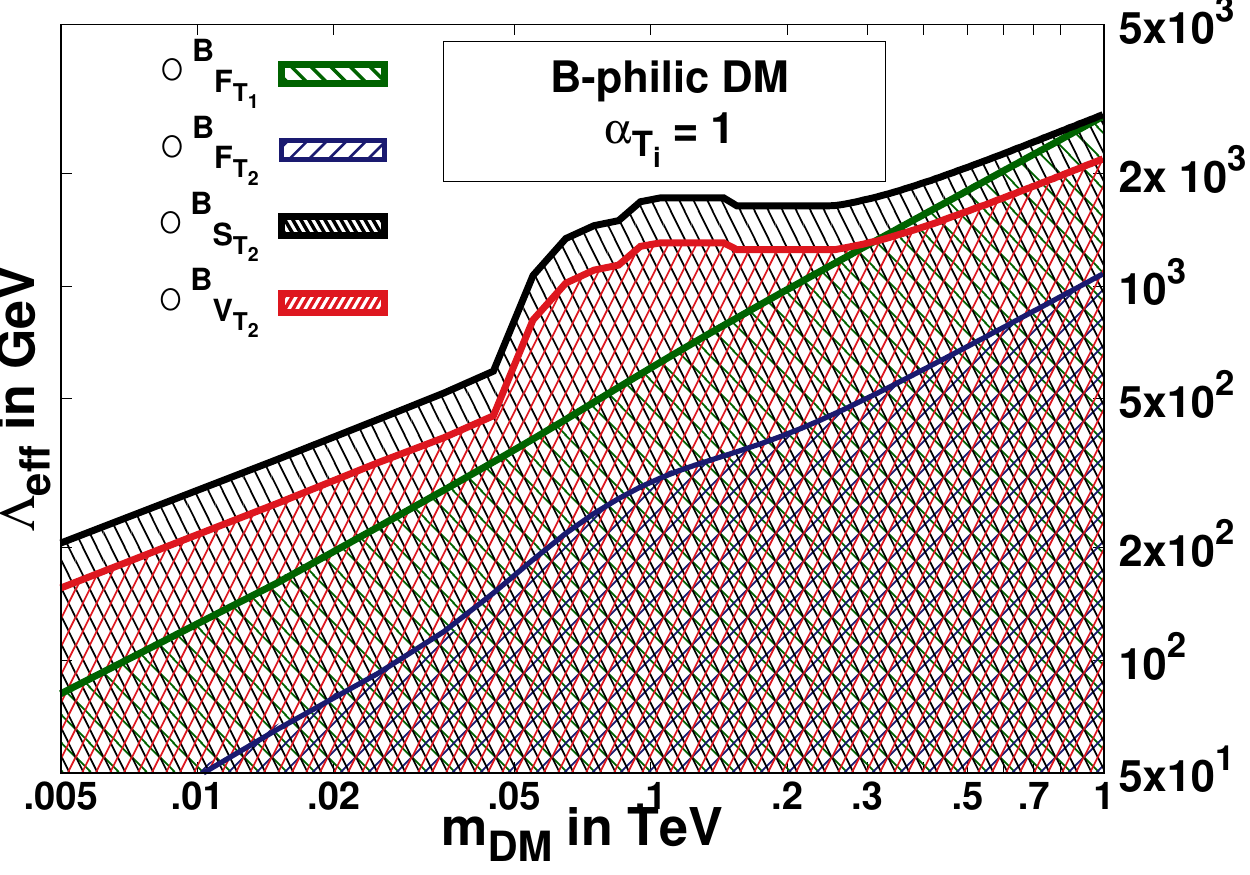}
\subcaption{\small \em{B-philic DM}}
\label{Bphilicrelicdensity}
\end{center}

	\caption{\small \em {Relic density contours are drawn using MadDM v.3.0   \cite{Ambrogi:2018jqj} satisfying $\Omega_{\rm DM}h^2$ = $0.1199 \pm 0.0022$   \cite{Ade:2015xua} in the plane defined by DM mass  and the respective  upper bound on cut-off corresponding to the Type-1 or 2 twist operators operators at fixed coupling $\alpha_{T_i}$ = 1. The  respective shaded regions depict the  allowed parameter space. Contours in  figure \ref{leptophilicrelicdensity} are depicted   assuming the universal lepton flavor couplings of effective  DM - SM lepton interactions.}}
\end{figure*}
\section{DM Phenomenology}
\label{sec:DMConstraints}
\subsection{Constraints from Relic Density}
\label{subsec:relic}
In this sub-section we discuss the
 relic abundance of {\it lepto-philic and $U(1)_Y$  gauge Boson-philic} DM and constrain the effective interactions from the predicted  DM relic density  ${\mathbf \Omega_{\rm DM}\mathbf h^2}$ of  $0.1138\pm .0045$ and $0.1199\pm 0.0022$  by WMAP   \cite{Komatsu:2014ioa} and Planck   \cite{Ade:2015xua} collaborations respectively. 
The current relic density for dark matter  is calculated  from the DM number density $n_{\rm DM}(t)$ by solving the Boltzmann equation 
\begin{equation}\small
  \frac{d}{dt}n_{\rm DM}(t)\,+\, 3\,{\mathbf H}(t)\,n_{\rm DM}(t)=-\left\langle\sigma_{\rm ann}\left\vert \vec v\right\vert\right\rangle
 [n_{\rm DM}(t)^2-n_{\rm DM}^{\rm eq}(t)^2]
\label{eqn:boltzeqn1}
\end{equation}
where ${\mathbf H}$ is the Hubble parameter, $\left\langle\sigma_{\rm ann}\left\vert \vec v\right\vert\right\rangle$
 is the thermal average of annihilation cross section multiplied by the relative
 velocity of dark matter pair. $n_{\rm DM}^{\rm eq}$ represents the dark matter number density
 in thermal equilibrium and is given by
\begin{equation}
n_{\rm DM}^{\rm eq}=\left[{\mathbf g}\left(\frac{m_{\rm DM} \,T}{2\pi}\right)^{\frac{3}{2}}\exp\left\{\frac{-m_{\rm DM}}{T}\right\}\right]^{1/2} \end{equation} where ${\mathbf g}$ is the degrees of freedom. 
To numerically compute the current DM relic density we need to calculate the thermally averaged
 DM annihilation cross sections. Since the freeze-out for thermal relics occurs when the  massive particle is non-relativistic {\it i.e.} $\left\vert\vec v\right\vert <<c$, we  make an expansion in $\left\vert \vec v\right\vert/ c$  and then $\langle\sigma\left\vert \vec  v
\right\vert\rangle$ can be approximated as $\langle\sigma\left\vert \vec  v\right\vert\rangle = a + b\left \vert\vec v\right\vert^2 + \mathcal{O}(\left\vert\vec v\right\vert^4)$.
\par Defining  the dark matter relic abundance as a ratio of  the thermal relic density $\rho_{\rm DM}$ and critical density of the universe $\rho_c=1.05373 \times 10^{-5}{\mathbf h^2}$  $\gev/(c^2 cm^3)$, where ${\mathbf h}$ is the dimensionless Hubble parameter and solving the  Boltzmann equation for the  thermal relic density  we  get
\bea
\Omega_{\rm DM}  {\mathbf h}^2&=&\frac{\pi\,\sqrt{{\mathbf g_{\rm eff}}(x_F)}}{\sqrt{90}}\frac{x_F\,T_0^3\,\mathbf g_0}{M_{Pl}\,\rho_c\,\langle\sigma_{ann} v\rangle\, {\mathbf g_{\rm eff}}(x_F)}\nonumber\\
&\approx& 0.12 \, \frac{x_F}{28} \, \frac{\sqrt{\mathbf g_{\rm eff}(x_F)}}{10}\, \frac{2\times10^{-26} cm^3/s}{\langle\sigma\left\vert \vec  v\right\vert\rangle}
\label{boltzmann7}
\eea

where $M_{Pl}$ and ${\mathbf g_{\rm eff}}$ are Planck mass and effective number of degrees of freedom near the freeze-out temperature $T_F=\frac{m_{\rm DM}}{x_F}$, where $x_F$ is given by
\bea
x_F&=&\log \left[c^\prime\,\left(c^\prime+2\right)\, \sqrt{\frac{45}{8}}\,\frac{\mathbf g_0 \,M_{Pl}\, m_{
\rm DM}\,\langle\sigma\left\vert \vec  v\right\vert\rangle}{2\,\pi^3\, \sqrt{x_F\, \mathbf g_{\rm eff}(x_F)}} \right]
\eea
\noindent where $c^\prime$ is a parameter of the order of one.
\par We have computed the thermal-averaged annihilation cross-section in the appendix for the Dirac fermion, a real scalar and real vector DM candidates. They are worked out in \ref{thermalAveragedCrosssection}. For $\tau^\pm$-philic (electro-philic) case,  all DM - leptons couplings except that of DM - $\tau^\pm$ ($e^\pm$)  identically vanish.

	\par  To compute relic density numerically, we have used MadDM \cite{Ambrogi:2018jqj} and MadGraph \cite{Alwall:2014hca}. We have generated the input model file required by  MadGraph using FeynRules \cite{Alloul:2013bka}, which calculates all the required couplings and Feynman rules by using the full Lagrangian given in equation \eqref{MainLag}. We scan over the DM mass range 10 - 1000 GeV whose relic density satisfy $\Omega_{\rm DM}\mathbf h^2 \le  $ 0.1199 \cite{Ade:2015xua}  for all vanishing interactions except one  non-zero coupling $\alpha_{T_i}$ fixed at unity. We study and plot the relic density contours in the plane defined by DM mass and the cut-off corresponding to four (two fermionic, one scalar and one Vector DM) lepto-philic, $\tau^\pm$-philic and $B$-philic operators in figures \ref{leptophilicrelicdensity}, \ref{tauphilicrelicdensity} and \ref{Bphilicrelicdensity} respectively. We observe a bump around $m_{\rm DM} \simeq$ 45-90 GeV in figure \ref{Bphilicrelicdensity}    due to opening up of $Z\ \gamma$  and $Z\ Z$ annihilation channels induced by $B$ philic DM operators. The increase in the total annihilation cross-section decreases  the number density of the DM and as a result for a fixed DM relic density and coupling constant the contour shows a bump in the $m_{\rm DM}$-$\Lambda$ plane. All points lying on the solid lines  in figure \ref{fig:relicdensity} satisfy $\Omega_{\rm DM}\mathbf h^2 \le  $ 0.1199 \cite{Ade:2015xua}. These points are also the allowed upper limit on the cut-off for a given DM mass and thus the shaded region enclosed by the corresponding solid line is the cosmologically allowed parameter region of the respective operator.  The lower value of the cut-off corresponding to the same DM mass  will lead to only partial contribution to relic density and therefore may survive in model where (a) more than one type of DM particles are allowed and/ or (b) switching more than one type of effective operators  simultaneously. We observe that the sensitivity of the cut-off for fermionic Type-1  operator increases with the varying DM mass. Among the Type-2 operators, scalar DM cut-off is found to be the most sensitive.  It is important to mention that  the relic density contours for a electro-philic DM will be almost same to that of the $\tau^{\pm}$-philic case.
\begin{figure*}
\centering
\begin{multicols}{2}
\includegraphics[width=0.49\textwidth,clip]{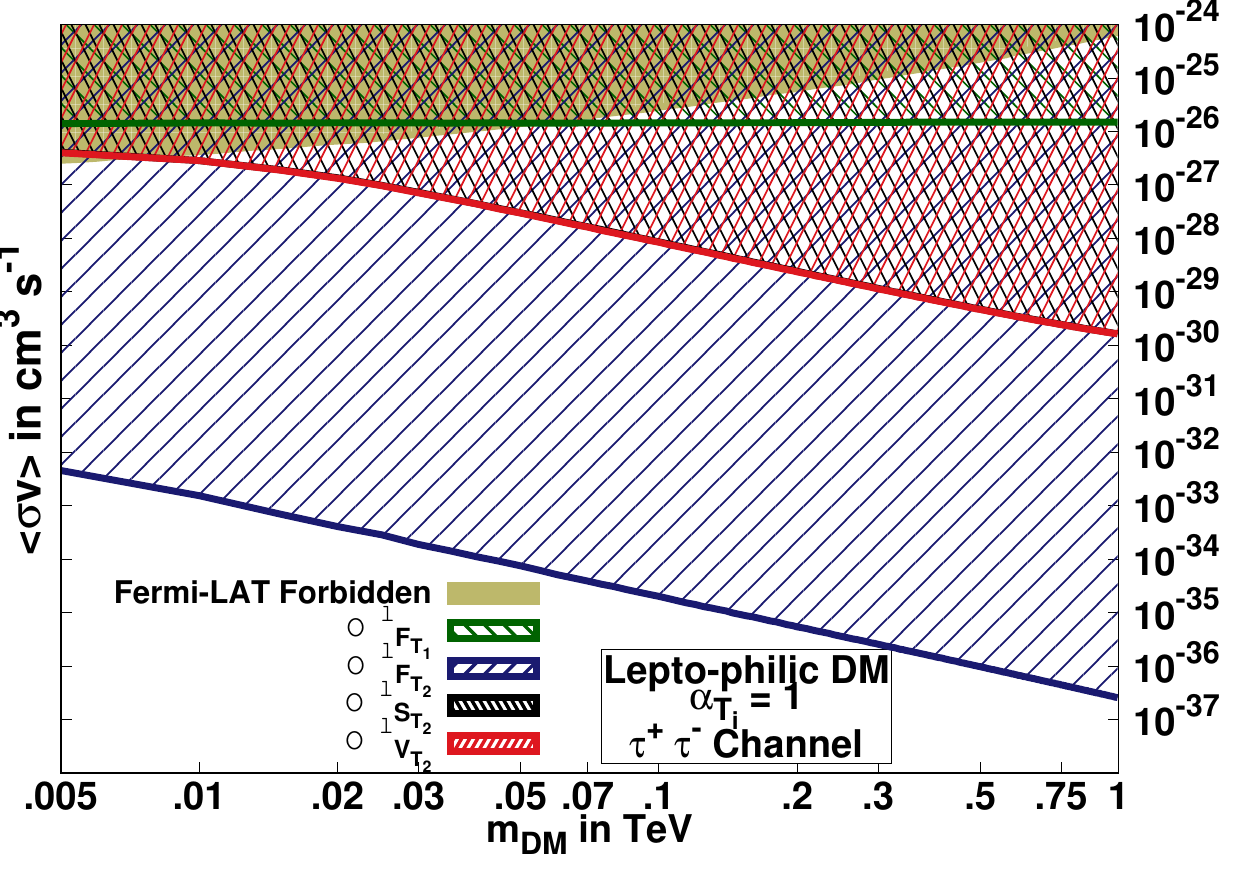}
\subcaption{\small \em{Lepto-philic DM annihilation to $\tau^+\tau^-$}}\label{leptoindirect}
\includegraphics[width=0.49\textwidth,clip]{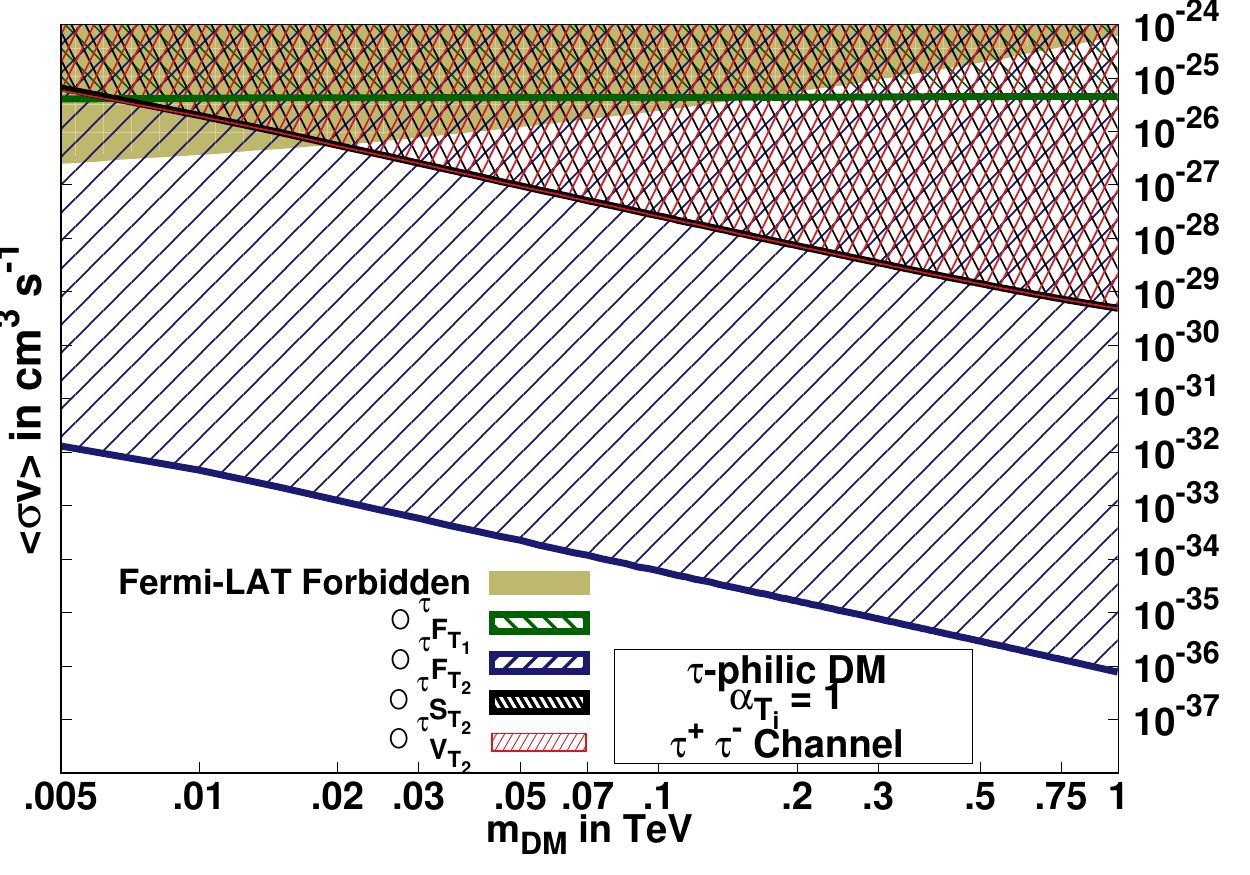}
\subcaption{\small \em{$\tau^\pm$-philic DM DM annihilation to $\tau^+\tau^-$}}\label{tauindirect}
\end{multicols}
\begin{center}
\includegraphics[width=0.49\textwidth,clip]{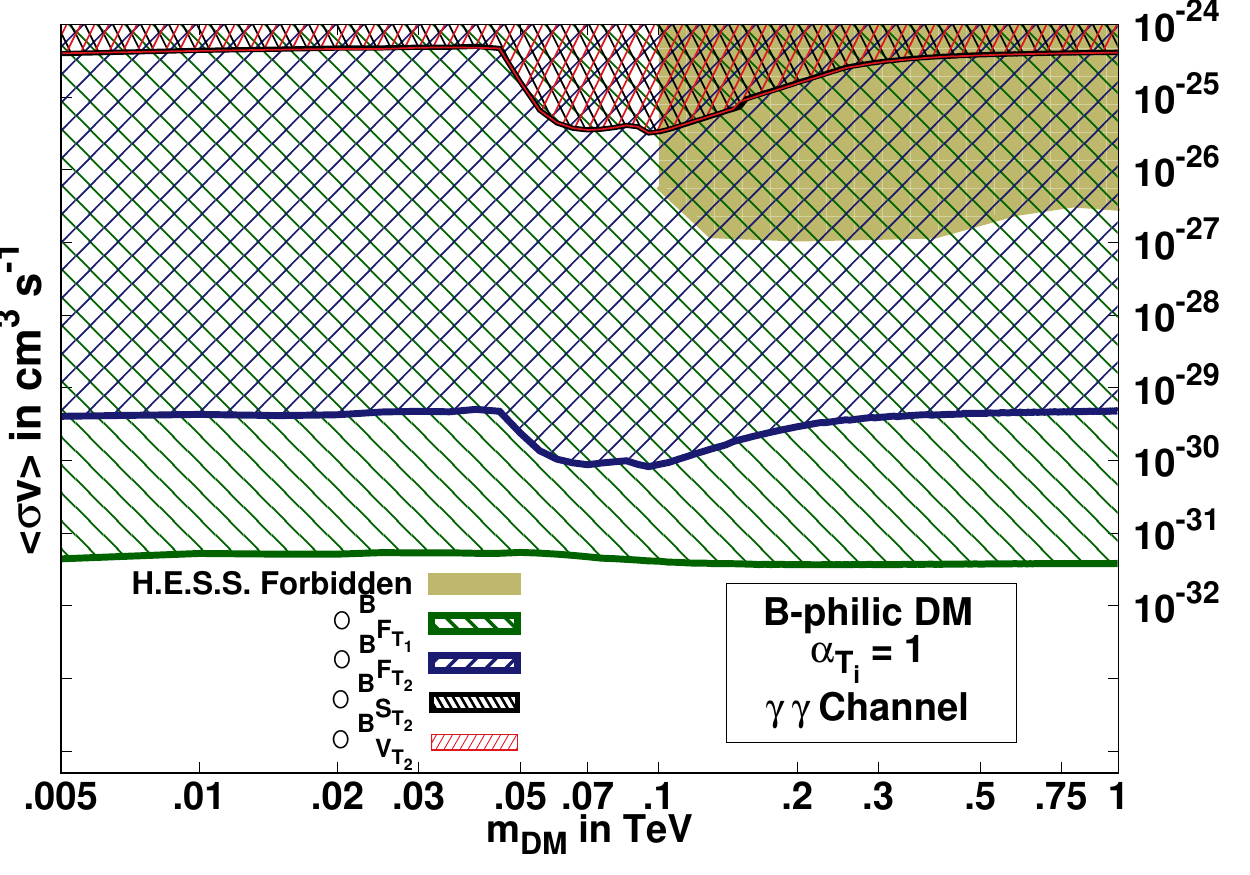}
\subcaption{\small \em{B-philic DM annihilation to $\gamma\,\gamma$}}\label{Bindirect}
\end{center}
  \caption{\small \em{Solid lines in all figures depict the variation of thermally averaged DM annihilation
      cross-sections {\it w.r.t.} the DM mass  at the fixed coupling $\alpha_{T_i}$ = 1 and  the  respective upper bound on the  cut-off  satisfying the relic density contribution. We plot the median of the
      DM annihilation cross section derived from a combined analysis
      of the nominal target sample for the $\tau^+\tau^-$ channel \cite{Fermi-LAT:2016uux} in figures \ref{leptoindirect}, \ref{tauindirect} and $\gamma\gamma$ \cite{Abramowski:2013ax} channel in figure \ref{Bindirect}  assuming a 100\%
branching fraction and thus restrict the respective allowed shaded region from above. Figure \ref{leptoindirect} shows the contours for the fermionic, scalar and vector DM interacting {\it via} Type-1 or 2 twist operators assuming the universal lepton flavor couplings with the DM. }}
\label{fig:indirectdetection} 
\end{figure*}
\subsection{Indirect Detection}
\label{subsec:InD}
Since the DM annihilation rate is
 proportional to the square of DM density,  therefore DM annihilation is likely to be propelled  in the over-dense region of the universe such as galactic center, dwarf spheroids and sun  generate high  flux of energetic light SM particles like charged hadrons, jets, the charged leptons {\it e.g.} electron, positron and  photon. Since the non-relativistic DM particles are colliding at rest with each other, the  energy of gamma-rays and the produced charged lepton is of the order of $ \sim m_{\rm DM}$. Indirect experiments which in general are either ground-based or satellite borne particle detectors  are sensitive to these characteristic fluxes of light SM  particles.  For example  FermiLAT (Large Area Telescope) is a space borne experiment designed to measure the tracks of electron-positron pairs which are produced when gamma-rays interact with the detector
 material (thin and high-Z foil) \cite{Ackermann:2015zua, TheFermi-LAT:2015kwa, Fermi-LAT:2016uux}, while HESS is the ground based cherenkov telescope geared to detect the gamma ray spectrum \cite{Abramowski:2013ax}.

\par In this section we calculate the thermally averaged 
 annihilation cross sections for fermionic, real scalar  and real vector Boson DM candidates to pair of charged leptons and photons. The analytical expressions for these cross-sections corresponding to the lepto-philic and $U(1)_Y$ gauge Boson B-philic operators are given in equations \eqref{ThAvLannxsecT1}-\eqref{ThAvLannxsecV} and  \eqref{ThAvBannxsecT1}-\eqref{ThAvBannxsecV} respectively. We have used 220 Km/s (average rotational velocity of galaxy) as the average velocity of the DM.

\par The thermal averaged DM annihilation cross-section is computed numerically for a given set of parameter $\left(m_{\rm DM},\, \alpha_{T_i}, \,\Lambda_{\rm  eff}\right)$ which satisfy the
relic density constraint from the Planck data as depicted in figure \ref{fig:relicdensity}. For the lepto-philic and $\tau^\pm$-philic operators we compute the dominant thermal averaged annihilation cross-section to $\tau^+\tau^-$ pair using equations \eqref{ThAvLannxsecT1}, \eqref{ThAvLannxsecT2}, \eqref{ThAvLannxsecS} and \eqref{ThAvLannxsecV} corresponding to  the fermionic DM Type-1 and Type-2,  scalar DM Type-2 and vector DM Type-2 induced operators. The  variation of the annihilation cross-section with the DM mass are depicted in figures \ref{leptoindirect} and \ref{tauindirect} respectively.  The solid lines in figures \ref{leptoindirect} and \ref{tauindirect}  are essentially the lower bound on the allowed annihilation cross-section satisfying the relic density constraints for a given DM mass. It is interesting to note that although the analytical expression for  the scalar and vector DM annihilation cross-section are not same but still we observe a complete overlap of the scalar and vector DM solid lines for $\alpha_{T_2}^{V^0}=\alpha_{T_2}^{\phi^0}$ which is an artifact of the two distinct  values of the respective effective cut-off that satisfy the same relic density for a given DM mass. These results are compared  with the upper bound on the allowed annihilation cross-section in $ \tau^+\,\tau^-$ channel obtained from the FermiLAT data \cite{Fermi-LAT:2016uux}. Thus the null experimental results   for the given mass range translate into the lower limits on the cut-off for the respective operators. 

 The DM annihilation cross-section to the pair of electrons induced by the electro-philic DM are identical to those depicted in figure \ref{tauindirect} for the respective operators.

\begin{figure}
\centering
       \includegraphics[width=0.5\textwidth,clip]{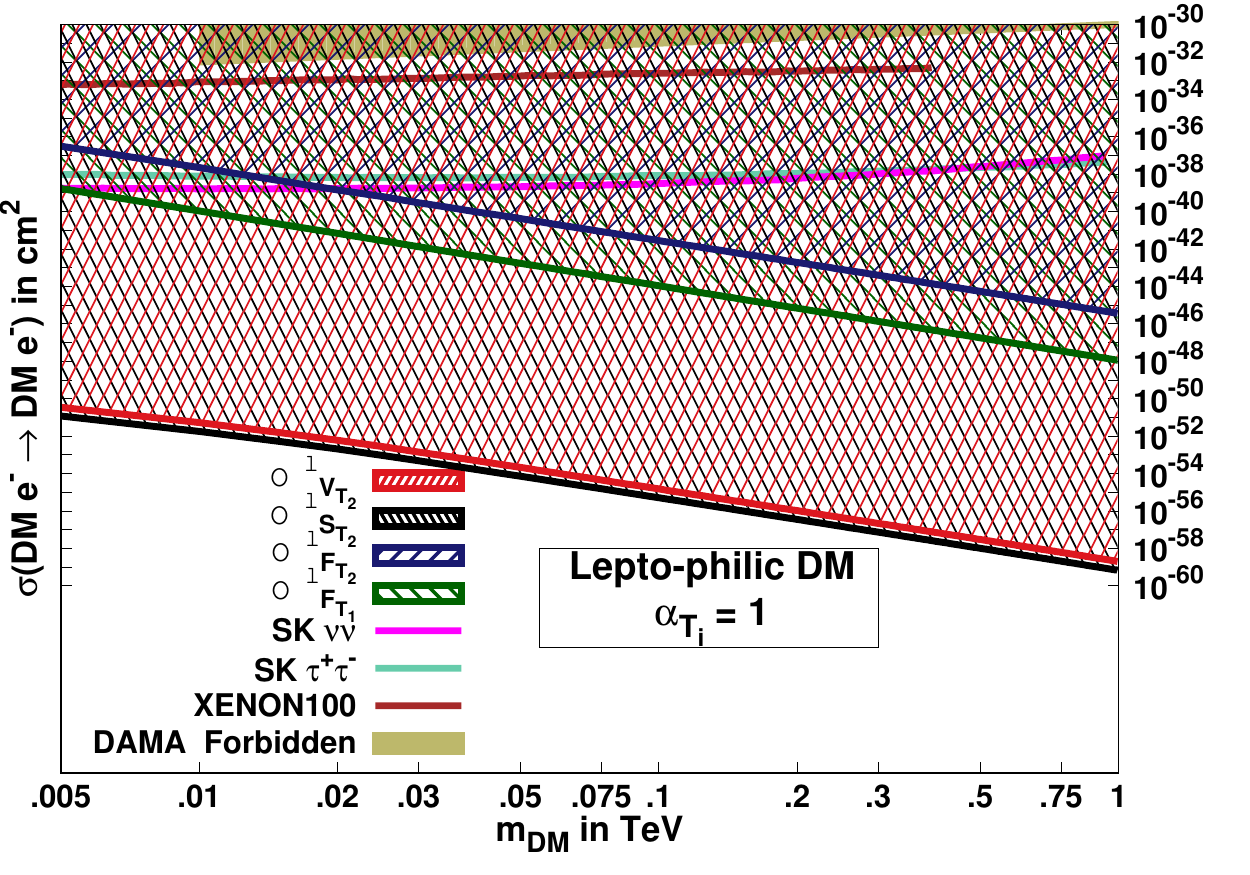}
   \caption{\small \em{Variation of DM - free electron elastic scattering 
      cross-sections {\it w.r.t.} the DM mass  at fixed lepton flavor universal coupling $\alpha_{T_i}$ = 1 and  the  respective upper bound on the  cut-off  satisfying the relic density contribution for lepto-philic operators. The exclusion plots from DAMA at 90\%  C.L. for the case of DM-electron scattering are also shown \cite{Kopp:2009et}. Bounds at 90\% C.L. are shown for XENON100 from inelastic DM-atom scattering \cite{Aprile:2015ade}. The dashed curves show the 90\% CL constraint from the Super-Kamiokande limit on neutrinos from the Sun, by assuming annihilation into $\tau^+\tau^-$ or $\nu\bar\nu$ \cite{Kopp:2009et}.}}
\label{fig:ddetection} 
\end{figure}
\par For $U(1)_Y$  gauge Boson B-philic DM, we look for the photon pair production channel and the  thermally averaged
 DM annihilation cross section is computed using equations \eqref{ThAvBannxsecT1}, \eqref{ThAvBannxsecT2}, \eqref{ThAvBannxsecS} and \eqref{ThAvBannxsecV}  corresponding to the fermionic DM Type-1 and 2,  scalar DM and vector DM Type-2 induced operators respectively.   Figure \ref{Bindirect} show the   annihilation cross-sections depict in  solid line for each case satisfying the relic density  and thereby  giving the lower bound on the cosmologically allowed annihilation cross-section. 
As in the case of lepto-philic and $\tau^\pm$-philic, we observe that the scalar and vector annihilation cross-section overlaps for the reason mentioned earlier. We compare our results with that obtained from the null observation at HESS  \cite{Abramowski:2013ax} for DM mass 100 GeV and above, which gives the upper lower bounds on the respective B-philic operators.

\subsection{DM-electron scattering}
\label{subsec:DDetection}
Direct detection experiments \cite{Bernabei:2013xsa, Bernabei:2018yyw, Aalseth:2012if, Angloher:2016rji, Agnese:2013rvf, Aprile:2016swn, Aprile:2017aty, Akerib:2016vxi, Cui:2017nnn}  look for the scattering of nucleon or atom by DM particles. These experiments are designed to measure the recoil momentum of the nucleons or atoms of the detector material. These  scattering can be broadly classified as (a) DM-electron scattering, (b) DM-atom scattering, and (c) DM-nucleus scattering.  Since the lepto-philic, $\tau^\pm$-philic and B-philic DM do not have direct interactions with quarks or gluons at tree level, therefore  we explore tree level DM-electron elastic scattering induced by the  lepto-philic operators only.
\par In this article, we restrict our study for those direct-detection processes which are realised at the tree level interactions of the lepto-philic DM operators with the free and bound electrons. 
\par Consider the elastic (inelastic) scattering of non-relativistic DM having
 four-momentum $k$ with  free (bound) electron 
 having four momentum $p$ to a final state DM and scattered electron having four momentum $k^\prime$ and $p^\prime$ respectively. In order to understand the realistic DM-electron scattering, we initiate our computation by giving the following leading contributions to  the  spin averaged matrix elements squared   corresponding to the scattering processes induced by fermionic, scalar and vector lepto-philic operators: 
\begin{subequations}
\begin{eqnarray}
	&&\small\overline{|{\cal M}_{free}^{\chi, T1}|^2}= \frac{\left(4\pi \alpha_{T1}\ (k\cdot  p)\right)2}{ \Lambda_{\rm eff}^6}\ 2 \ \bigg[\left[ m_{DM}^2\ -\ k\cdot k' \right]\nn\\
	&&\small\times\ {\cal F}_1\nn+ \frac{p^2\ k\cdot k'\ (m_e^2\ +\ p \cdot p')}{4\ (k\cdot\ p)^2}\ +\ \frac{k'\cdot p\ (m_e^2-2p\cdot p')}{k\cdot p}\nn\\
	&&\small-\frac{p^2\ k'\cdot p'}{2 (k\cdot p)} -\frac{p^2\ (k\cdot p')\ (k'\cdot p)}{(k\cdot p)^2}\ +\ \frac{m_e^2\ (k\cdot k')}{m_{DM}^2}\nn\\
	&&\small-\ \frac{k\cdot k'}{m_{DM}^2} (p^2\ k\cdot p' + \ 2\ p\cdot p')\ + \frac{2\ (k\cdot p)\ (k'\cdot p')}{m_{DM}^2}\nn\\
	&&\small\ \hskip 4cm +\ \frac{6\ (k\cdot p')\ (k'\cdot p)}{m_{DM}^2}\bigg] \nn\\
	&& \text{where}\nn\\
	&&\small{\cal F}_1= \bigg[ \frac{2\ m_e^2}{m_{DM}^2}\ +\frac{p^2\ (k\cdot p')}{m_{DM}^2}\ (k\cdot p)\ -\frac{p\cdot p'}{m_{DM}^2}\nn\\
	&&\small \hskip 3cm -\ \frac{3p^2\ (p\cdot p')}{8\ (k\cdot p)^2}  +\frac{5\ m_e^2\ p^2}{8\ (k\cdot p)^2} \bigg]\label{DDmat1}\\
	&&\small\overline{|{\cal M}_{free}^{\chi, T2}|^2}= \frac{\left(4\pi \alpha_{T2}\, m_{\rm DM}\, (k\cdot p)\right)^2 {\cal F}^\prime}{ 2 \Lambda_{\rm eff}^6}\ 4 \left(\frac{k\cdot  k^\prime}{m_{\rm DM}^2} +1 \right)\nn\\
	&&\label{DDmat2}\\
	&&\small\overline{|{\cal M}_{free}^{\phi^0, T2}|^2}=\frac{ \left(4\pi \alpha_{T2}\ (k\cdot p)\right)^2}{ \Lambda_{\rm eff}^4}\,\,\,\times\,\,\, {\cal F}^\prime\label{DDmat3}\\ 
	&&\small\overline{|{\cal M}_{free}^{V^0, T2}|^2}= \frac{ \left(4\pi \alpha_{T2}(k\cdot  p)^2\right)^2{\cal F}^\prime }{3 \Lambda_{\rm eff}^4}\ \left(\frac{(k\cdot  k')^2}{m_{\rm DM}^4}  +2 \right)   \nn\\
	&&{\rm where}\nn\\
	\label{DDmat4}\nn\\
	&&\small{\cal F}^\prime = \bigg[\frac{m_e^2}{m_{\rm DM}^2}  -  \frac{p^2}{m_{\rm DM}^2} \frac{k\cdot  p^\prime}{k\cdot  p}-2  \frac{p\cdot  p^\prime}{m_{\rm DM}^2} +4 \frac{k\cdot  p}{m_{\rm DM}^2} \frac{k\cdot  p^\prime}{m_{\rm DM}^2}\nn\\
	&&\small \hskip 4cm  +\frac{1}{8} \frac{p^2\,\left(p\cdot  p^\prime + m_e^2\right)}{(k\cdot  p)^2}\bigg]
\end{eqnarray}
\end{subequations}
 The corresponding scattering cross-sections of DM with the free electron at rest are given as    
\begin{subequations}
\begin{eqnarray}
\sigma_{T1}^{\chi e}&=& 36 \pi \alpha_{T_1}^2 \  \frac{m_e^4}{\Lambda_{\rm  eff}^6}\\
\sigma_{T2}^{\chi e}&=& 36 \pi \alpha_{T_2}^2 \ \frac{m_e^4}{\Lambda_{\rm  eff}^6}\\
\sigma_{T2}^{\phi^0 e}&=&9\pi \frac{\alpha_{\phi^0}^2}{\Lambda_{\rm  eff}^4} \ \frac{m_e^4}{m_{\phi^{0}}^2}\\
\sigma_{T2}^{V^0 e}&=& 9\pi \frac{\alpha_{V^0}^2}{\Lambda_{\rm  eff}^4} \ \frac{m_e^4}{m_{V^{0}}^2}
\end{eqnarray}
\end{subequations}
\par  In figure \ref{fig:ddetection} we plot the DM - free electron elastic scattering
 cross-section with varying  DM mass depicted in solid lines for Fermionic Type-1 and 2,  scalar Type-2 \& vector Type-2 twist-2 lepto-philic operators. The cross-section is computed for the coupling fixed at the unity and the corresponding  cut-off which satisfy the relic density $\Omega_{\rm DM}h^2$ = 0.119 \cite{Ade:2015xua} for a given DM mass. Although the analytical expressions of the scattering cross-sections corresponding to the  Type I and Type II twist interactions of the fermionic DM  are similar but we observe the two distinct solid lines corresponding to these contributions in figure \ref{fig:ddetection} because of the  different  values of the respective cut-offs contributing to the same relic density for a given DM mass when $\alpha_{T_1}=\alpha_{T_2}$. Same reason holds for the observed distinguishable contributions from scalar and vector DM scattering cross-sections respectively in figure \ref{fig:ddetection}  for 
$\alpha_{T_2}^{V^0}=\alpha_{T_2}^{\phi^0}$ although the corresponding analytical expressions are same. These results are then compared with the null results of DAMA/LIBRA   \cite{Bernabei:2013xsa, Bernabei:2018yyw}
 at 90\%  confidence level for DM-electron scattering and
 XENON100   \cite{Aprile:2016swn, Aprile:2017aty} at 90\% confidence level for inelastic
 DM-atom scattering. 

It is important to note that electro-philic DM -free electron scattering cross-sections corresponding to the respective operators computed using the upper bound on the cut-off obtained for a given DM mass from relic density constraints as shown in figure \ref{tauphilicrelicdensity} and unity coupling strength,  will be slightly higher than those shown in the figure \ref{fig:ddetection}. 
\par The DM - free electron scattering corresponding to the $\tau^\pm$-philic and $B$-philic operators occurs at the one loop level and therefore are further suppressed. On the same note, due to the absence  of the tree level DM - quark interactions, DM-nucleon scattering induced by the lepto-philic and $U(1)_Y$-philic twist-2 operators are either one or two loop(s) suppressed, however it dominates over the DM-free (bound) electron scattering.

\subsubsection{Inelastic scattering : Effect of bound electrons}
The  non-relativistic DM $\left(E_{DM}\sim m_{DM}\right)$ essentially collides the bound electron 
 of fixed energy $E_e=m_e- E_B^{nl}$ (here $E_B^{nl}$ is the binding energy of
 electron in $l^{\rm th}$ orbital of  $n^{\rm th}$ shell ) and momentum distribution $\vec{p}$ and then finally ejects the electron from the atom with the energy $E_R+m_e-E_{B}^{nl}$, where $E_R$ is the  recoil energy  of the scattered electron. 
The inelastic DM - electron differential scattering cross-section {\it w.r.t.} the recoil energy   in the lab frame is computed to be
 \bea
\frac{d\sigma_{nlm}}{dE_R} &=& \frac{\overline{\left\vert{\cal M}_{nlm}\right\vert^2}}{32\pi\ E_{DM}\ E_e\ v_{rel}\left\vert\vec{k}+\vec{p}\right\vert }\nn\\
&=& \frac{\left\vert\chi_{nl}\left(\left\vert\vec{p}\right\vert\right)\right\vert^2 \left\vert Y_{lm}(\theta,\phi)\right\vert^2\,\overline{\left\vert{\cal M}_{free}\right\vert^2} }{32\pi\ E_{DM}\ E_e\ v_{rel}\left\vert\vec{k}+\vec{p}\right\vert } 
 \eea
where $\chi_{nl}\left(\left\vert\vec{p}\right\vert\right) $ is 
 the $l^\text{th}$   radial momentum space   wave-function for $n^\text{th}$ shell   and   $Y_{lm}\left(\theta,\phi\right)$ is the  angular wave function associated with the bound electron. Summing over all possible shells (which depends on the detector material),
\bea 
\sum_{nlm} \frac{d\sigma_{nlm}}{dE_R} &=&\sum_{nl} \frac{\left\vert\chi_{nl}\left(\left\vert\vec{p}\right\vert\right)\right\vert^2\,\overline{\left\vert{\cal M}_{free}\right\vert^2} }{32\pi\ E_{DM}\ E_e\ v_{rel} \left\vert\vec{k}+\vec{p}\right\vert} \frac{(2l+1)}{4\pi}\nn\\
\label{dsigma2}
\eea
Following the  prescription given in the appendixes {\bf B} and {\bf C}  of reference \cite{Kopp:2009et}  we estimate  the total event rate as
\bea
  \frac{dR}{dE_R}&=& \frac{\rho_0\ n_{\color{red}T}}{m_{DM}\ }\ \int \frac{d^3p}{(2\pi)^3}\
\int d^3v \ v_{rel}\ f_\odot(\vec{v})\ \sum_{nlm} \frac{d\sigma_{nlm}}{dE_R}\nn\\
\eea
where $\rho_0$ , $n_{\color{red}T}$ , $f_\odot(\vec{v})$ and $v_{rel}$ are local density of DM ($\approx 0.3$ GeV/cm$^3$ ), number of target particles per unit mass, velocity distribution of DM in lab frame and relative velocity of DM and bound electron respectively. Substituting \eqref{dsigma2} and using {\bf NaI}  as the detector material \cite{Bernabei:2018yyw}, we compute the event rate as
\bea
  \frac{dR}{dE_R}&=& \frac{\rho_0}{64\pi\ m_{DM}^3\ m_e\ (m_I+m_{Na}) }\nn\\
  &&\hskip 2cm \sum_{n,l} \epsilon_{nl} \,\, \eta(v_{min})\,\, \overline{|{\cal M}_{free}|^2}
  \eea
  where $\eta(v_{min})\equiv\int d^3v\ \frac{f\odot (\vec{v})}{v}\ \theta(v-v_{min})$ with $v_{min} \approx \frac{E_R}{|\vec{p}|} + \frac{|\vec{p}|}{m_{DM}}$ and $\epsilon_{nl}$ is the 
suppression factor 
\bea
\epsilon_{nl}&=& \sqrt{2m_e\ (E_R-E_B^{nl})}\ (2l+1)\nn\\
&&\hskip 2cm \int \frac{dp}{(2\pi)^3}\left\vert\vec{p}\right\vert\ \left\vert\chi_{nl}\left(\left\vert\vec{p}\right\vert\right)\right\vert^2
\label{supp}
\eea
For the detector signal with the deposited energy  $\sim E_R$ = 2 - 4 keV, which is sensitive to the DAMA/LIBRA experiment, the integral in equation \eqref{supp} is found to be maximum for the electron corresponding to the  $3s$ orbital of Iodine  $\sim 10^{-2}$ MeV$^{-1}$ for {\bf NaI} crystals. Substituting  the binding energy of $3s$ electron of Iodine $\sim 1$ keV, the suppression factor  becomes   $\sim\ 10^{-6}$. 
\par It is important to mention that  the event rates from lepto  and $U(1)_Y$-philic induced DM-nucleon scattering   which are  suppressed  by  $\sim \left(\alpha_{\rm em}\, Z/\pi\right)^2$, where $Z$ is charge of the nucleus at one loop order, are however do not have any wave-function suppression.

\begin{figure*}[tbh]
\centering
	\begin{multicols}{2}
\includegraphics[width=0.49\textwidth,clip]{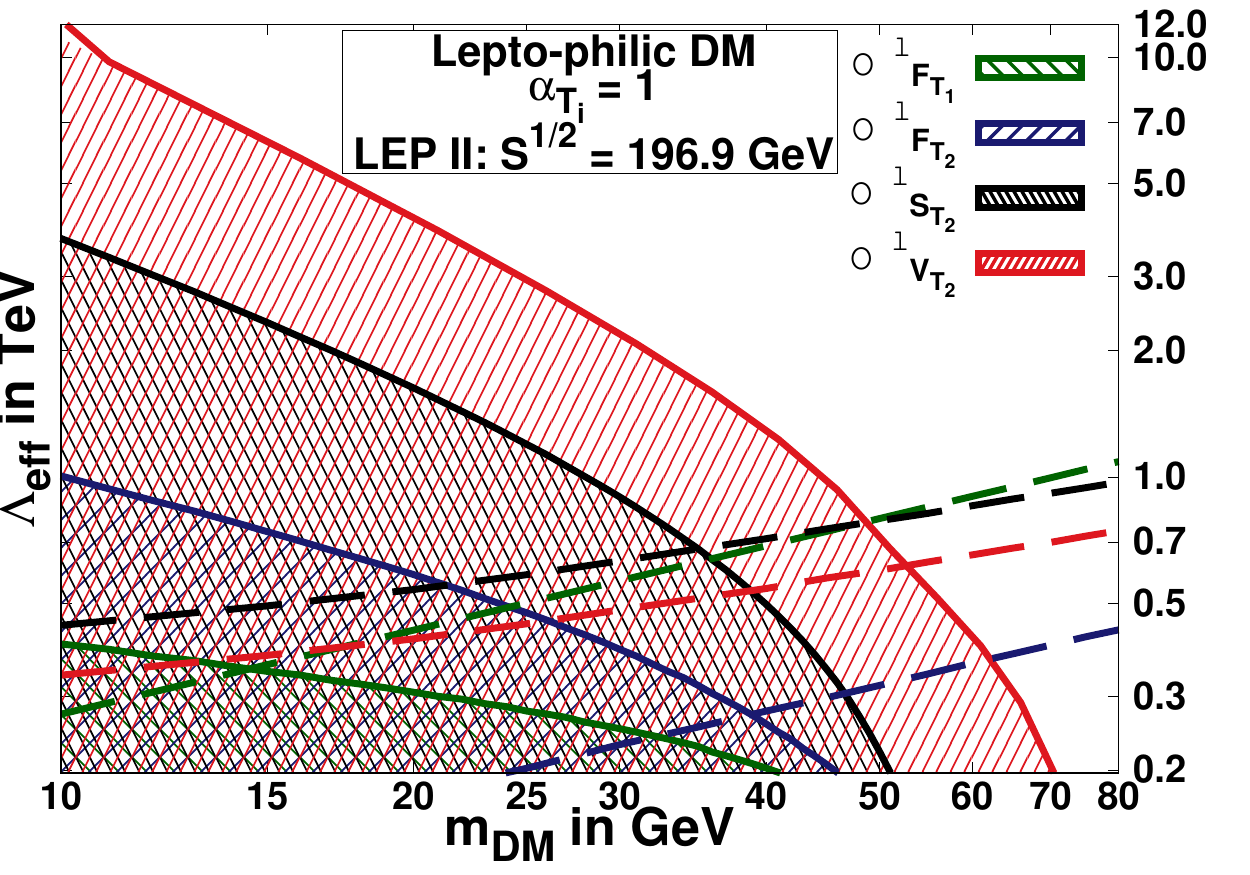}
\subcaption{Lepto-philic}\label{lepto95CLLEP}
\columnbreak
		\includegraphics[width=0.49\textwidth,clip]{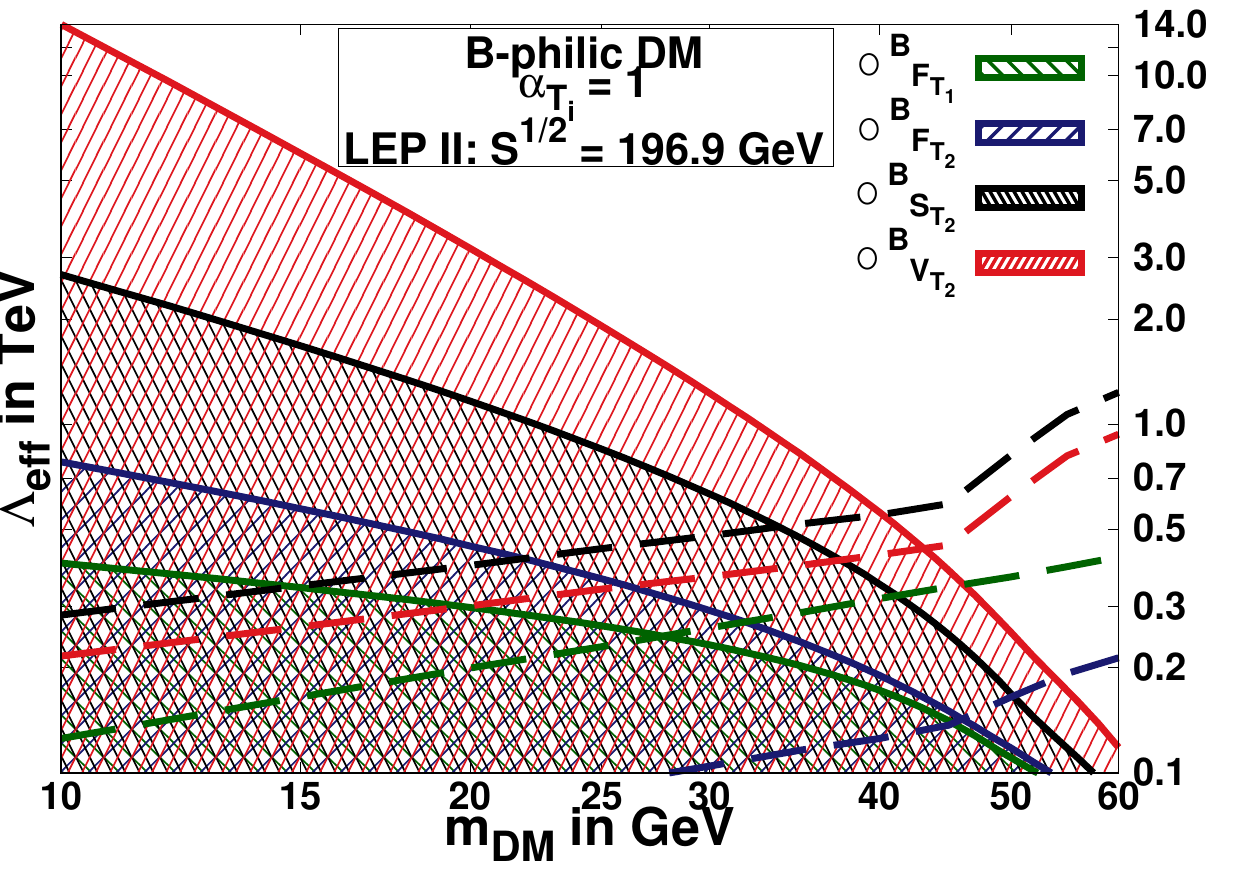}\\
 \subcaption{B-philic}\label{B95CLLEP}
	\end{multicols}
        \caption{\small \em{Solid lines depict the contours in the plane defined by DM mass and the kinematic reach of the cut-off $\Lambda_{\rm  eff}$ for $e^+e^-\to {\rm DM\, pairs} + \gamma^\star \to \,\,\not\!\!\!E_T + q_{i}\bar q_{i} $ at fixed coupling $\alpha_{T_i}$ = 1, $\sqrt{s}$ = 196.9 GeV and an integrated luminosity of 679.4 pb$^{-1}$, satisfying the constraint $\delta\sigma_{\rm tot}$  = .032 pb obtained from combined analysis of DELPHI and L3 \cite{Schael:2013ita}. The enclosed shaded region corresponding to each solid line are forbidden by LEP observation. The regions below the  colored dashed lines corresponding to respective four operators satisfy  the relic density constraint $\Omega_{\rm DM}h^2 \le$ 0.1199 $\pm$ 0.0022.}}  
        \label{95CLLEP}
\end{figure*}

\begin{figure*}
\centering
	\begin{multicols}{2}
\includegraphics[width=0.49\textwidth,clip]{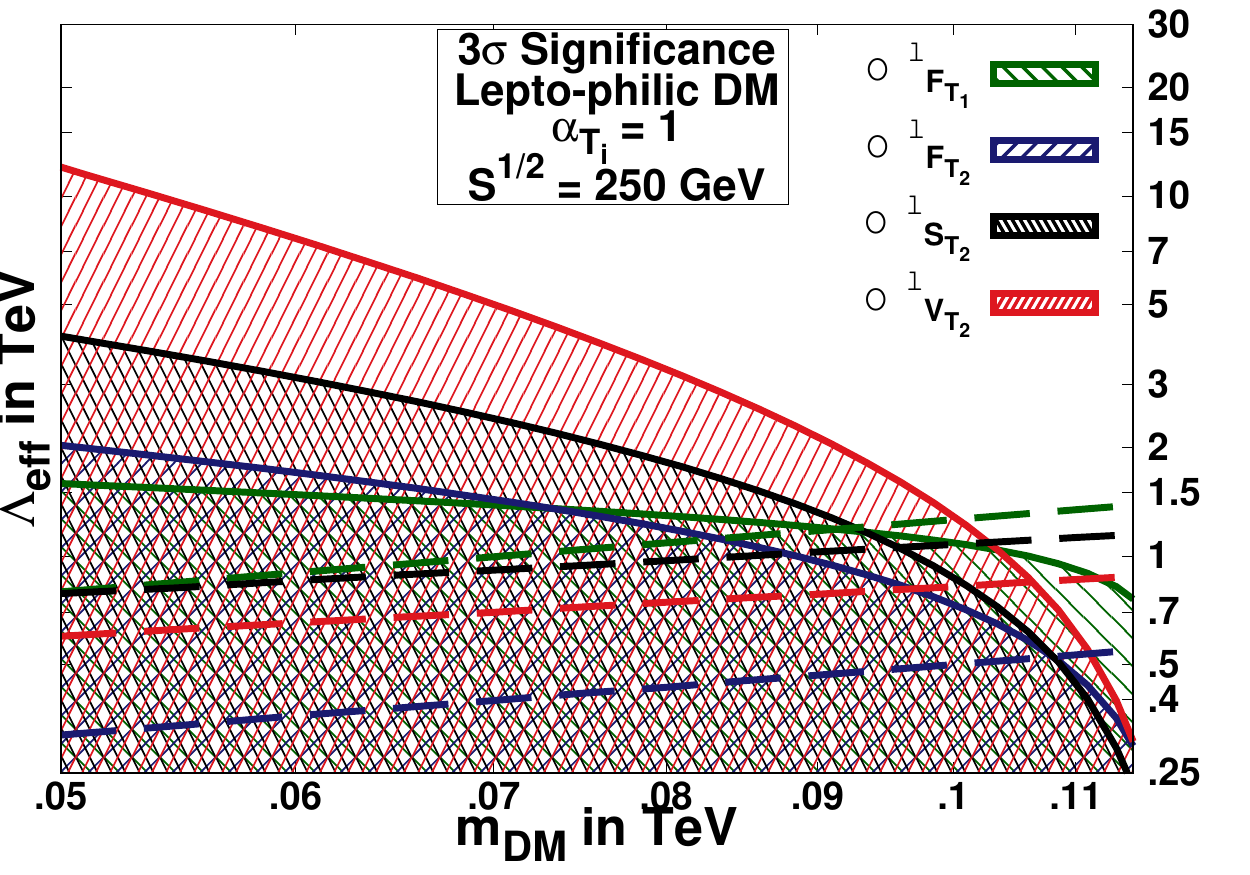}
\subcaption{}\label{lepto3sigma.25TeV}
\includegraphics[width=0.49\textwidth,clip]{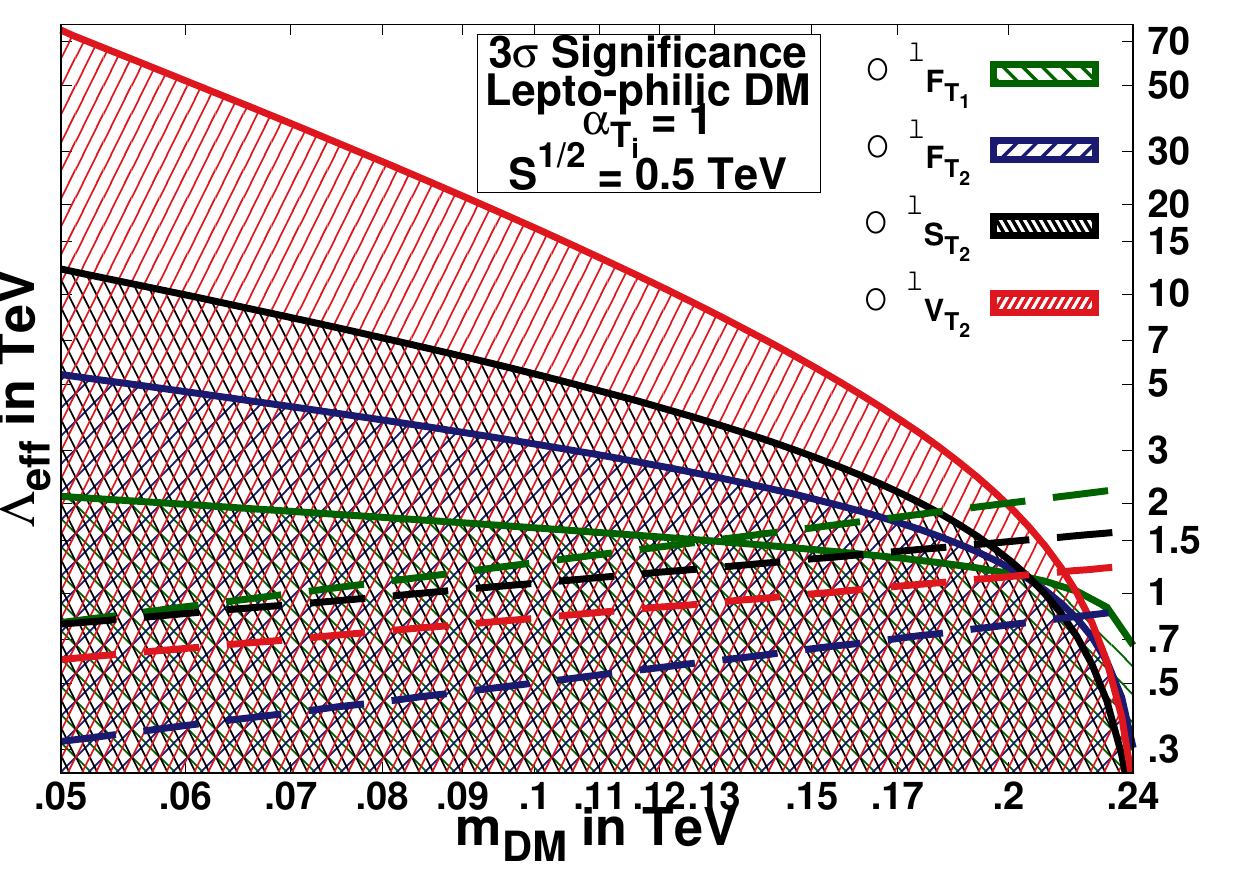}
\subcaption{}\label{lepto3sigma.5TeV}
\includegraphics[width=0.49\textwidth,clip]{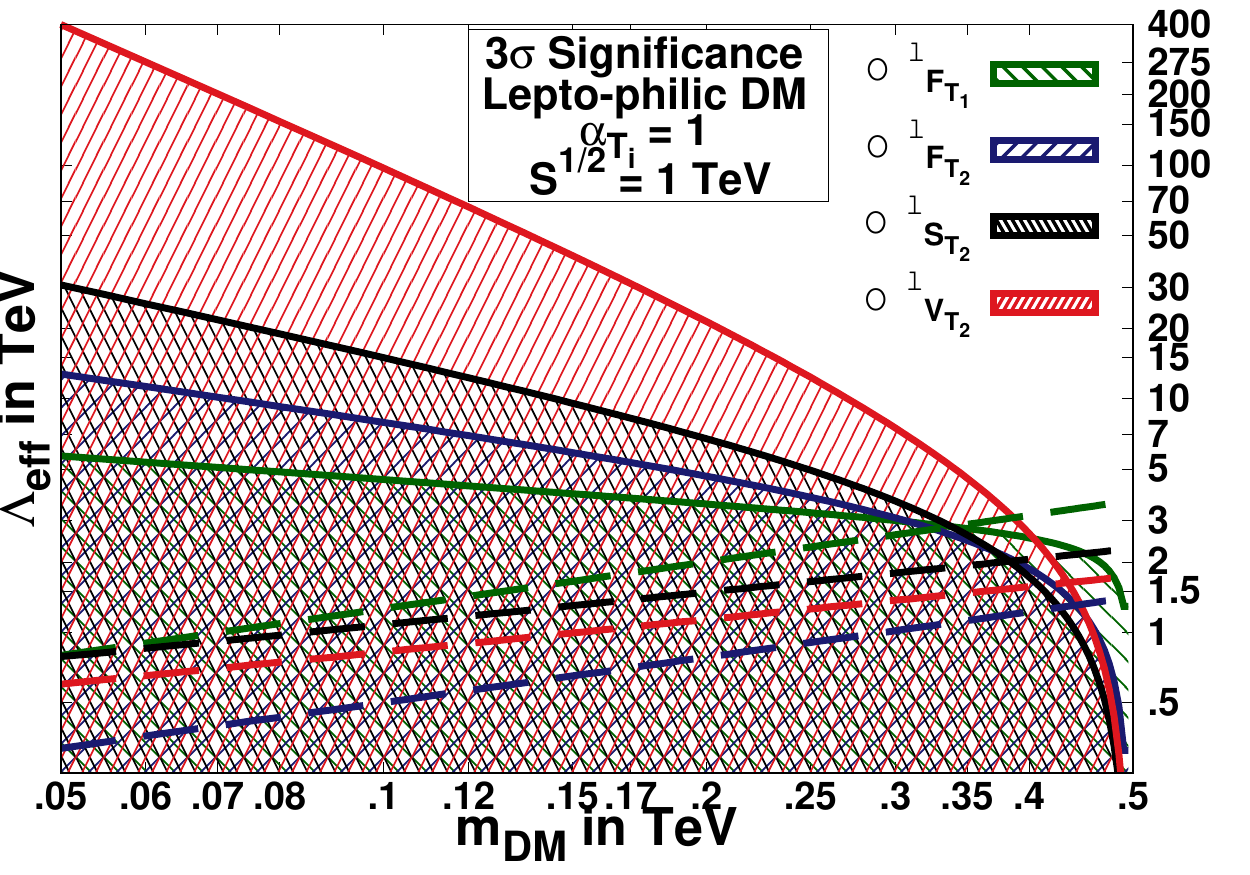}
\subcaption{}\label{lepto3sigma1TeV}
\columnbreak
\includegraphics[width=0.49\textwidth,clip]{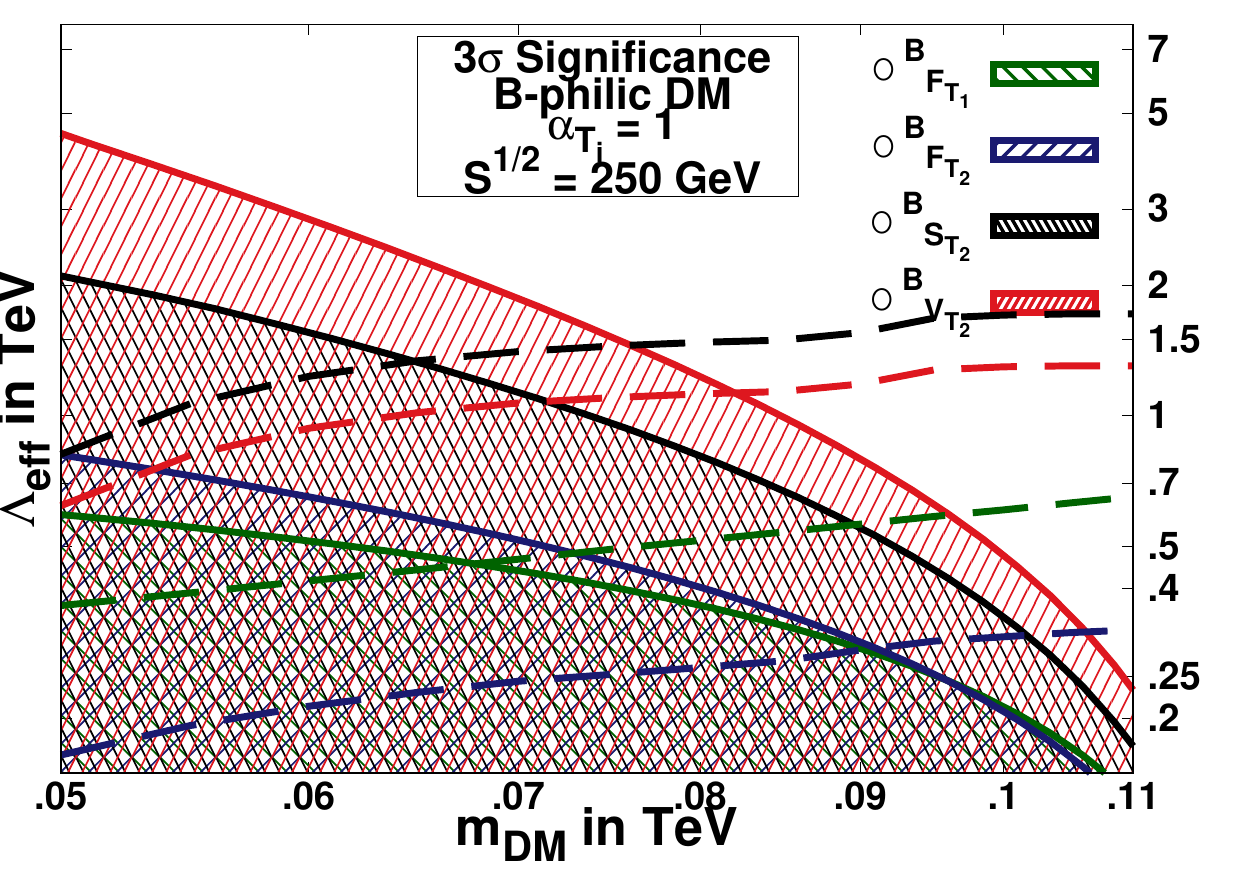}
\subcaption{}\label{B3sigma.25TeV}
\includegraphics[width=0.49\textwidth,clip]{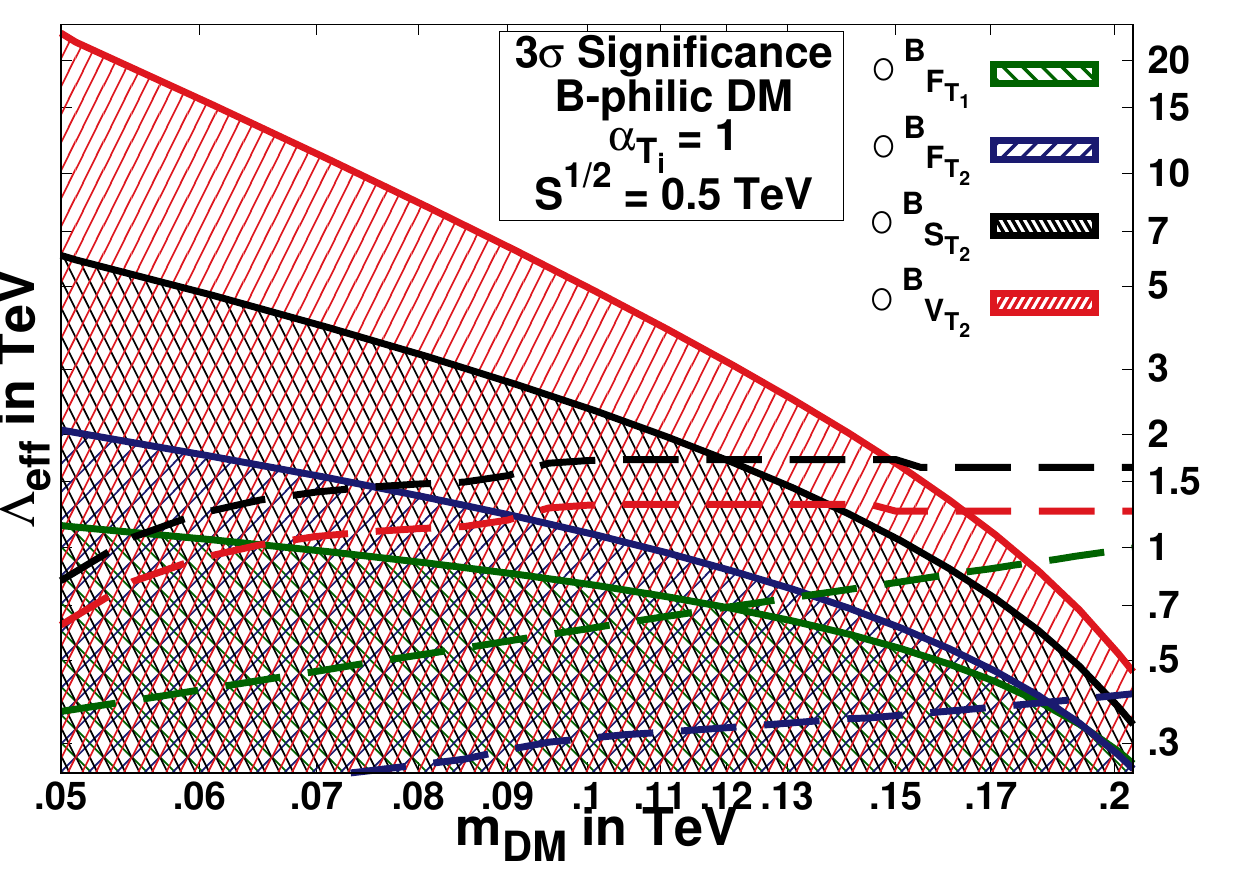}
 \subcaption{}\label{B3sigma.5TeV}
\includegraphics[width=0.49\textwidth,clip]{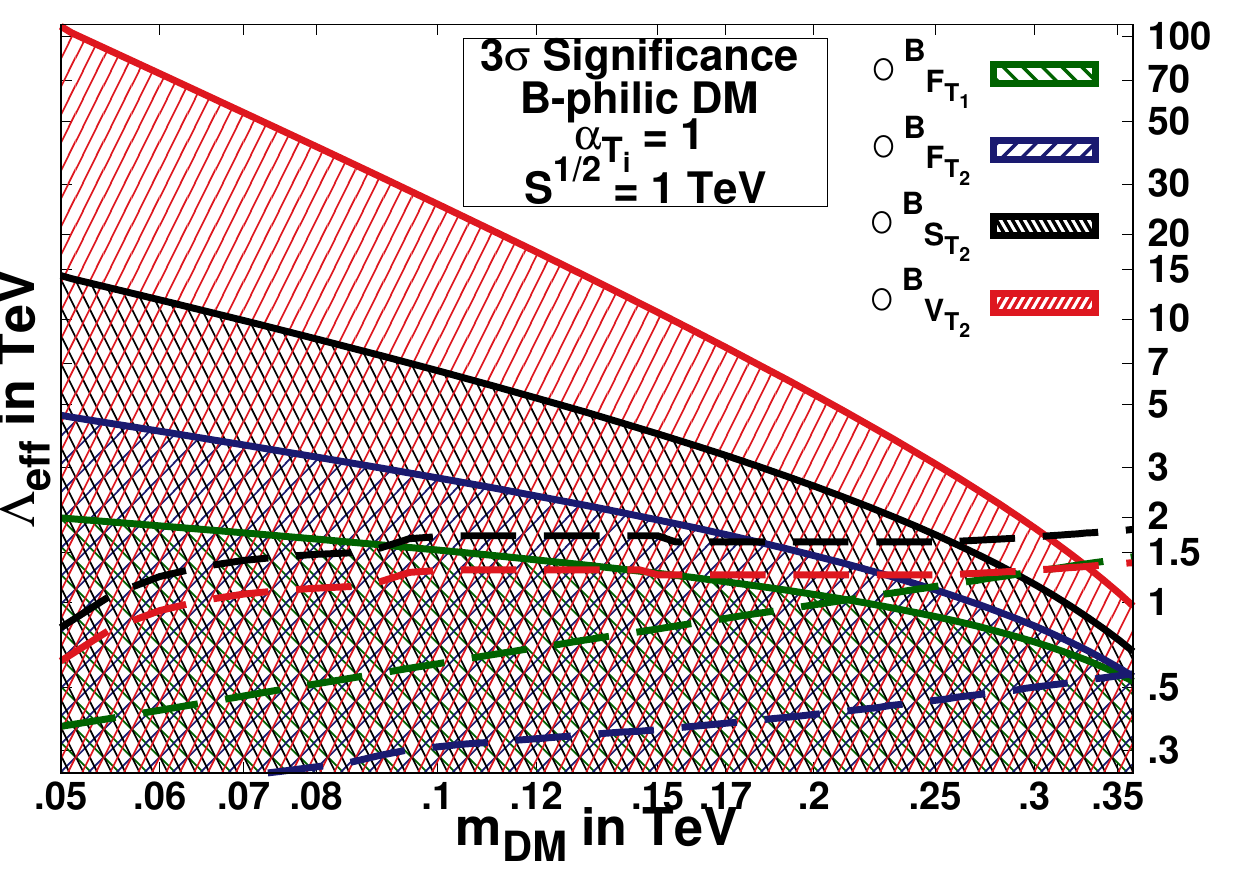}
 \subcaption{}\label{B3sigma1TeV}
	\end{multicols}
	\caption{\small \em{Solid lines depict $3\sigma$ efficiency contours at fixed coupling $\alpha_{T_i}$ = 1 for DM pair production process  with mono-photon ($e^+e^-\to \not\!\!\!E_T + \gamma $) in the plane defined by DM mass and the kinematic reach of the cut-off $\Lambda_{\rm  eff}$ corresponding to  (a) $\sqrt{s}$ = 250 GeV; ${\cal L}$ = 250 fb$^{-1}$ in \ref{lepto3sigma.25TeV} and \ref{B3sigma.25TeV}, (b) $\sqrt{s}$ = 500 GeV; ${\cal L}$ = 500 fb$^{-1}$ in \ref{lepto3sigma.5TeV} and \ref{B3sigma.5TeV}  and (c)  $\sqrt{s}$ = 1 TeV; ${\cal L}$ = 1 ab$^{-1}$ in \ref{lepto3sigma1TeV} and \ref{B3sigma1TeV}. The shaded region corresponding to each solid line are likely to probed  by ILC with greater than $3\sigma$ efficiency. The regions below the  colored dashed lines corresponding to respective four operators satisfy  the relic density constraint $\Omega_{\rm DM}h^2 \le $ 0.1199 $\pm$ 0.0022. }}
        \label{3sigmaeff}
\end{figure*}

\begin{figure*}
        \centering
\begin{multicols}{2}
\includegraphics[width=0.43\textwidth,clip]{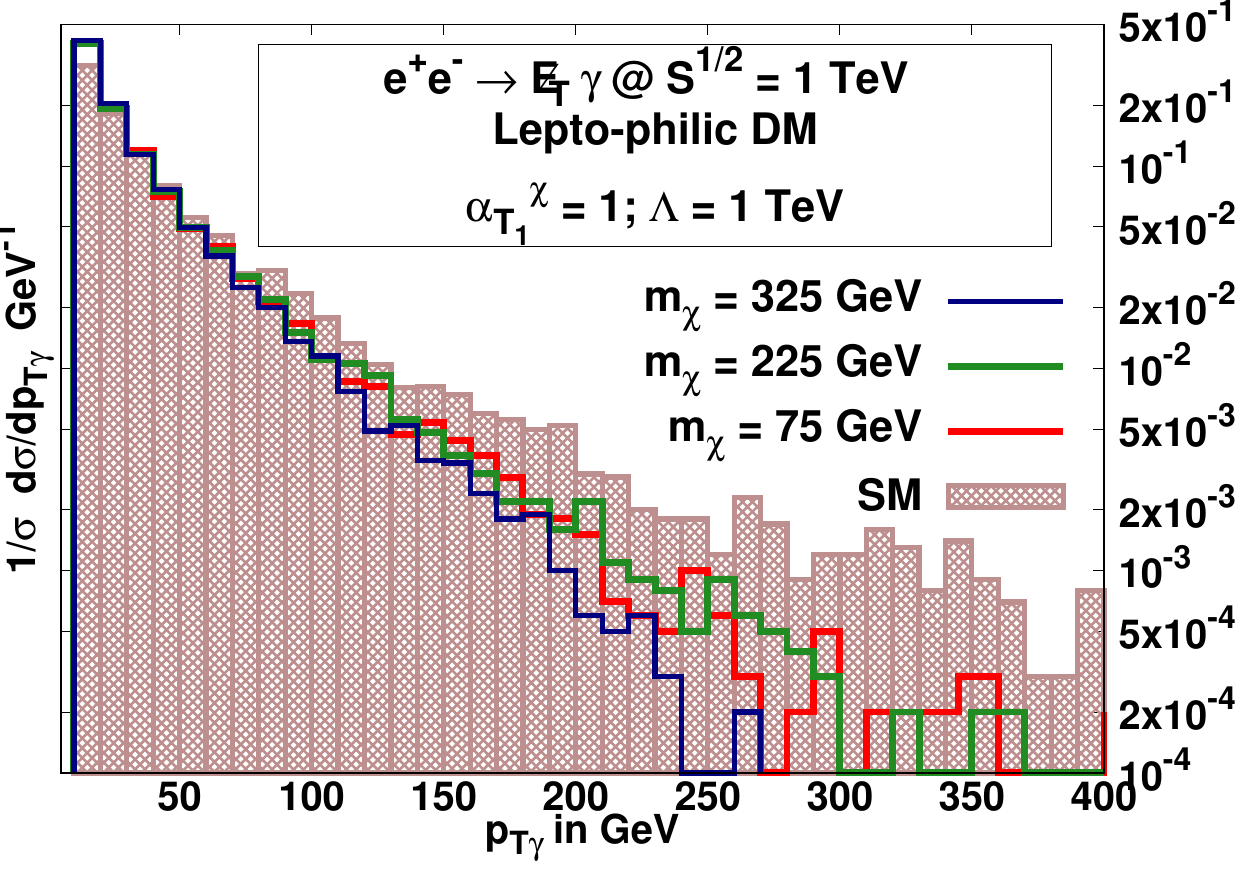}
\subcaption{}\label{fermiT1ptdist}
\includegraphics[width=0.43\textwidth,clip]{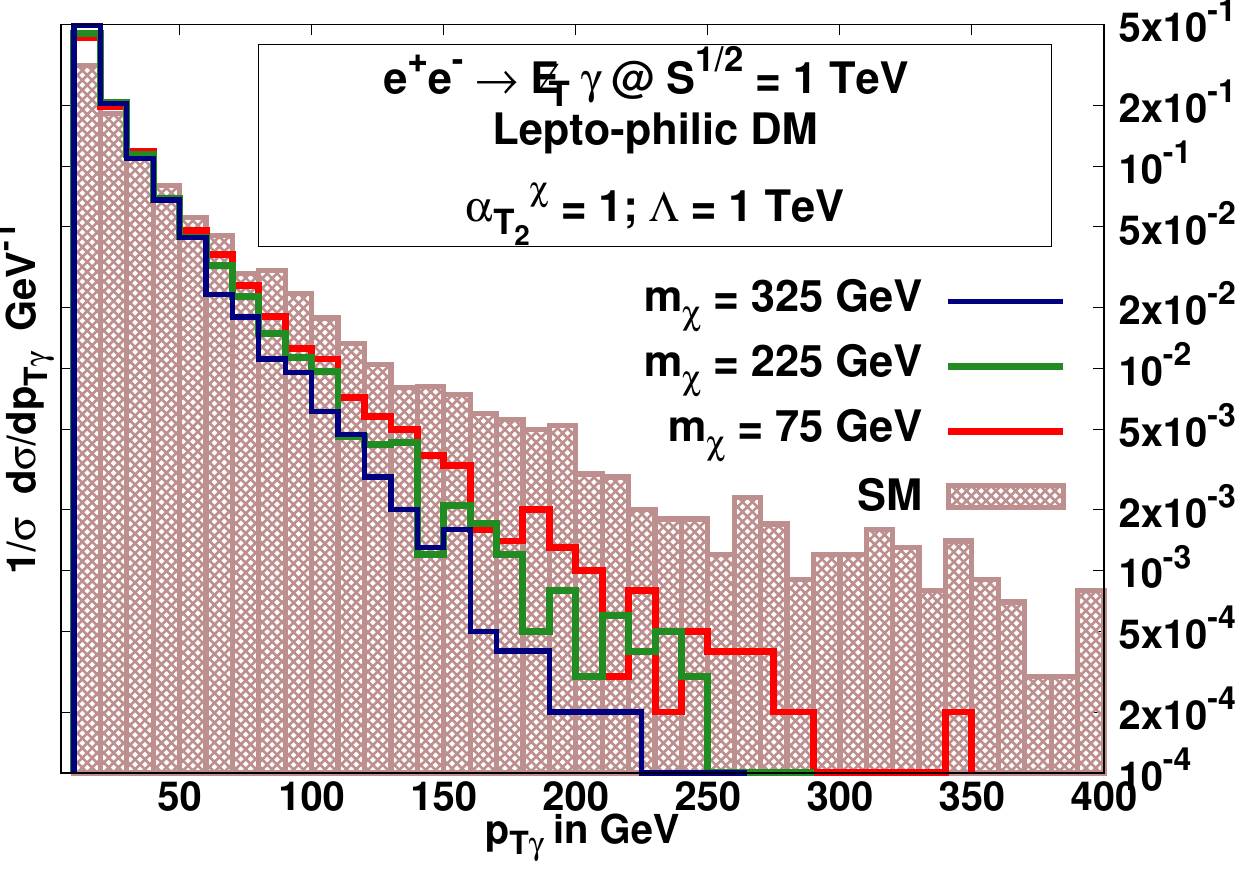}
\subcaption{}\label{fermiT2ptdist}
\includegraphics[width=0.43\textwidth,clip]{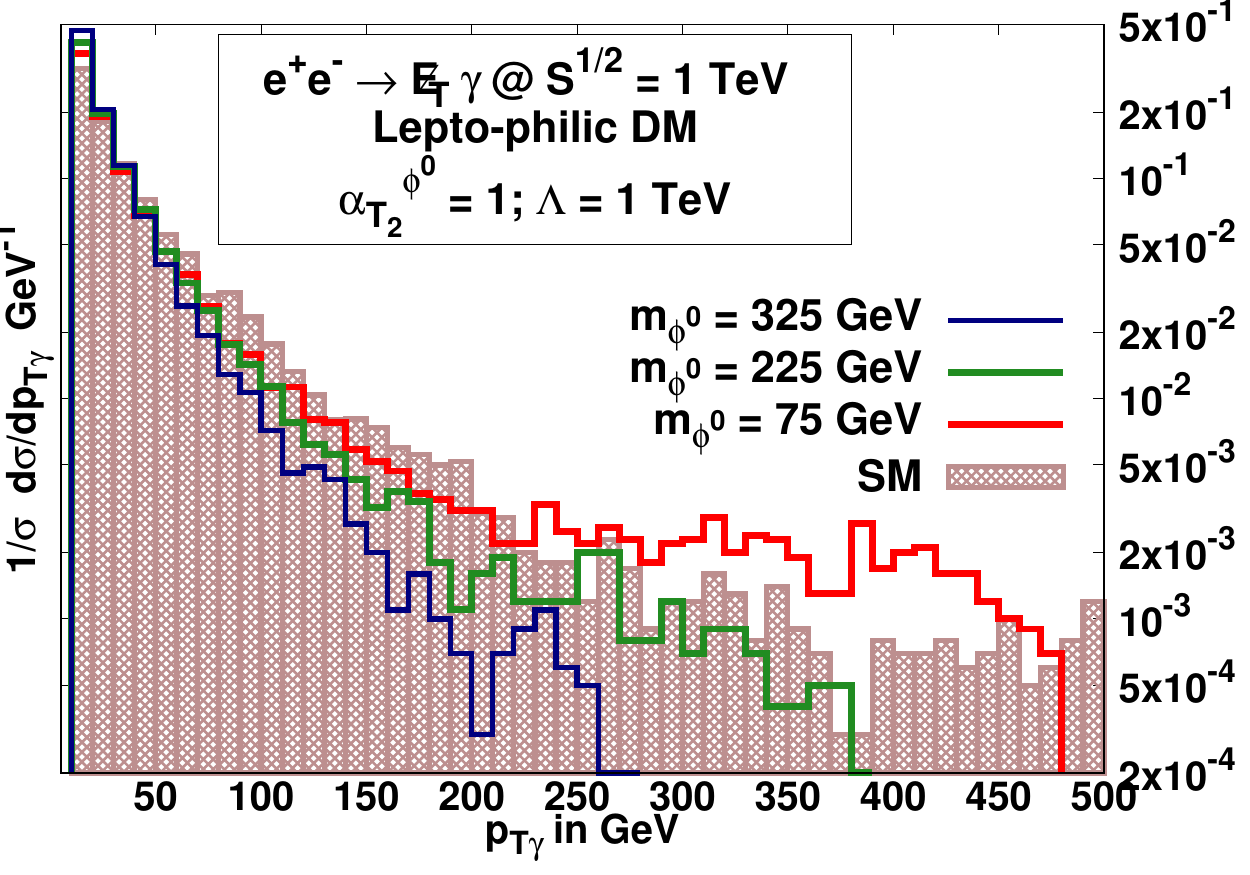}
\subcaption{}\label{scalT2ptdist}
\includegraphics[width=0.43\textwidth,clip]{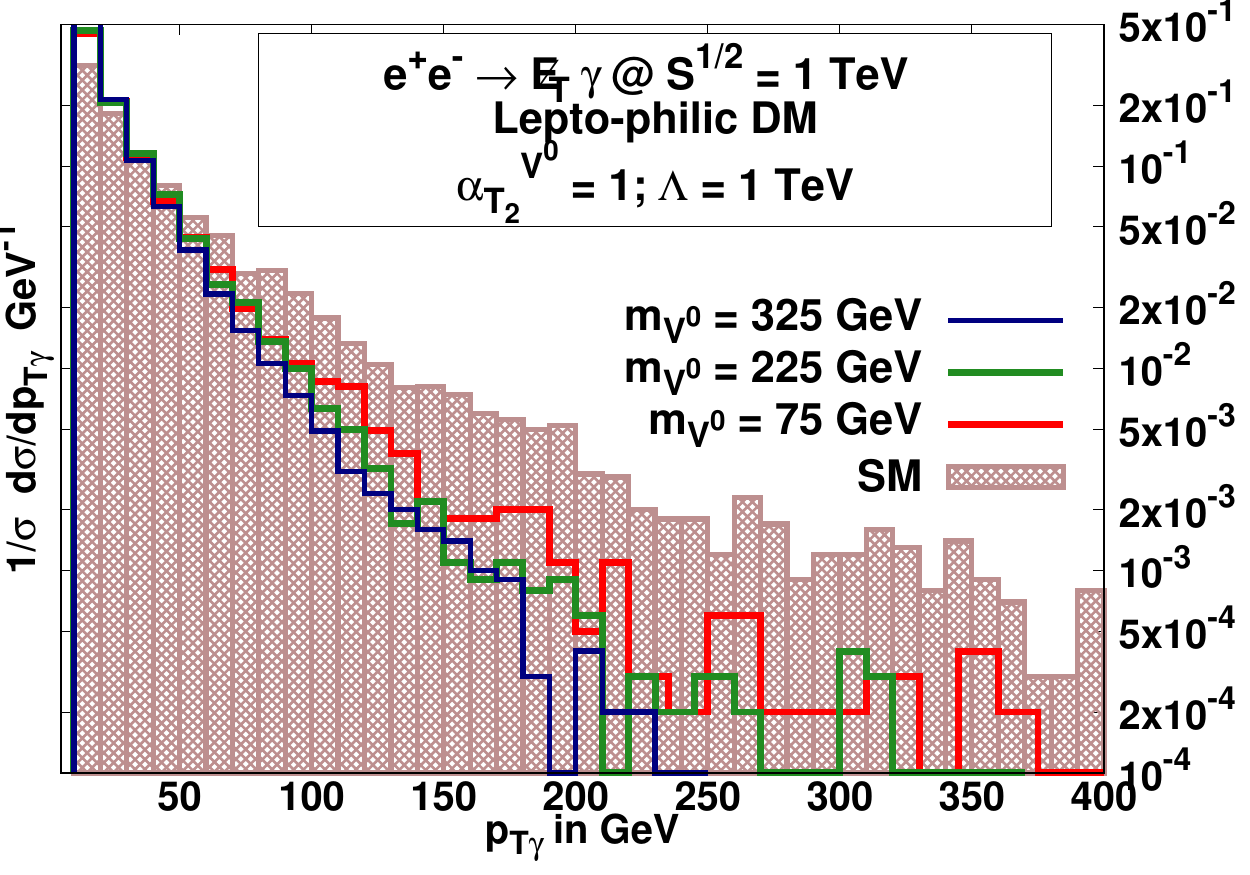}
\subcaption{}\label{vecT2ptdist}
\columnbreak
\includegraphics[width=0.43\textwidth,clip]{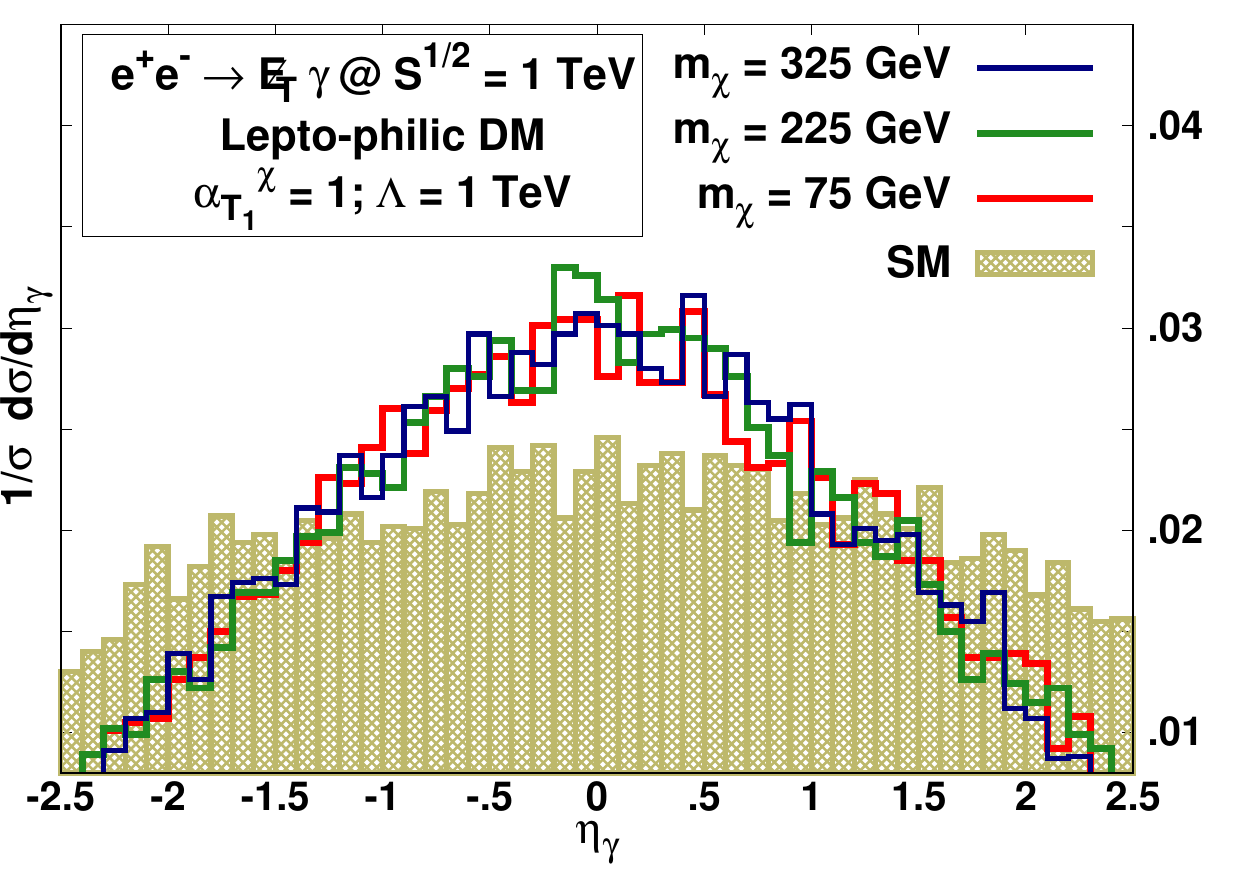}
\subcaption{}\label{fermiT1etadist}
\includegraphics[width=0.43\textwidth,clip]{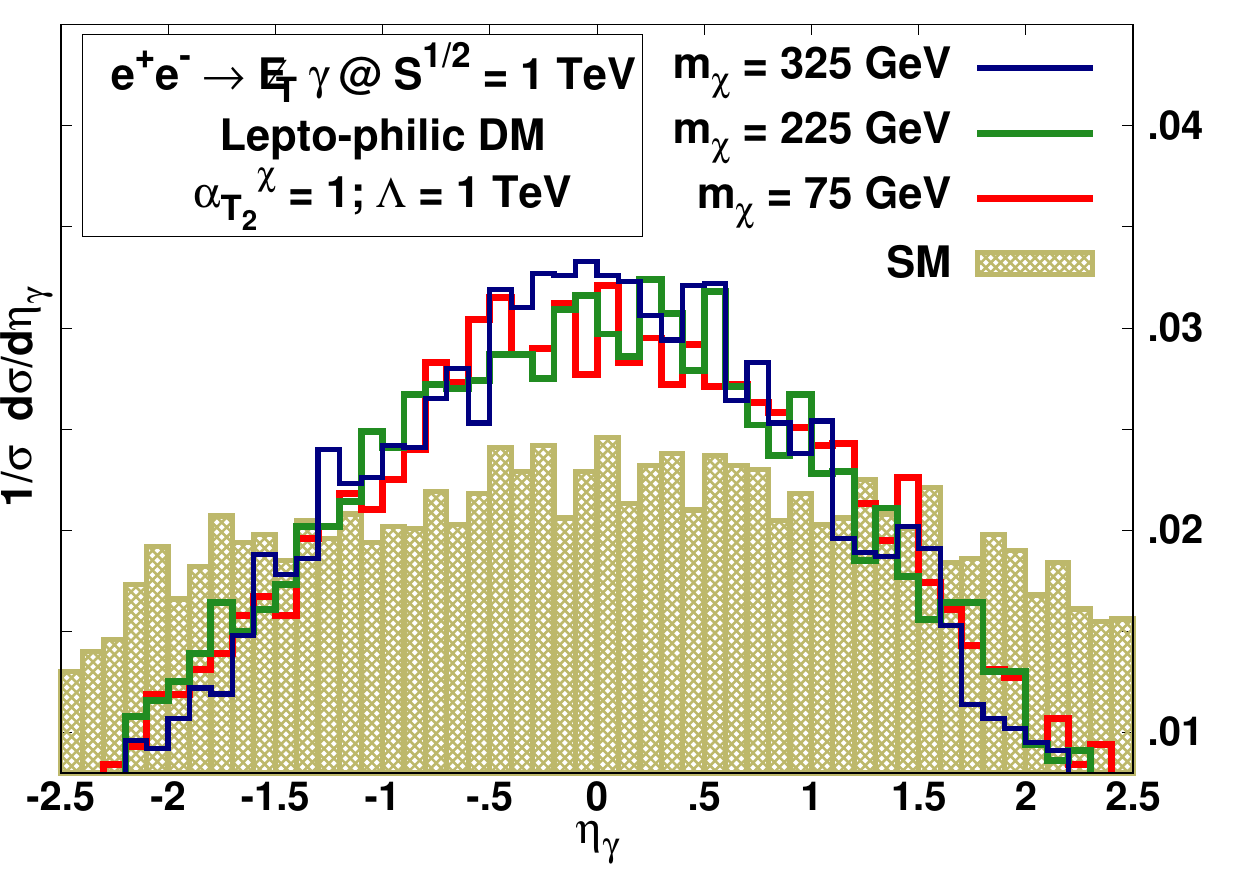}
\subcaption{}\label{fermiT2etadist}
\includegraphics[width=0.43\textwidth,clip]{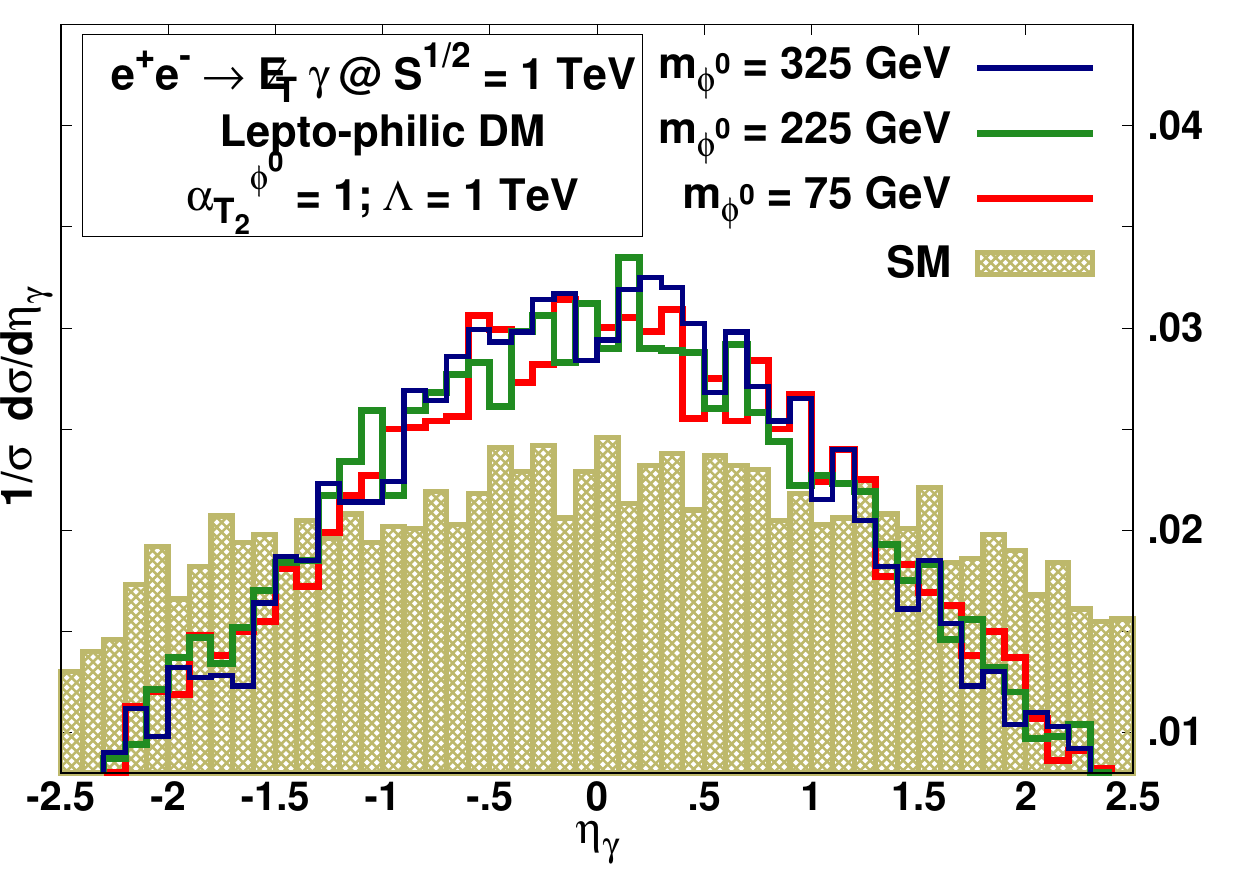}
\subcaption{}\label{scalT2etadist}
\includegraphics[width=0.43\textwidth,clip]{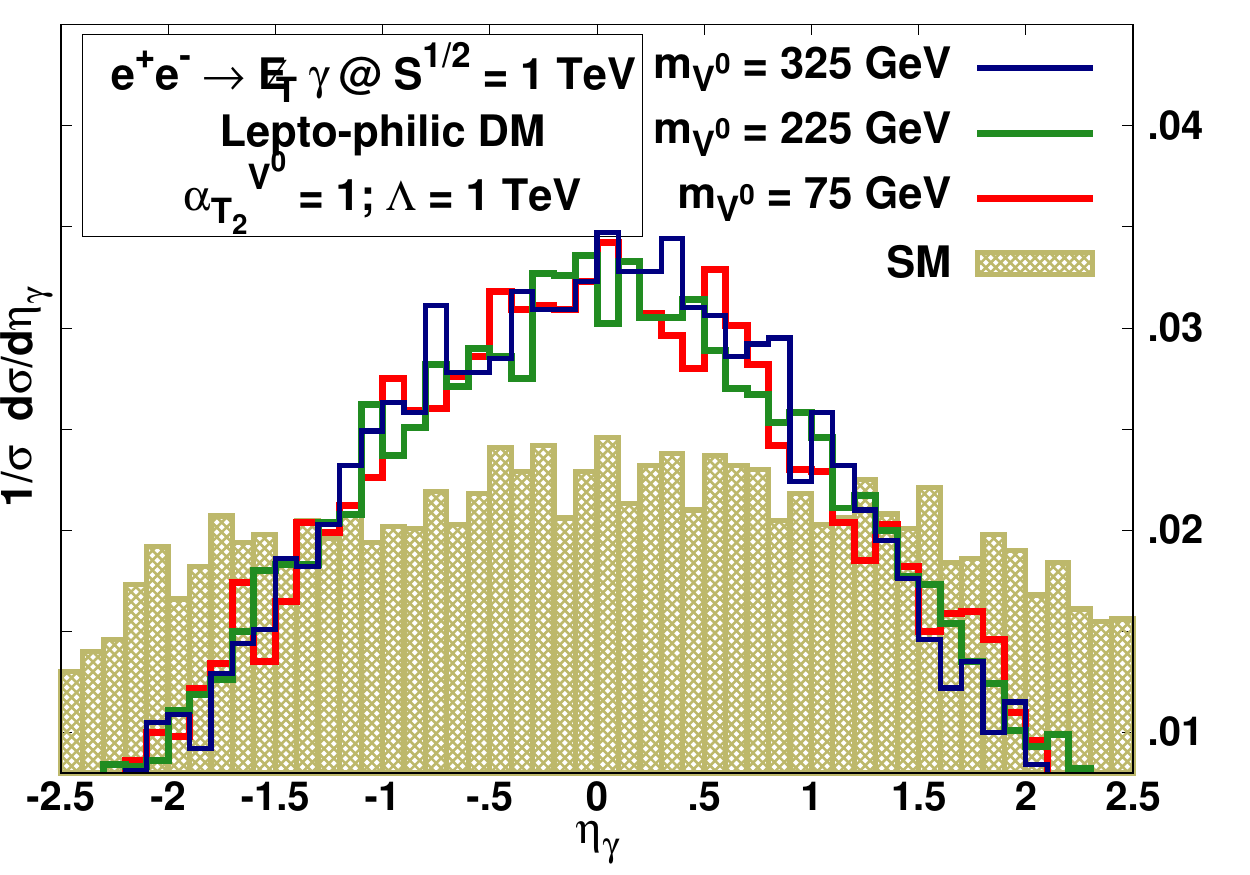}
\subcaption{}\label{vecT2etadist}
\end{multicols}
        \caption{\small \em{Normalized 1-D differential  cross-sections   {\it w.r.t.}  $p_{T_\gamma}$ (bin width 10 GeV) and $\eta_\gamma $ (bin width 0.1) corresponding to the  SM processes (shaded histograms) and   those induced by  lepto=philic operators  at the three representative values of DM masses: 75, 225 and 325 GeV respectively.}}
        \label{fig:distrVpgs}
\end{figure*}

\section{Collider sensitivity of effective operators}
\label{sec:collider}
\subsection{LEP Constraints on the effective operators}
\label{subsec:LEPCons}
We investigate the constraints on the lepto-philic and B-philic effective operators from the existing results and observations from LEP data. We compute the cross-section for the process $e^+e^-\to \gamma^\star + \, {\rm DM \, pair}$, and compare with the combined analysis from DELPHI and L3 collaborations   for $e^+e^-\to \gamma^\star + Z \to q_{i}\bar q_{i} + \nu_{l_j}\bar\nu_{l_j}$ at $\sqrt{s}$ = 196.9 GeV and an integrated luminosity of 679.4 pb$^{-1}$, where $q_i\equiv u,\,d,\,s$ and $\nu_{l_j}\equiv \nu_e,\,\nu_{\mu},\nu_\tau$. The  measured cross-section from the combined analysis for the said process is found to be .055 pb along with the measured statistical error $\delta\sigma_{\rm stat}$, systematic error $\delta\sigma_{\rm syst}$ and total  error $\delta\sigma_{\rm tot}$ of .031 pb, .008 pb and .032 pb respectively \cite{Schael:2013ita}.  Therefore, contribution due to an additional channel  containing the final states DM pairs and resulting into the missing energy along with two quark jets can be constrained from the observed $\delta\sigma_{\rm tot}$.

\par In figures \ref{lepto95CLLEP} and \ref{B95CLLEP}, we plot the 95\% C.L. solid line contours satisfying the cross-section observed $\delta\sigma_{\rm tot}$ .032 pb  corresponding to the lepto-philic and B-philic operators in the two dimensional plane defined by the DM mass and the lower bound on the cut-off  at the fixed value of respective  coupling $\alpha_{T_i}$. The respective shaded regions in figure \ref{95CLLEP} are disallowed by the combined LEP analysis. Thus  the phenomenologically interesting DM mass range $\lesssim $ 50 GeV is completely disfavored by the LEP experiments. 
\begin{figure*}
        \centering
\begin{multicols}{2}
\includegraphics[width=0.43\textwidth,clip]{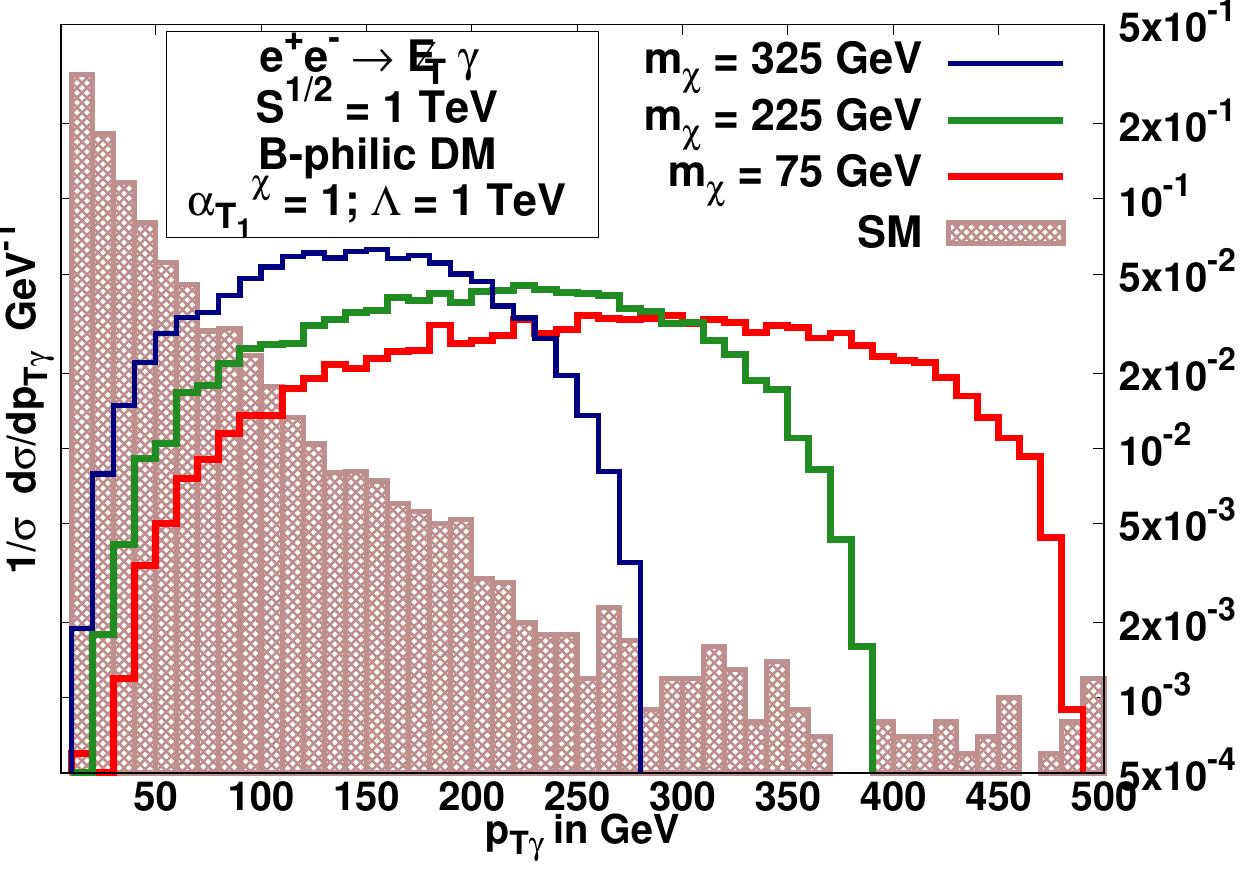}
\subcaption{}\label{BfermiT1ptdist}
\includegraphics[width=0.43\textwidth,clip]{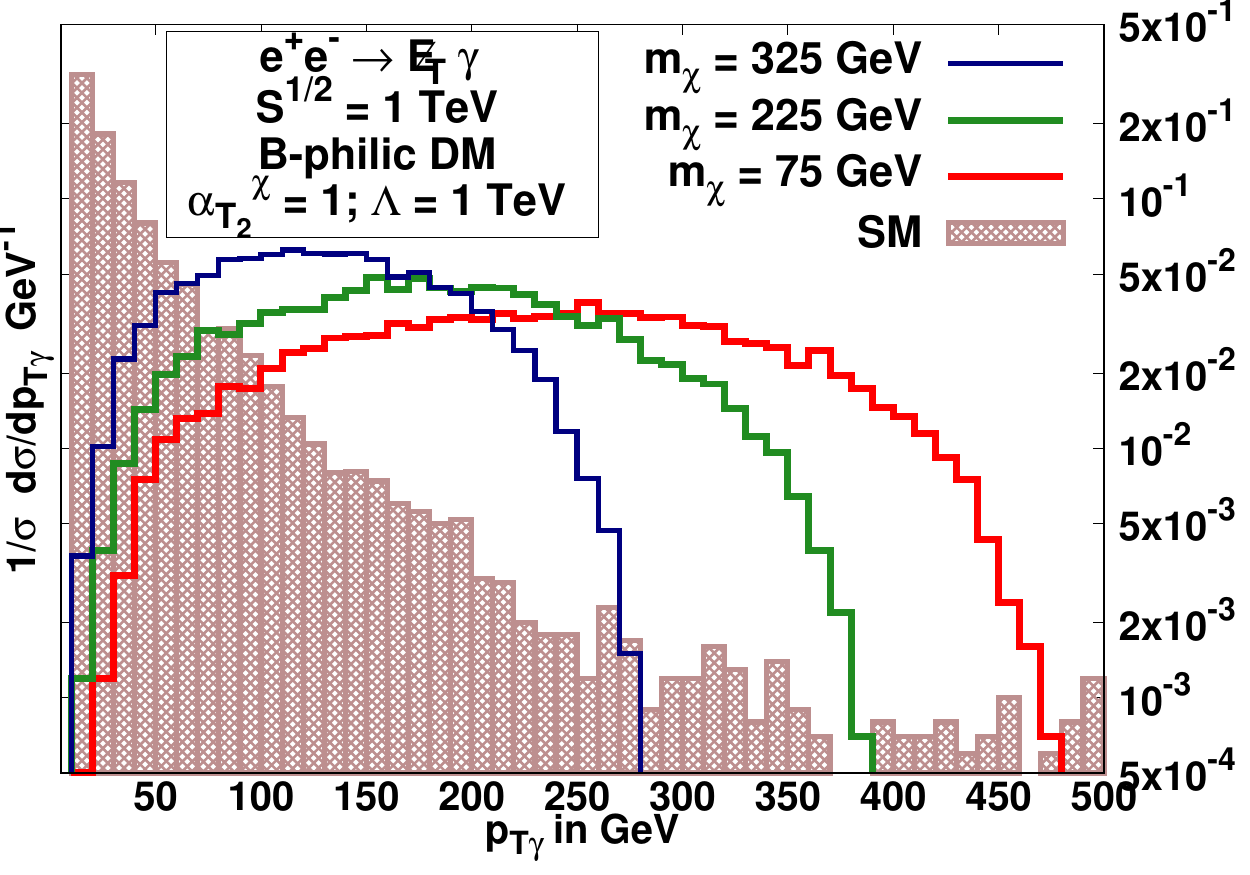}
 \subcaption{}\label{BfermiT2ptdist}
\includegraphics[width=0.43\textwidth,clip]{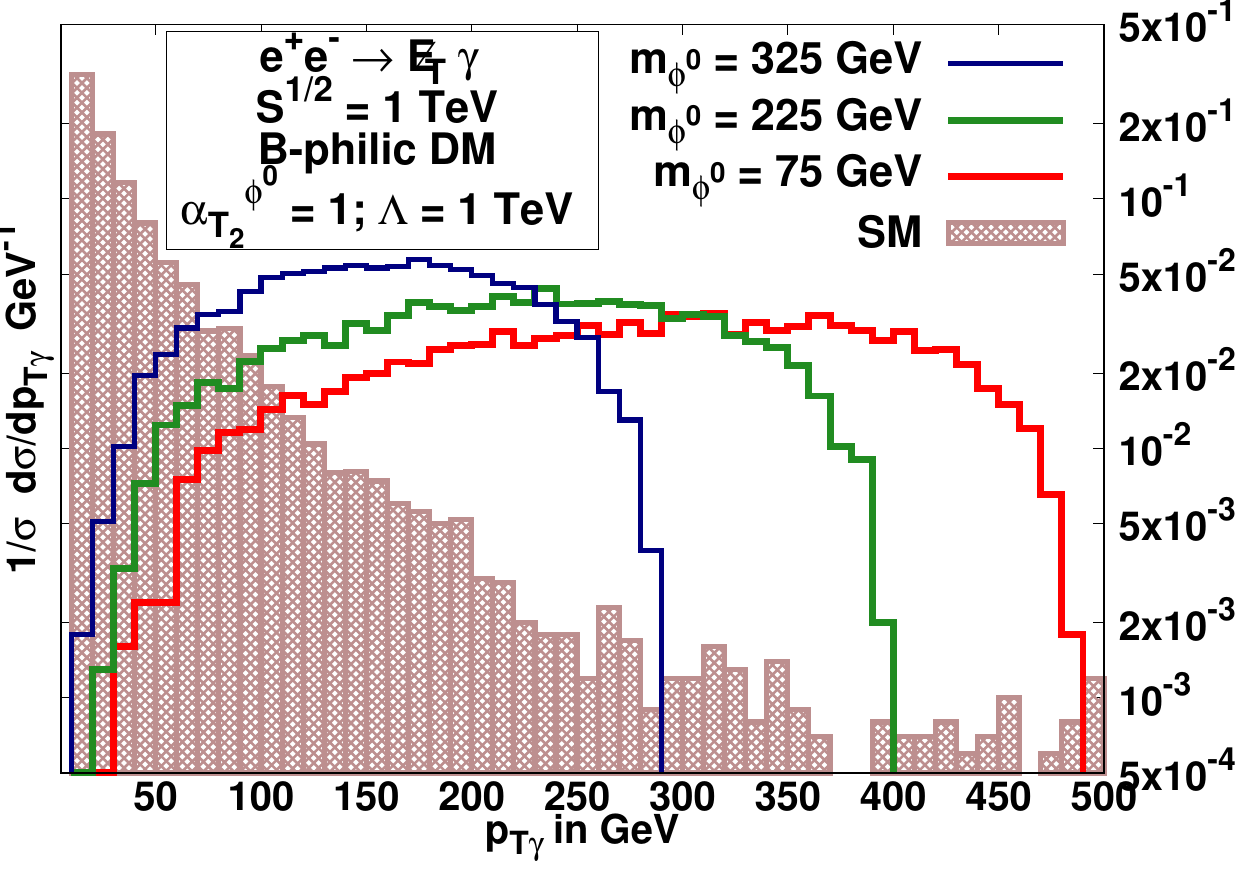}
 \subcaption{}\label{BscalT2ptdist}
\includegraphics[width=0.43\textwidth,clip]{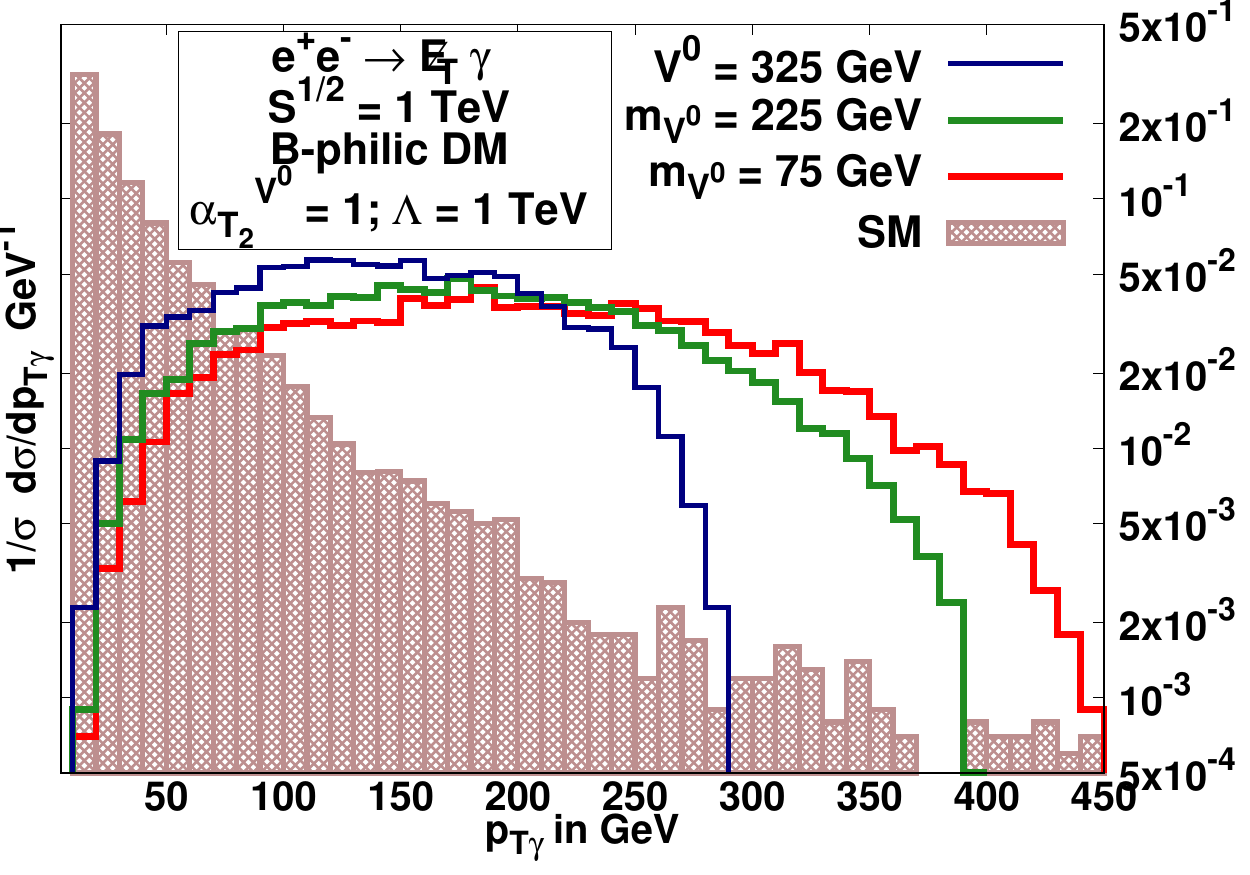}
\subcaption{}\label{BvecT2ptdist}
\columnbreak
\includegraphics[width=0.43\textwidth,clip]{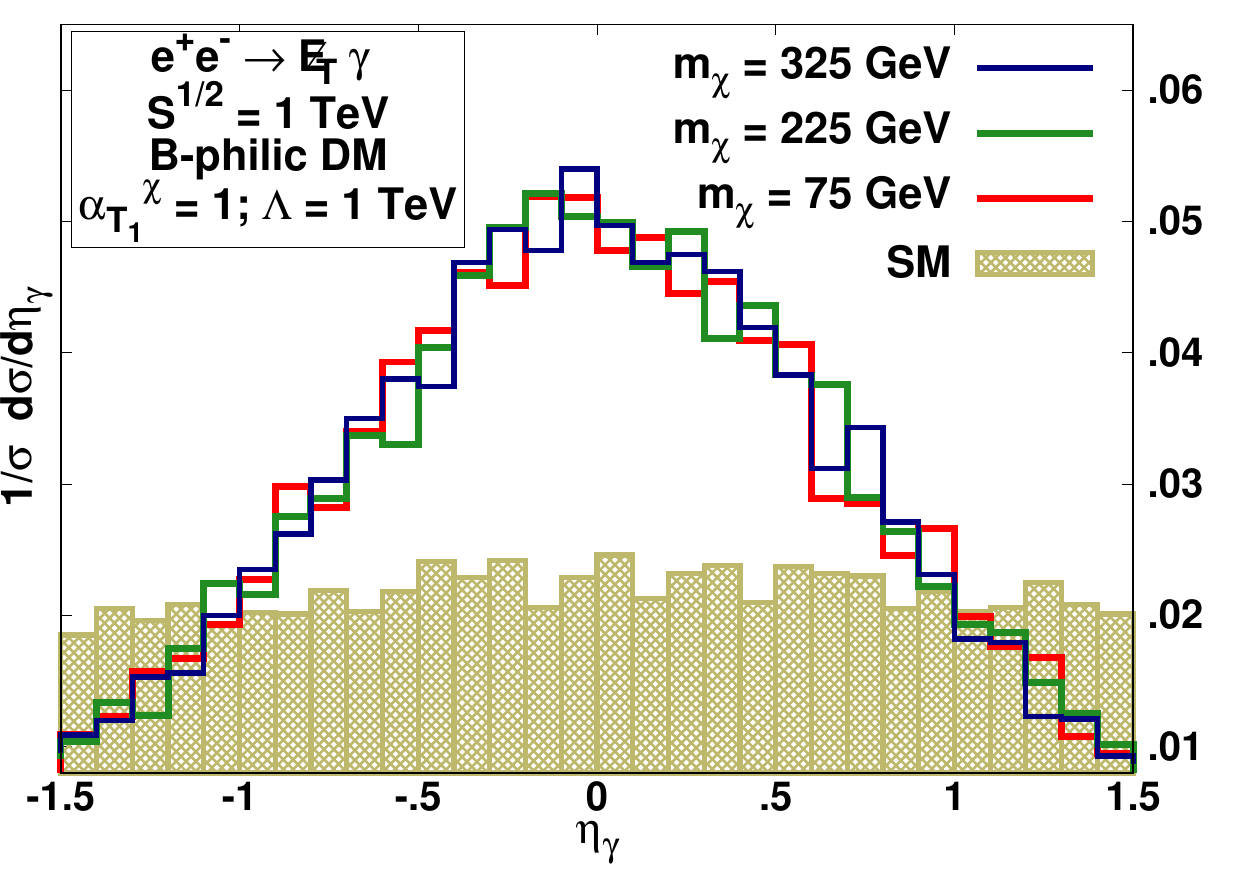}
 \subcaption{}\label{BfermiT1etadist}
\includegraphics[width=0.43\textwidth,clip]{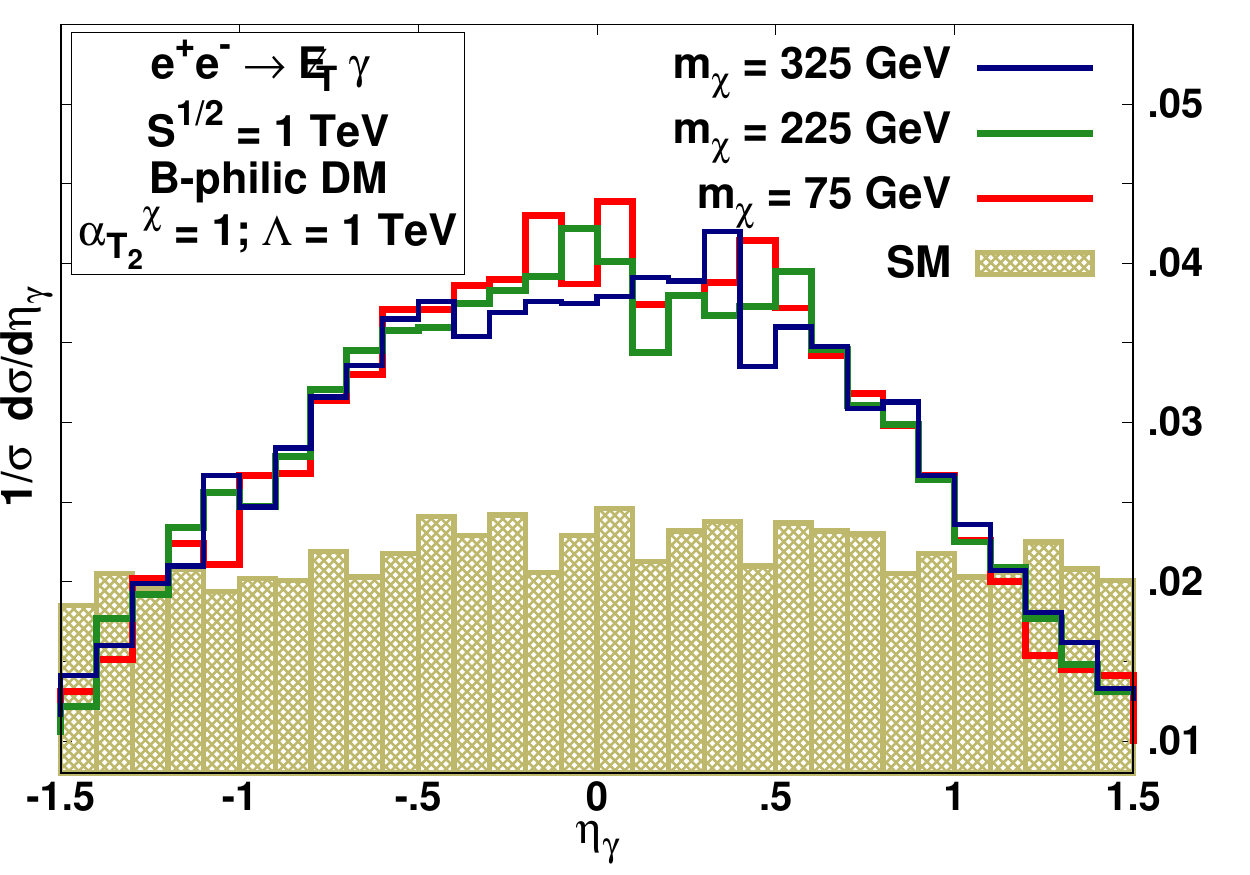}
\subcaption{}\label{BfermiT2etadist}
\includegraphics[width=0.43\textwidth,clip]{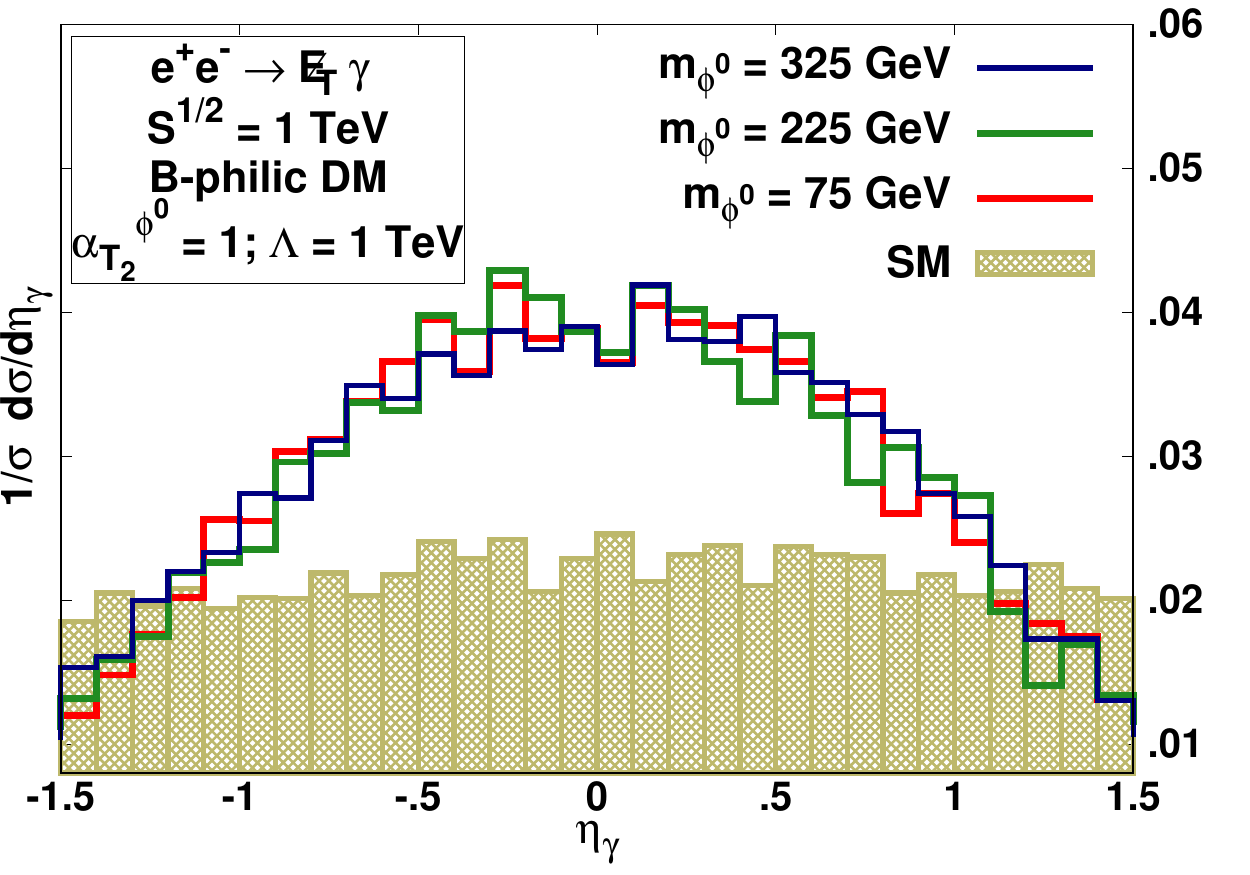}
 \subcaption{}\label{BscalT2etadist}
\includegraphics[width=0.43\textwidth,clip]{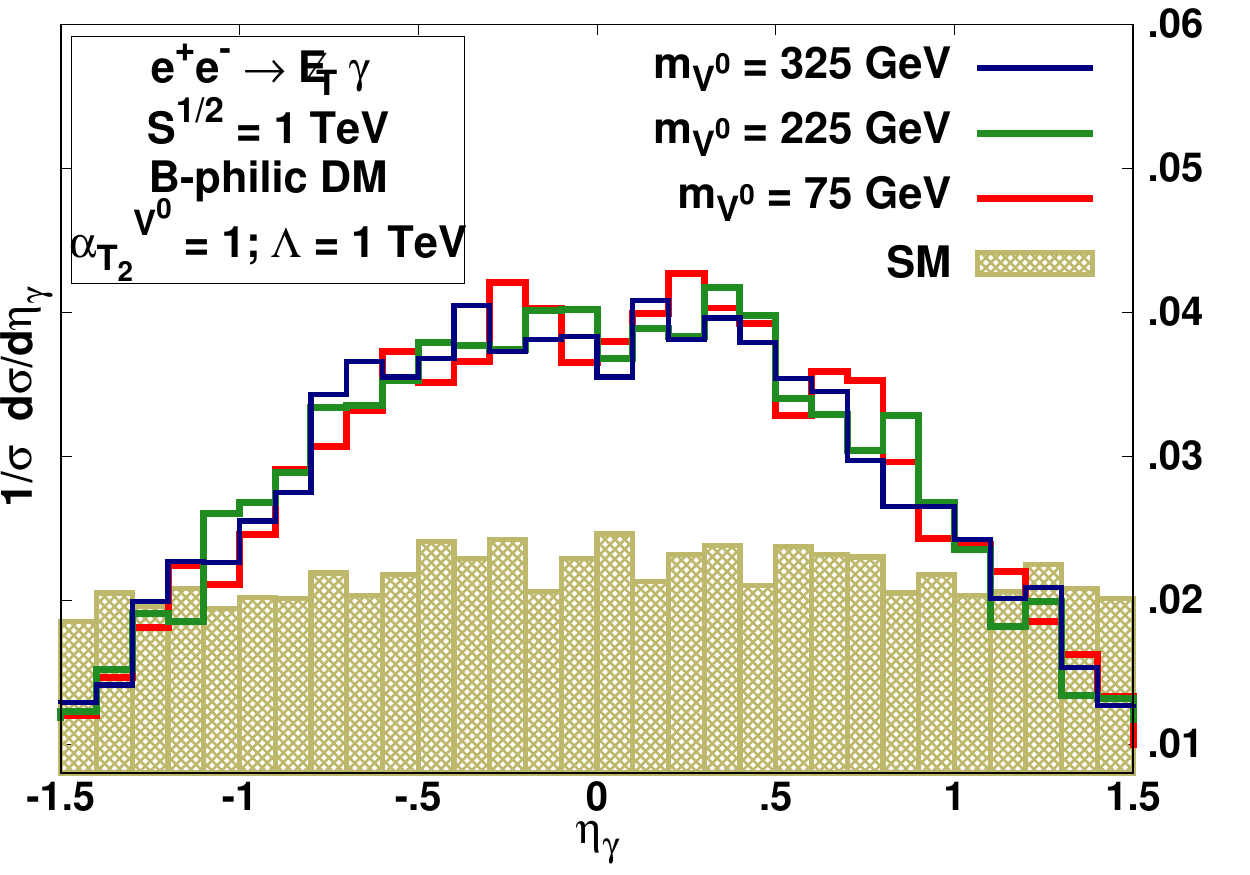}
  \caption{}\label{BvecT2etadist}
\end{multicols}
\caption{\small \em{Normalized 1-D differential  cross-sections   {\it w.r.t.}  $p_{T_\gamma}$ (bin width 10 GeV) and $\eta_\gamma $ (bin width 0.1) corresponding to the  SM processes (shaded histograms) and   those induced by  B-philic operators  at the three representative values of DM masses: 75, 225 and 325 GeV respectively.}}
        \label{fig:B_distrVpgs}
\vskip -2cm
\end{figure*}
\subsection{Constraints from LHC Observations}
Some of the interesting signatures of DM are missing energy $\not\!\!E_T$ on transverse plane with Jets, photons, $Z^0$ or any other visible SM particles. For lepto-philic and B-philic DM cases we consider $ p p \rightarrow \left(Z^0\to l^+\ l^-\right) + \not\!\! E_T$  and $p\ p\ \rightarrow\ \not\!\!E_T\ \gamma$ processes at the LHC respectively \cite{Kahlhoefer:2017dnp,Boveia:2018yeb}. The recent analysis of $p\, p\to l^+l^-l^+l^-$ at  $\sqrt{s}$ = 8 TeV for both CMS \cite{CMS:2012bw} and ATLAS \cite{Aad:2014wra} showed that the measured events  are consistent with SM and an integrated luminosity of 20.7 fb$^{-1}$. However, in reference \cite{Bell:2014tta} the authors have shown that the contribution of lepto-philic DM dimension six operators can be strongly constrained at $\sqrt{s}$ = 14 TeV  for the DM mass $\le$ 250 GeV. We believe the enhancement in the centre of mass energy and luminosity will enable both the detectors  to probe the sensitivity of the lepto-philic twist operators.  

\par Recently the ATLAS Collaboration \cite{Aaboud:2017dor} has reported
 the events containing an energetic photon and large missing transverse momentum for BSM searches at $\sqrt{s}=13$ TeV and an integrated luminosity of $36.1$ fb$^{-1}$ which agrees with SM predictions within the systematic and statistical uncertainty.  The SM background contribution  arises from $p\ p \rightarrow Z (\rightarrow \nu \bar{\nu})\ \gamma$, $p\ p \rightarrow W (\rightarrow l \nu)\ \gamma$, $p\ p \rightarrow Z (\rightarrow l\ l)\ \gamma$ and $p\ p \rightarrow \gamma + jets$ processes. The events containing fake photons coming from electrons and jets are also included in the analysis. The experimental observations constrain the contribution of the effective operators and put an upper bound on dis-allowed $\Lambda_{eff}$  for DM coupling fixed at unity and a given DM mass due to non-observation of any appreciable change in SM predicted Events.

\subsection{$\slash\!\!\! \!E_T$ + Mono-photon signals at ILC}
\begin{table*}\footnotesize
\centering
\begin{tabular*}{\textwidth}{c|@{\extracolsep{\fill}} ccccc|}\hline\hline
	&\textit{ILC-250}&\textit{ILC-500}&\textit{ILC-1000}\\
	$\sqrt{s} \left( \textit{in GeV}\right )$& 250 & 500 & 1000  \\
	$L_{int} \left( \textit{in $fb^{-1}$}\right )$ & 250 & 500 & 1000  \\
	$\sigma_{bg} \,(pb)$ &1.07 &  1.48 & 2.07 \\\hline\hline
\end{tabular*}
\caption{\small \em{Accelerator parameters as per Technical Design Report \cite{Behnke:2013lya, Behnke:2013xla}.  $\sigma_{bg}$ is the background cross section for $e^-\,e^+\,\rightarrow\,\sum \nu_i\,\bar{\nu}_i\,\gamma$ process computed using the  selection cuts defined in section \ref{basiccuts}}}
\label{table:accelparam}
\end{table*}


\begin{table*}\footnotesize
\centering
\begin{tabular}{c}
ILC Parameters: $\sqrt{s}$ = 1  TeV;  L = 1000  fb$^{-1}$\\\hline
\end{tabular}
\begin{tabular}{||p{2cm}|p{6cm}||p{6cm}||}\hline\hline
\textbf{Process:}&  $e^+\ e^- \rightarrow \not\!E \ \gamma^*\ (\gamma^*\rightarrow j j)$ & $e^+\ e^- \rightarrow \not\! E \ \gamma$\\
&&\\
 \textbf{Cuts:}&  $\left\vert m_{j_1\ j_2}\right\vert \le \sqrt{m_Z^2 - 5 \Gamma_Z m_Z}$  &   $\frac{2E_\gamma}{\sqrt{s}}\ \not\!\epsilon\ [0.98\ ,\ 0.99]$  \\
& $\not \!\! E_T \ge \sqrt{m_Z^2 + 5 \Gamma_Z m_Z} $ & \\
&&\\ \hline
\end{tabular}
\begin{tabular}{||p{2cm}|p{1.71cm}|p{1.71cm}|p{1.71cm}||p{1.7cm}|p{1.71cm}|p{1.71cm}||}\hline\hline
\underline{\bf Operators} & \underline{\bf 75 GeV} &\underline{\bf 225 GeV}  &\underline{\bf 325 GeV} &\underline{\bf 75 GeV}&\underline{\bf 225 GeV} &\underline{\bf 325 GeV}\\ 
${\mathbf {\cal O}^L_{F_{T_1}}}$&3.4&2.3&1.9&3.9&2.7&2.3\\
${\mathbf {\cal O}^L_{F_{T_2}}}$&8.2&3.5&2.3&9.7&4.2&2.8\\
${\mathbf {\cal O}^L_{S_{T_2}}}$&9.2&4.6&2.4&20.2&5.7&3.1\\
${\mathbf {\cal O}^L_{V_{T_2}}}$&135.5&12.2&4.3&175.4&16.1&5.8\\\hline
${\mathbf {\cal O}^B_{F_{T_1}}}$&1.8&1.1&0.7&2.5&0.9&0.6\\
${\mathbf {\cal O}^B_{F_{T_2}}}$&3.7&1.4&0.8&3.4&1.2&0.7\\
${\mathbf {\cal O}^B_{S_{T_2}}}$&10.2&2.4&1.1&9.1&2.1&0.9\\
${\mathbf {\cal O}^B_{V_{T_2}}}$&9.5&4.7&1.6&46.6&4.0&1.4\\\hline
\end{tabular}
\caption{\small \em{$3\sigma$ significance upper bound on $\Lambda_{\rm eff}$ in TeV with respective $\alpha_{T_i}$ fixed at unity for given three choices of $m_{\rm DM} \equiv$ 75, 225 and 325 GeV  respectively. }}
\label{table:3sigma}
\end{table*}

We consider following DM production processes along-with on/ off shell photon at the proposed ILC,  
for the DM mass range  $\sim$ 50 - 500 GeV: (a) $e^+\,e^-\,\rightarrow \,\chi\,\bar{\chi}\,\gamma/\,\gamma^\star$,  (b) $e^+\,e^-\,\rightarrow \,\phi^0\,\phi^0\,\gamma /\,\gamma^\star$,  and (c) $e^+\,e^-\,\rightarrow \,V^0\,V^0\,\gamma /\,\gamma^\star$. The  dominant SM background  for $e^+e^-\to \not \!\! E_T + \gamma/\,\gamma^\star$  signature comes from $Z^0\gamma$ production process: $e^+\,e^-\,\rightarrow\,Z^0+\gamma/\,\gamma^\star \to \sum \nu_i\,\bar{\nu}_i + \gamma/\,\gamma^\star$. 
\par The analysis for the background and the signal processes corresponding to the accelerator parameters as conceived  in the {\it Technical Design Report for ILC} \cite{Behnke:2013lya, Behnke:2013xla} and given in Table \ref{table:accelparam} is performed by simulating SM backgrounds and the DM signatures using Madgraph \cite{Alwall:2014hca} and  the model file generated by FeynRules \cite{Alloul:2013bka}. We impose the basic selection cuts  $\not \!\! E_T \ge \sqrt{m_Z^2 + 5 \Gamma_Z m_Z} $ and $\left\vert m_{j_1\ j_2}\right\vert \le \sqrt{m_Z^2 - 5 \Gamma_Z m_Z}$ to reduce the backgrounds for the process  $e^+\ e^- \rightarrow \not\!E \ \gamma^*\ (\gamma^*\rightarrow j j)$. On the same note, we impose the following cuts to reduce the backgrounds for the DM pair production in association with mono-photon:
\begin{itemize}
 \item  Transverse momentum of photon $p_{T_{\gamma}} \geq$ 10 GeV,  
 \item  Pseudo-rapidity of photon is restricted as $\left\vert\eta_\gamma\right\vert\leq$ 2.5,
 \item  dis-allowed recoil photon energy against on-shell $Z^0$ \\ 
$ \frac{2\,E_\gamma}{\sqrt{s}}$\ \ \  $\not\!\epsilon\ \ \ \left[0.8,0.9\right]$,\   $\left[0.95,0.98\right]$ and  $\left[0.98,0.99\right]$    for $\sqrt{s}$ = 250 GeV, 500 GeV and 1 TeV respectively.
\label{basiccuts}
\end{itemize}

\par As a first step towards preliminary analysis we study the significance ${\cal S}$ for the DM production processes, defined as 
\bea
{\cal S}=\frac{N_{S}}{\sqrt{N_B+\left(\delta_{\rm sys} N_B\right)^2}}
\eea
where $N_S$ is the number  of DM  with mono-photon events, $N_B$ is
 the number of SM  background events and $\delta_{\rm sys}$  is the systematic error. We compare the three sigma significance of the DM pair production with associated with On/ Off-shell  photons, 
for three representative values of DM mass 75, 225,  and 325 GeV respectively. We give the kinematic reach of $\Lambda_{\rm eff}$ corresponding to all cases of scalar, fermionic and vector DM based on the $3\sigma$ efficiency  in table \ref{table:3sigma} and find that mono photon signatures gives the better kinematic reach of $\Lambda_{\rm eff}$   for a given coupling and mass of the DM. Therefore, we restrict our analysis for the DM pair production with associated with mono-photons. 

\par The  $3\,\sigma$ sensitivity contours in $\Lambda_{\rm eff}-m_{\rm DM}$ plane are drawn for the DM production cross-sections with $\alpha_{T_i}$ = 1 and  conservative $\delta_{\rm sys} \sim $ 1\% in figures \ref{lepto3sigma.25TeV}, \ref{B3sigma.25TeV}, \ref{lepto3sigma.5TeV},  \ref{B3sigma.5TeV}, \ref{lepto3sigma1TeV} and  \ref{B3sigma1TeV}. Figures   \ref{lepto3sigma.25TeV} and   \ref{B3sigma.25TeV} correspond  to lepto-philic and B-philic operators respectively for the proposed ILC at $\sqrt{s}$ = 250 GeV at an integrated luminosity of 250 fb$^{-1}$, figures   \ref{lepto3sigma.5TeV} and   \ref{B3sigma.5TeV} correspond  to lepto-philic and B-philic operators respectively for the proposed ILC at $\sqrt{s}$ = 500 GeV at an integrated luminosity of 500 fb$^{-1}$, and figures \ref{lepto3sigma1TeV} and  \ref{B3sigma1TeV} respectively depict the same for $\sqrt{s}$ = 1 TeV at an integrated luminosity of 1 ab$^{-1}$. The shaded region of parameter space associated with each contour can be explored by the proposed collider at ${\cal S}\ge 3$.  Thus, we get the kinematic reach on the cut-off scale $\Lambda_{\rm  eff}$  at ILC for all relevant twist-2 lepto-philic and $U(1)_Y$  gauge Boson B-philic induced DM operators.

\subsection{ Differential Cross-sections and $\chi^2$ Analysis}
The photon transverse momentum ($p_{T_\gamma}$) and photon
 pseudo-rapidity ($\eta_\gamma$) are found to be  most sensitive kinematic
 observables for the process $e^+e^-\to \slash\!\!\! \! E_T + \gamma$. 
To study the shape profile and its mass dependence we generate the  normalized one dimensional distribution for the SM background processes and signals  for the fermionic, real scalar and real vector DM candidates, keeping  the respective effective coupling constant  to be unity and rest to zero. We plot the normalized differential cross-sections for fixed Cut-Off scale $\Lambda_{\rm  eff}$ = 1 TeV  {\it w.r.t.}  $p_{T_\gamma}$ and  $\eta_\gamma$ induced by lepto-philic (i) Type-1  fermionic DM operators  in figures \ref{fermiT1ptdist} and \ref{fermiT1etadist} respectively, (ii)   Type-2 fermionic DM operators  in figures  \ref{fermiT2ptdist} and \ref{fermiT2etadist}  respectively, respectively, (ii)   Type-2 Scalar DM operators  in figures  \ref{scalT2ptdist} and \ref{scalT2etadist}  respectively, and  (iv) Type-2 vector  DM operators in figures \ref{vecT2ptdist} and \ref{vecT2etadist} respectively. Each panel depict  three shape profiles of the differential distribution corresponding to three choices of  DM masses  75, 225 and 325 GeV respectively. Shaded rosy-brown and dark khaki histograms  depict the normalized differential distributions {\it w.r.t.}  $p_{T_\gamma}$ in figures  \ref{fermiT1ptdist}, \ref{fermiT2ptdist}, \ref{scalT2ptdist},  \ref{vecT2ptdist}  and {\it w.r.t.}  $\eta_\gamma$  in figures  \ref{fermiT1etadist}, \ref{fermiT2etadist}, \ref{scalT2etadist},  \ref{vecT2etadist} respectively for the background processes.
\begin{figure*}
 	\centering
\begin{multicols}{2}
\includegraphics[width=0.49\textwidth,clip]{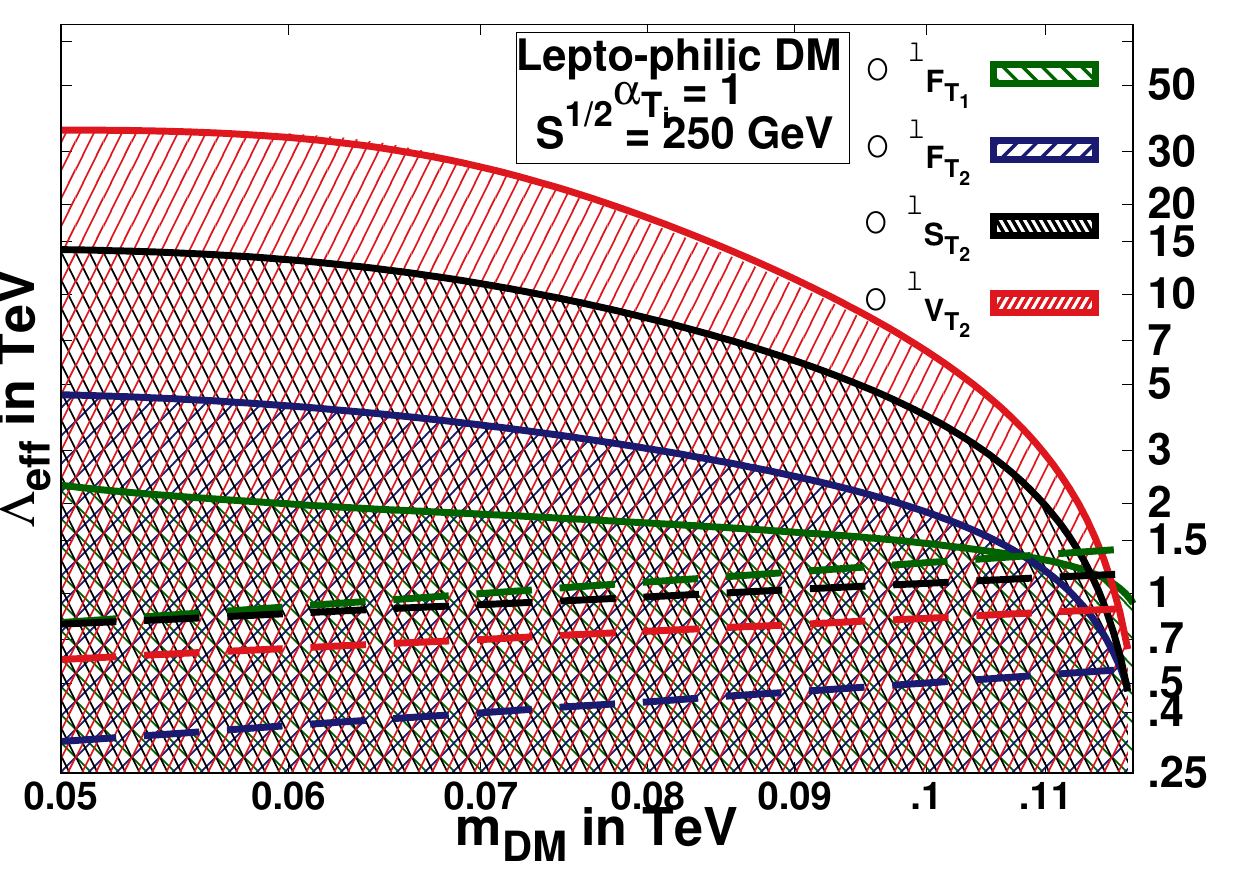}
 \subcaption{}\label{Lchisqr3sigma.25TeV}
\includegraphics[width=0.49\textwidth,clip]{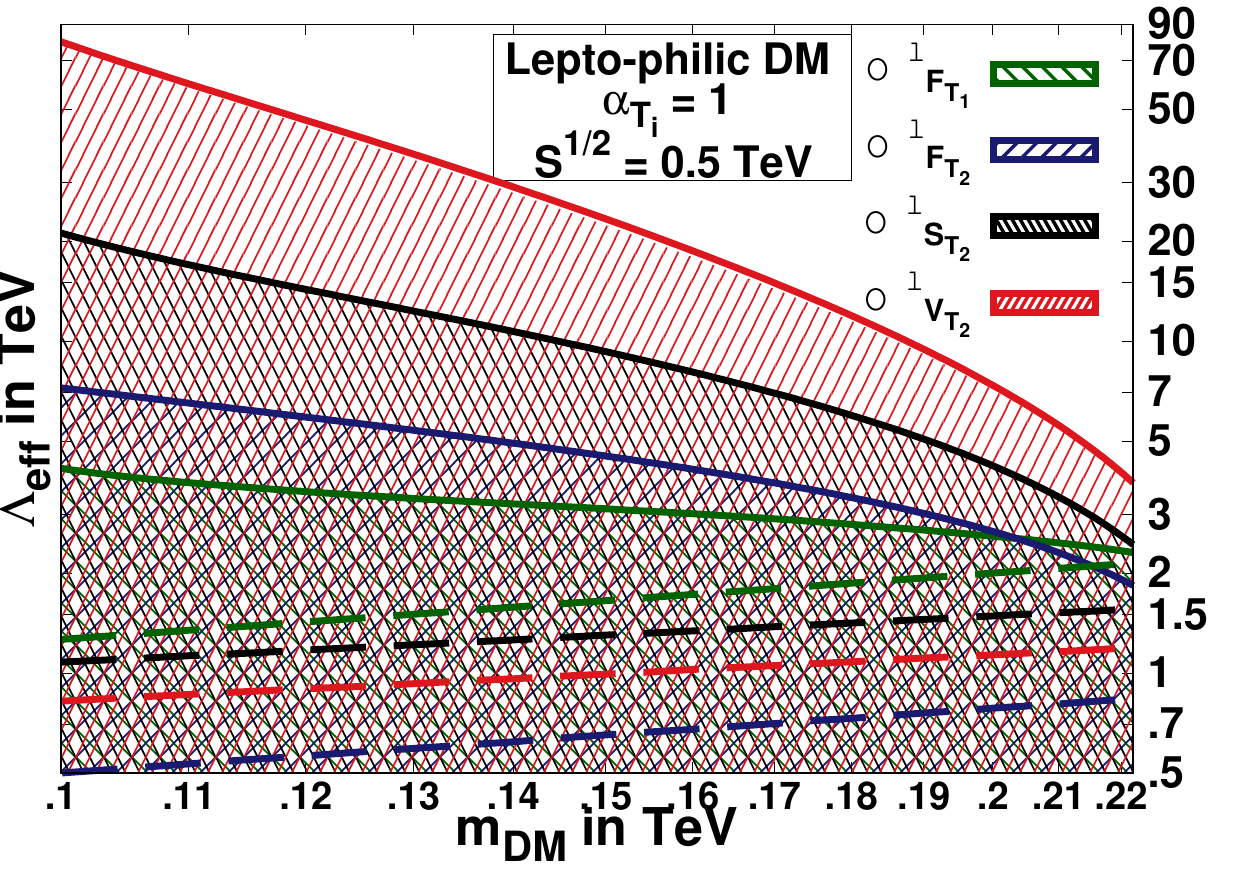}
\subcaption{}\label{Lchisqr3sigma.5TeV}
\includegraphics[width=0.49\textwidth,clip]{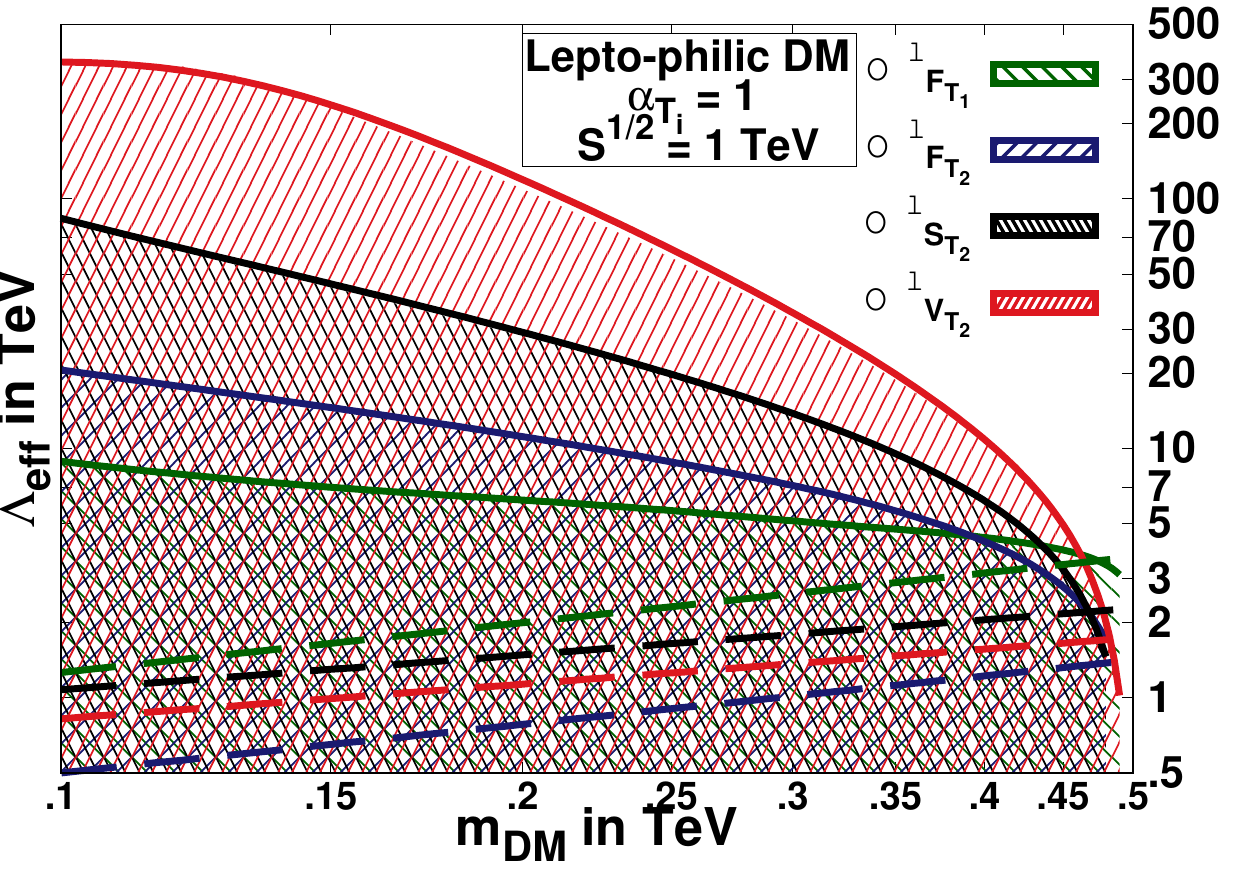}
\subcaption{}\label{Lchisqr3sigma1TeV}
\columnbreak
\includegraphics[width=0.49\textwidth,clip]{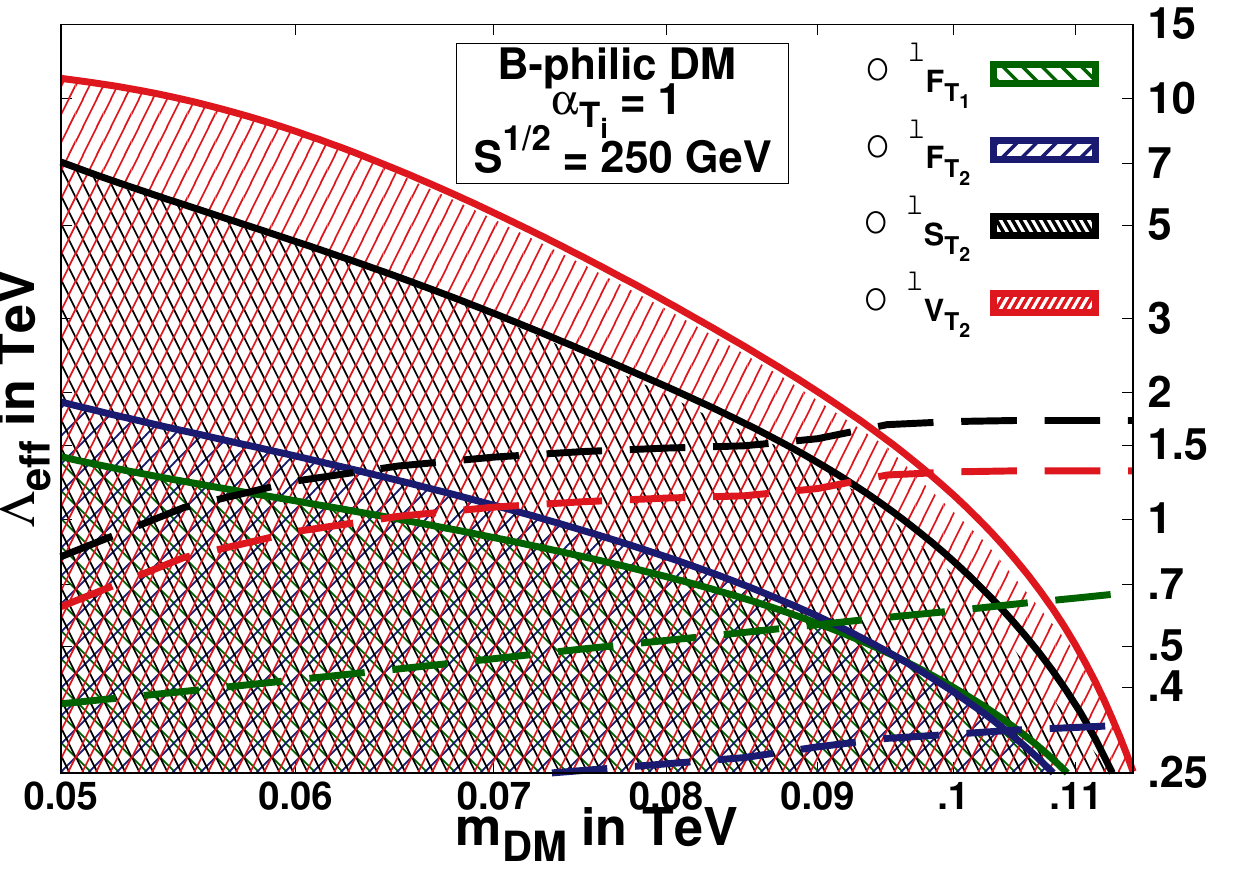}
\subcaption{}\label{Bchisqr3sigma.25TeV}
\includegraphics[width=0.49\textwidth,clip]{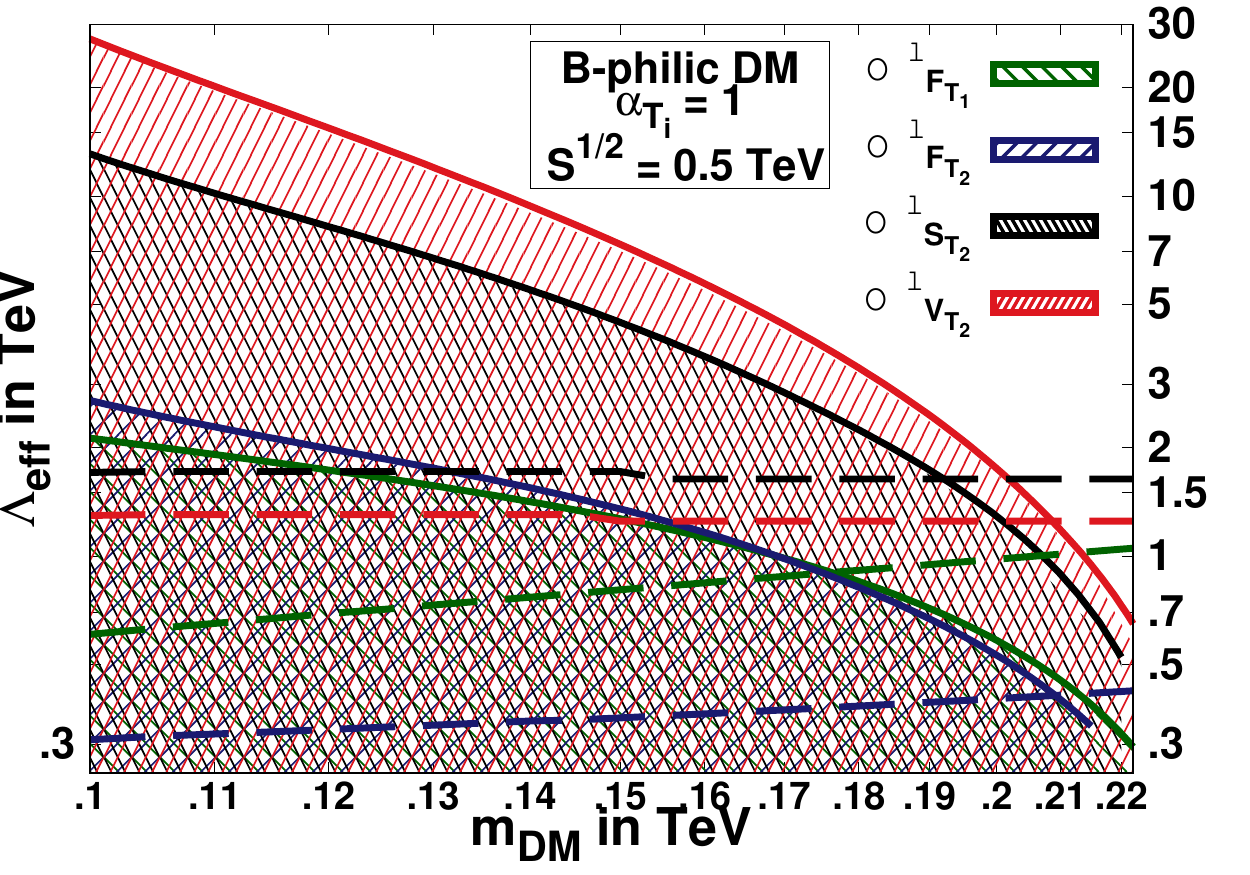}
\subcaption{}\label{Bchisqr3sigma.5TeV}
\includegraphics[width=0.49\textwidth,clip]{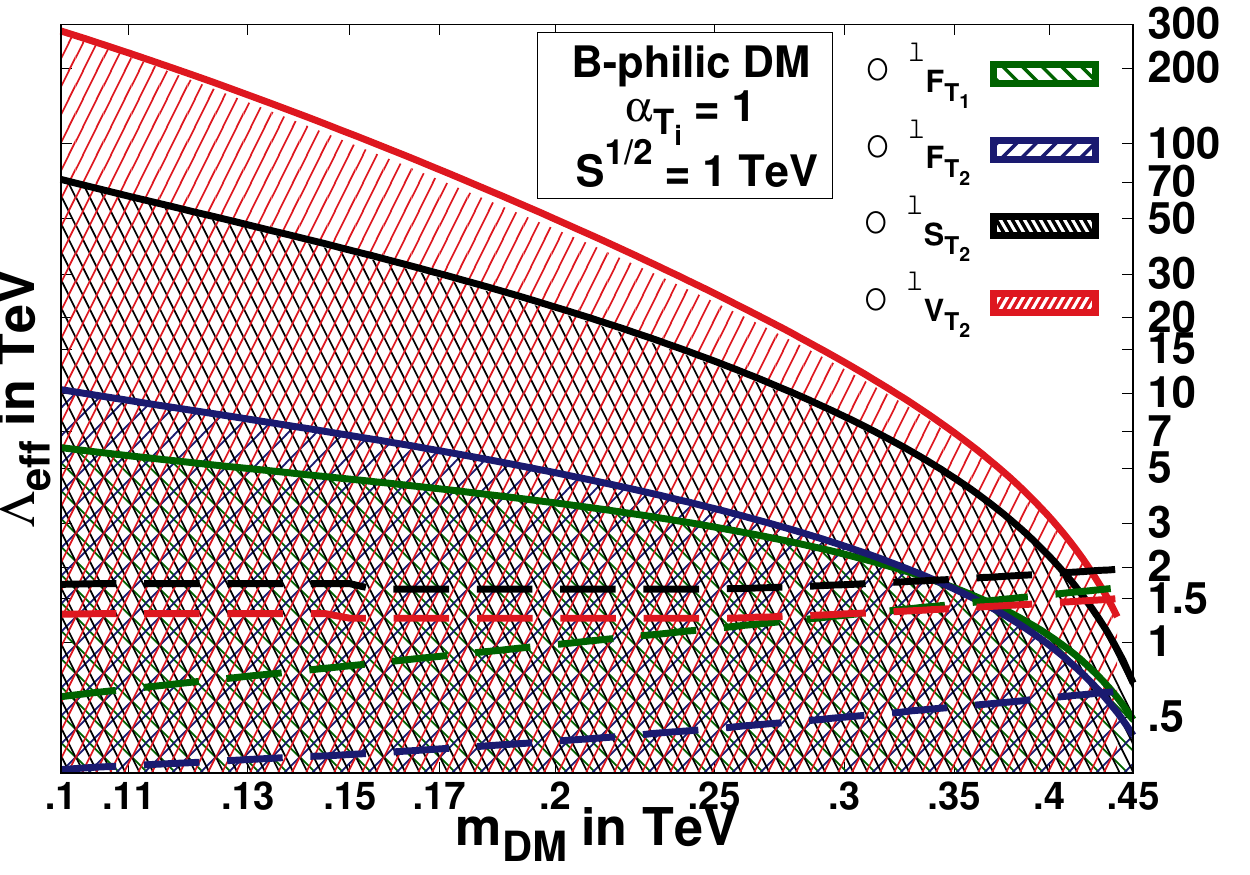}
\subcaption{}\label{Bchisqr3sigma1TeV}
\end{multicols}
	\caption{\small \em{Solid lines depict $3\sigma$ with 99.73 \% C.L. contours in the $m_{\rm DM}-\Lambda_{\rm  eff}$  plane from the $\chi^2$ analyses of the $e^+e^-\to \slash\!\!\! \!E_T +\gamma$ signature at the proposed ILC  designed for (i) $\sqrt{s}$ = 250 GeV with an integrated luminosity 250 fb$^{-1}$ in figures \ref{Lchisqr3sigma.25TeV} and \ref{Bchisqr3sigma.25TeV}, (ii) $\sqrt{s}$ = 500 GeV with an integrated luminosity 500 fb$^{-1}$ in figures \ref{Lchisqr3sigma.5TeV} and \ref{Bchisqr3sigma.5TeV}, and (iii) $\sqrt{s}$ = 1 TeV with an integrated luminosity 1 ab$^{-1}$ in figures \ref{Lchisqr3sigma1TeV} and \ref{Bchisqr3sigma1TeV} respectively. Each figure contain four contours corresponding to the twist-2 Type-1 fermionic and Type-2 fermionic, scalar and vector operators respectively. The enclosed shaded region corresponding to the respective contour is accessible for discovery with $\ge$ 99.73\% C.L.  The regions below the  colored dashed lines corresponding to respective four operators satisfy  the relic density constraint $\Omega_{\rm DM}h^2 \le $ 0.1199 $\pm$ 0.0022.}}
\label{fig:chisq}
\end{figure*}
%
\par Repeating the same exercise for the B-philic operators, we depict the shape profile of the normalised differential distributions {\it w.r.t.}  $p_{T_\gamma}$ and  $\eta_\gamma$ corresponding to three choices of DM masses 75, 225 and 325 GeV induced by  Type-1 fermionic, Type-2 fermionic, scalar and vector DM  {\it w.r.t.} $p_{T_\gamma}$ in figures  \ref{BfermiT1ptdist}, \ref{BfermiT2ptdist}, \ref{BscalT2ptdist}, \ref{BvecT2ptdist} and {\it w.r.t.} $\eta_\gamma$  in figures \ref{BfermiT1etadist}, \ref{BfermiT2etadist}, \ref{BscalT2etadist}, \ref{BvecT2etadist}     respectively. The differential distributions for the background SM processesses \textit{w.r.t. } $p_{T_\gamma}$ and $\eta_{\gamma}$ are depicted in all of these figures and are shown in shaded rosy-brown and dark khaki histograms respectively.

\par  We note that $p_{T_\gamma}^{max}$ decreases with increase in DM mass.  The  shape of normalized distributions are  comparatively more sensitive  {\it w.r.t.} DM masses in case of the B-philic operators. This suggests that  for the $B$-philic operators induced  interactions, imposition of  DM mass dependent dynamical cut can  minimize the background and enhance the significance.

\par However, to enhance the sensitivity of the $\Lambda_{\rm  eff}$ at a fixed coupling $\alpha_{T_i}$ = 1 {\it w.r.t.} DM masses, we  compute the $\chi^2$  with the double differential distributions of  kinematic observables  $p_{T_\gamma}$ and $\eta_\gamma$ corresponding to  the background and signal processes for  (i) 50 GeV $\le m_{\rm DM} \le$ 125 GeV  at $\sqrt{s} $ = 250 GeV and an integrated luminosity of 250 fb$^{-1}$, (ii)  100 GeV $\le m_{\rm DM} \le$ 250 GeV  at $\sqrt{s} $ = 500 GeV and an integrated luminosity of 500 fb$^{-1}$,  and (iii)  100 GeV $\le m_{\rm DM} \le$ 500 GeV at $\sqrt{s} $ = 1 TeV and an integrated luminosity of 1 ab$^{-1}$.
The $\chi^2$ is   defined as

\begin{eqnarray}
	\chi^2&\equiv&\chi^2 \left(m_{\rm DM},\, \alpha_{T_i},\,\Lambda_{\rm  eff} \right)\nn\\
	&&\hskip -1cm =\sum_{j=1}^{n_1}\sum_{i=1}^{n_2} \left [ \frac{\frac{\Delta N_{ij}^{NP}}{\left(\Delta p_{T_\gamma}\right)_i\, \left(\Delta \eta_\gamma\right)_j}}{\sqrt{ \frac{\Delta N_{ij}^{SM+NP}} {\left(\Delta p_{T_\gamma}\right)_i\, \left(\Delta \eta_\gamma\right)_j} +\delta_{\rm sys}^2\left\{ \frac{\Delta N_{ij}^{SM+NP}}{\left(\Delta p_{T_\gamma}\right)_i\, \left(\Delta \eta_\gamma\right)_j}\right\}^2} }\right ]^2\nonumber\\
\end{eqnarray}

where $\Delta N_{ij}^{NP}$ and $\Delta N_{ij}^{SM+NP}$  are the number of differential New Physics 
 and  total  events respectively  in the two dimensional   $\left[\left(\Delta p_{T_\gamma}\right)_i-\left(\Delta \eta_\gamma\right)_j\right]^{\rm th}$ grid. Here $\delta_{\rm sys}$ represents
 the total systematic error in the measurement.

\par We consider only one effective operator at time with the fixed coupling constant of unity and adopted a conservative value for the systematic error to  be $1\%$. We simulate the two-dimension differential distributions using the collider parameters as given in Table \ref{table:accelparam} and choosing the basic selection cuts. In addition we
  impose DM mass dependent dynamical cuts to minimize the background for $U(1)_Y$  gauge Boson B-philic induced interactions, which translates into an acceptance dynamical cut on the photon energy
\bea
 E_\gamma\le \frac{s-4m_{\rm DM}^2}{2\sqrt{s}}.
\eea
\par We plot the $3\sigma $ contours at 99.73\% C.L. in the $m_{\rm DM}-\Lambda_{\rm  eff}$ for lepto-philic and B-philic operators in figures \ref{Lchisqr3sigma.25TeV} and \ref{Bchisqr3sigma.25TeV} respectively corresponding to $\sqrt{s}$ = 250 GeV with an integrated luminosity of 250 fb$^{-1}$, figures \ref{Lchisqr3sigma.5TeV} and \ref{Bchisqr3sigma.5TeV} correspond to $3\sigma $ contours at 99.73\% C.L. in the $m_{\rm DM}-\Lambda_{\rm  eff}$ for lepto-philic and B-philic operators for $\sqrt{s}$ = 500 GeV with an integrated luminosity of 500 fb$^{-1}$. We also give the $3\sigma $ contours at 99.73\% C.L. for an upgraded high luminosity ILC operating at $\sqrt{s}$ = 1 TeV with an integrated luminosity of 1 ab$^{-1}$ in figures \ref{Lchisqr3sigma1TeV} and \ref{Bchisqr3sigma1TeV} corresponding to interactions induced by lepto-philic and B-philic operators respectively. 

\par We observe that the kinematic reach of the $\Lambda_{\rm  eff} $ is enhanced 5-6 times in comparison to that obtained from the naive significance analysis. The B-philic operators showed better response to the $\chi^2$ analysis based on the double differential distributions which was expected from their one dimensional  distribution shown in figure \ref{fig:B_distrVpgs}.

\section{Summary and Outlook}
\label{sec:summary}
The recent constraints
derived from the observation on the dwarf spheroidal
satellite galaxies in Fermi-LAT \cite{Ackermann:2015zua, TheFermi-LAT:2015kwa,Fermi-LAT:2016uux}, excess in electron/positron channel around 10 GeV  at PAMELA \cite{Adriani:2013uda, Adriani:2008zr}, excess in flux  of electrons/positrons around 400-500  GeV at ATIC \cite{Panov:2006kf} and PPB-BETS \cite{PPB-BETS} balloon  experiments and  exclusion of quark channels  by AMS-02 data \cite{Aguilar:2014mma,Aguilar:2016kjl}  hints toward the existence of non-baryonic DM. This implies that the direct detection experiments have to be sensitive  on the  recoil momentum of the atom or an electron in DM - atom and/ or DM - electron scattering respectively due to suppressed loop-level interactions of DM with the quarks in the nucleon. Characterization for such  lepto-philic and electro-weak gauge Boson B-philic DM particles  are likely to be difficult and challenging at the LHC and therefore it becomes imperative to probe the sensitivity of the associated DM pair production channels at the proposed lepton collider ILC. Motivated by  these observations and restrictions,  we have explored the viable alternative stable non-baryonic spin 1/2, 0 and 1 DM particles $\sim$ 10 - 1000 GeV, contributing to the relic density through their super-weak interactions with twist-2 leptonic and $U(1)_Y$ gauge Boson currents in a model independent approach.  In this article, we have considered the super-symmetric and Extra-Dimensional models inspired effective second rank twist interactions  of the leptons and gauge Bosons with the spin 1/2, 0 and 1 DM candidates.  

We have listed a minimal set of the twist-2 operators  corresponding to  lepto-philic and $U(1)_Y$ gauge Boson tensor currents in section \ref{sec:eff_int} which couples to the tensor currents generated by the bi-linears of  the DM fields. These DM operators contribute 
\par We have analytically calculated the thermalized annihilation cross-sections for the fermionic, scalar and vector DM induced by lepto-philic \eqref{ThAvLannxsecT1} - \eqref{ThAvLannxsecV} and B-philic  \eqref{ThAvBannxsecT1} - \eqref{ThAvBannxsecV} operators respectively, which  are in agreement numerically  with that of  MadDM. The  relic density contours satisfying the PLANCK observations depict  the upper bound on $\Lambda_{\rm eff}$ for fixed coupling $\alpha_{T_i}$  = 1 in the $m_{\rm DM}-\Lambda_{\rm eff} $ as shown in figures \ref{leptophilicrelicdensity}, \ref{tauphilicrelicdensity} and \ref{Bphilicrelicdensity} for the lepto-philic, $\tau^\pm$-philic and B-philic DM interactions respectively.  Using these upper bounds on $\Lambda_{\rm eff}$ for a given $m_{\rm DM}$, we estimated the thermally averaged annihilation indirect detection cross-section for lepto-philic and $\tau^\pm$-philic or electro-philic in figures \ref{leptoindirect} and \ref{tauindirect}  respectively are compared with that obtained from Fermi-LAT \cite{Ackermann:2015zua, TheFermi-LAT:2015kwa,Fermi-LAT:2016uux}, while thermally averaged annihilation indirect detection cross-section for B-philic DM shown in figure \ref{Bindirect}  is compared with the observations from H.E.S.S. data \cite{Abramowski:2013ax}. We find that the present experimental limits in the respective searches not only favours the allowed parameter space from the relic density, but  also constraints the DM  model by providing the lower bound on the  ${\Lambda_{\rm eff}}_{\rm min}$ for a given DM mass at fixed coupling $\alpha_{T_i}$. 
\par We have computed the elastic DM - free electron direct detection scattering cross-section analytically only for lepto-philic induced interactions and depicted in figure \ref{fig:ddetection} as the $\tau^\pm$-philic and B-philic DM interactions do not have any tree level interactions either with the atom or the nucleon. Although  the contribution of the loops are suppressed but they need to bee investigated for the complete study of the twist-2 operators. On superimposing  inelastic DM - atom  scattering cross-section from DAMA \cite{Kopp:2009et}. XENON100T \cite{Aprile:2015ade}  we observe that the parameter space allowed by the relic density  is shrunk and we get a  conservative  lower limit on the cut-off  at fixed coupling $\alpha_{T_i}$ = 1 for a given DM mass. We have analysed the bound state effects of the electron and derived the analytical expressions for the event rate \cite{Kopp:2009et}.

\par Next, we probed and compared the 3-$\sigma$ efficiency of DM production processes  $e^+ e^- \to \,\,\not\!\!\!E_T +\,\gamma^\star\to \,\,\not\!\!\!E_T + l^+l^-$ and $e^+ e^- \to \,\,\not\!\!\!E_T +\,\gamma$ at ILC induced through twist-2 interactions of lepto-philic and B-philic interactions for $m_{\rm DM}\,\sim$ 50 - 400 GeV as shown in table \ref{table:3sigma}.  The $3\sigma$ significance contours for the dominannt  DM pair production in association with mono-photon  at 99.73\% C.L. are drawn {\it w.r.t.} SM background in figures \ref{lepto3sigma.25TeV} and \ref{B3sigma.25TeV} for lepto-philic and B-philic respectively at $\sqrt{s}$ = 250 GeV and an integrated luminosity of 250 fb$^{-1}$, in figures \ref{lepto3sigma.5TeV} and \ref{B3sigma.5TeV} for lepto-philic and B-philic respectively at $\sqrt{s}$ = 500 GeV and an integrated luminosity of 500 fb$^{-1}$ and in figures \ref{lepto3sigma1TeV} and \ref{B3sigma1TeV} respectively for $\sqrt{s}$ = 1 TeV and 1 ab$^{-1}$, with basic kinematic cuts in Table \ref{table:accelparam}. We improve the sensitivity of the $\Lambda_{\rm eff}$ by minimizing the $\chi^2$ using  the optimal variable technique on the 2-D distributions {\it w.r.t.} ${p_T}_\gamma$ and $\eta_\gamma$ for the three stages of the proposed collider (i)  at $\sqrt{s}$ = 250 GeV with an integrated luminosity of 250 fb$^{-1}$, (ii) at $\sqrt{s}$ = 500 GeV with an integrated luminosity of 500 fb$^{-1}$ and (iii) at $\sqrt{s}$ = 1 TeV with an integrated luminosity of 	1 ab$^{-1}$. The three sigma contours for $\chi^2$ analysis in $m_{\rm DM}-\Lambda_{\rm eff}$ plane are drawn in figures \ref{Lchisqr3sigma.25TeV} and \ref{Bchisqr3sigma.25TeV} corresponding to the lepto-philic and B-philic respectively for case (i) and similarly, contours corresponding to case (ii) and case (iii) are shown in figures \ref{Lchisqr3sigma.5TeV}, \ref{Bchisqr3sigma.5TeV} and \ref{Lchisqr3sigma1TeV}, \ref{Bchisqr3sigma1TeV} respectively.
\par We hope this study will be useful in studying the physics potential of the ILC in
context to dark matter searches.

\vskip 5mm
\begin{acknowledgements}
 HB and SD thank  Mihoko Nojiri and Mamta Dahiya for discussions and suggestions throughout the work. HB acknowledges the CSIR-JRF fellowship. HB and SD acknowledge the partial financial
 support from the CSIR grant No. 03(1340)/ 15/ EMR-II. SD thanks the Theory Division, KEK, for an excellent hospitality where this problem was conceived.
\end{acknowledgements}
\appendix
\section{Thermal averaged annihilation cross-sections}
\label{thermalAveragedCrosssection}

The fermionic, scalar and
 vector DM pair  annihilation cross sections  to SM $l^+l^-$ pairs of mass $m_l$ induced by the {\it lepto-philic} twist-2 operators are given by
\bea
&&\small \sigma^{\rm ann}_{T_1}\left(\chi\bar{\chi}\to l^+l^-\right) =\frac{\pi\alpha_{T_1}^2}{s}\,\left(\frac{m_\chi^2}{\Lambda_{\rm eff}^6}\right)\sqrt{\frac{s-4m_l^2}{s-4m_\chi^2}}\nonumber\\
&&\small \times\,\bigg[16\,m_l^2\,m_\chi^2+32\,m_\chi^4 +\left( 9\, m_l^2\,m_\chi^2-\frac{11}{3}\, m_l^4 + \frac{14}{3} \,m_\chi^4\right)\left\vert \vec  v\right\vert^2  \bigg]\nn\\
\label{LannxsecT1}\\
&&\small \sigma^{\rm ann}_{T_2}\left(\chi\bar{\chi}\to l^+l^-\right)  =\,\,\,\, \frac{\pi \alpha_{T_2}^2}{s}   \left(\frac{m_\chi^2}{\Lambda_{\rm eff}^6}\right) \sqrt{\frac{s-4m_l^2}{s-4m_\chi^2}}\,m_l^2\nn\\
&&\small\hskip 5cm \left(m_\chi^2-m_l^2\right)\left\vert \vec  v\right\vert^2 \label{LannxsecT2}\\
&&\small \sigma^{\rm ann}_{S}\left(\phi^0\,{\phi^0}\to l^+l^-\right) =\frac{\pi \alpha_{\phi^0}^2}{s} \left(\frac{m_{\phi^0}^2}{\Lambda_{\rm eff}^4}\right) \sqrt{\frac{s-4m_l^2}{s-4m_{\phi^0}^2}}\nn\\
&&\small \times\,\, \left[2\,m_l^2\,-\frac{2m_l^4}{m_{\phi^0}^2} + \left(\frac{4}{3}\frac{m_l^4}{m_{\phi^0}^2}\,+\frac{11}{6} \,m_l^2\,+\frac{16}{3} \,m_{\phi^0}^2 \right) \left\vert \vec  v\right\vert^2   \right]\nn\\
&&\label{LannxsecS}\\
&&\small \sigma^{\rm ann}_{V}\left(V^0\,{V^0}\to l^+l^-\right) =\frac{\pi \alpha_{V^0}^2}{s} \left(\frac{m_{V^0}^2}{\Lambda_{\rm eff}^4}\right)\, \sqrt{\frac{s-4m_l^2}{s-4m_{V^0}^2}}\nn\\
&&\small \times \,\, \left[\frac{2}{3}\,m_l^2-\frac{2}{3}\,\frac{m_l^4}{m_{V^0}^2}\,+\, \left( \frac{2}{9}\,\frac{m_l^4}{m_{V^0}^2}\,+\,\frac{5}{6}\,m_l^2\,+\,\frac{16}{9} m_{V^0}^2 \right)\left\vert \vec  v\right\vert^2   \right]\nn\\
\label{LannxsecV}
\eea

The  fermionic, scalar and
 vector DM pair  annihilation cross sections  to photon pairs induced by the {\it $U(1)_Y$ Boson B-philic} twist-2 operators are given by
\bea
&&\small \sigma^{\rm ann}_{T_1}\left(\chi\bar{\chi}\to \gamma\,\gamma\right)=\frac{8\pi\alpha_{T_1}^2}{3s}\, \cos^2\theta_{\rm W}\,\frac{m_\chi^6}{\Lambda_{\rm eff}^6}\sqrt{\frac{s}{s-4m_\chi^2}} \,\, \,\left\vert \vec  v\right\vert^2\nn\\
&&\label{BannxsecT1}\\
&&\small\sigma^{\rm ann}_{T_2}\left(\chi\bar{\chi}\to \gamma\,\gamma\right)=\frac{\pi \alpha_{T_2}^2}{s}\,\cos^2\theta_{\rm W}\,\frac{m_\chi^6}{\Lambda_{\rm eff}^6}\sqrt{\frac{s}{s-4m_\chi^2}} \, \,\,\left\vert \vec  v\right\vert^2\nn\\
&&\label{BannxsecT2}\\
&&\small\sigma^{\rm ann}_{S} \left(\phi^0\,{\phi^0}\to \gamma\,\gamma\right)= \frac{2\pi \alpha_{\phi^0}^2}{s} \,\cos^2\theta_{\rm W}\,\frac{m_{\phi^0}^4}{\Lambda_{\rm eff}^4}\nn\\
&&\hskip 3cm \small\sqrt{\frac{s}{s-4m_{\phi^0}^2}} \left[ \,1\,+\, \frac{2}{3}\,\,\, \left\vert \vec  v\right\vert^2 \right]\label{BannxsecS} \\
&&\small\sigma^{\rm ann}_{V}\left(V^0\,{V^0}\to \gamma\,\gamma\right) = \frac{2\pi \alpha_{V^0}^2}{3 s} \,\cos^2\theta_{\rm W}\,\frac{m_{V^0}^4}{\Lambda_{\rm eff}^4}\nn\\
&&\hskip 3cm \small\sqrt{\frac{s}{s-4m_{V^0}^2}} \small\left[ \,1\,+\, \left\vert \vec  v\right\vert^2 \right] \label{BannxsecV}
\eea
\noindent where $\theta_{\rm W}$ is the Weinberg mixing angle.
\par The DM relic density    is given in terms of
 thermally averaged  DM annihilation cross sections $\left\langle\sigma_{ann}\left\vert\vec v\right\vert\right\rangle$  in equation \eqref{boltzmann7}. To compute the same we  express the relative velocity of DM pair in the laboratory frame $\left\vert \vec  v\right\vert$  in terms of c.m. energy $\sqrt{s}$ as
\bea
\left\vert \vec v\right\vert =\frac{\sqrt{s\, (s-4\,m_{\rm DM}^2)}}{s-2\, m_{\rm DM}^2}.
\eea
Since $v<<c$, for non-relativistic DM we expand  $s\,=\,4\,m_{\rm DM}^2+m_{\rm DM}^2\,\left\vert \vec  v\right\vert^2+\frac{3}{4}\,m_{\rm DM}^2\,\left\vert \vec  v\right\vert^4+\mathcal{O}\left(\left\vert \vec  v\right\vert^6\right)$ and compute  the thermally averaged annihilation cross sections
 for lepto-philic and $U(1)_Y$  gauge Boson-philic  DM respectively.
\par Thermal averaged  annihilation cross-sections corresponding to the cross-sections given in equations \eqref{LannxsecT1} - \eqref{LannxsecV} for the lepto-philic operators are given respectively as
\begin{eqnarray}
&&\small\left\langle\sigma_{T_1}^{\rm ann} \left\vert \vec  v\right\vert\right\rangle \left(\chi\bar{\chi}\to l^+l^-\right)=\pi \alpha_{T_1}^2\,\left(\frac{m_\chi^4}{\Lambda_{\rm eff}^6}\right)\sqrt{1-\frac{m_l^2}{m_\chi^2}}\nn\\
&&\hskip 3cm  \small\times\left[ 1+ \frac{m_l^2}{2m_\chi^2} -\frac{34}{96} \frac{6}{x_F} \right]\label{ThAvLannxsecT1} \\
&&\small\left\langle\sigma_{T_2}^{\rm ann} \left\vert \vec  v\right\vert \right\rangle\left(\chi\bar{\chi}\to l^+l^-\right)=\frac{\pi \alpha_{T_2}^2}{2}\left(\frac{m_l^2m_\chi^2}{\Lambda_{\rm eff}^6}\right)\nn\\
&&\hskip 3cm \small\times\left[1-\frac{m_l^2}{m_\chi^2}\right]^{\frac{3}{2}} \frac{6}{x_F}\label{ThAvLannxsecT2} \\
&&\small\left\langle\sigma_{S}^{\rm ann} \left\vert \vec  v\right\vert \right\rangle \left(\phi^0\,{\phi^0}\to l^+l^-\right)= \pi\alpha_{\phi^0}^2\left(\frac{m_{\phi^0}^2}{\Lambda_{\rm eff}^4}\right)\sqrt{1-\frac{m_l^2}{m_{\phi^0}^2}}\nn\\
&&\   \small\times\left[\, \frac{m_l^2}{m_{\phi^0}^2}- \frac{m_l^4}{m_{\phi^0}^4} + \left( \frac{8}{3} + \frac{11}{12} \frac{m_l^2}{m_{\phi^0}^2}+ \frac{2}{3} \frac{m_l^4}{m_{\phi^0}^4}  \right) \frac{6}{x_F} \,\right]\nn\\
\label{ThAvLannxsecS}\\
 &&\small\left\langle\sigma_{V}^{\rm ann} \left\vert \vec  v\right\vert \right\rangle \left(V^0\,{V^0}\to l^+l^-\right)= \frac{8\pi\alpha_{V^0}^2}{9}\left(\frac{m_{V^0}^2}{\Lambda_{\rm eff}^4} \right) \sqrt{1-\frac{m_l^2}{m_{V^0}^2}}\nn\\
&&\small\times\left[ \frac{3}{8}\frac{m_l^2}{m_{V^0}^2}-\frac{3}{8}\frac{m_l^4}{m_{V^0}^4} + \left( 1+\frac{15}{32}\frac{m_l^2}{m_{V^0}^2}+\frac{1}{8}\frac{m_l^4}{m_{V^0}^4} \right)  \frac{6}{x_F} \right]\nn\\
\label{ThAvLannxsecV}
\end{eqnarray}
Similarly, the annihilation cross-sections given in equations \eqref{BannxsecT1}-\eqref{BannxsecV} for the $U(1)_Y$ gauge Boson B-philic operators are thermalized to give the following thermal averaged annihilation cross-sections:
\begin{eqnarray}
&&\small\left\langle\sigma_{T_1}^{\rm ann} \left\vert \vec  v\right\vert\right\rangle \left(\chi\bar{\chi}\to \gamma\,\gamma\right) =\frac{4\pi \alpha_{T_1}^2}{3}\, \cos^2\theta_{\rm W}\,\left(\frac{m_\chi^4}{\Lambda_{\rm eff}^6}\right)\,\frac{6}{x_F} \label{ThAvBannxsecT1}\\
&&\small\left\langle\sigma_{T_2}^{\rm ann} \left\vert \vec  v\right\vert \right\rangle \left(\chi\bar{\chi}\to \gamma\,\gamma\right)=\frac{4\pi \alpha_{T_2}^2}{2}\, \cos^2\theta_{\rm W}\,\left(\frac{m_\chi^4}{\Lambda_{\rm eff}^6}\right) \frac{6}{x_F} \label{ThAvBannxsecT2}\\
&&\small\left\langle\sigma_{S}^{\rm ann} \left\vert \vec  v\right\vert \right\rangle \left(\phi^0\,{\phi^0}\to \gamma\,\gamma\right)= \pi \alpha_{\phi^0}^2\, \cos^2\theta_{\rm W}\,\left(\frac{m_{\phi^0}^2}{\Lambda_{\rm eff}^4}\right)\nn\\
&&\hskip 5cm \small\left[\, 1+ \frac{1}{6} \, \frac{6}{x_F} \,\right] \label{ThAvBannxsecS}\\
&&\small\left\langle\sigma_{V}^{\rm ann} \left\vert \vec  v\right\vert \right\rangle \left(V^0\,{V^0}\to \gamma\,\gamma\right)= \frac{\pi \alpha_{V^0}^2}{3}\, \cos^2\theta_{\rm W}\,\left(\frac{m_{V^0}^2}{\Lambda_{\rm eff}^4} \right)\nn\\
&&\hskip 5cm \small\left[ 1+ \frac{1}{2}\,  \frac{6}{x_F} \right] \label{ThAvBannxsecV}
\end{eqnarray}



\end{document}